\documentclass[twoside]{article}
\usepackage{qic}

\newcommand{\beq}{\begin{eqnarray}}
\newcommand{\eeq}{\end{eqnarray}}

\newcommand{\pd}[2]{\frac{\partial^2 {#1}}{\partial {#2}^2}}

\begin{document}
\bibliographystyle{unsrt}
\setlength{\textheight}{8.0truein}    

\runninghead{Electrode Configurations  $\ldots$}
            {J.P.Home and A.M.Steane}

\normalsize\textlineskip
\thispagestyle{empty}
\setcounter{page}{1}

\copyrightheading{0}{0}{2003}{000--000}

\vspace*{0.88truein}

\alphfootnote

\fpage{1}

\centerline{\bf
ELECTRODE CONFIGURATIONS FOR FAST SEPARATION OF TRAPPED IONS}
\vspace*{0.37truein}
\centerline{\footnotesize
J. P. HOME}
\vspace*{0.015truein}
\centerline{\footnotesize\it Centre for Quantum Computation, 
Department of Atomic and Laser Physics, Clarendon Laboratory, Parks Road}
\baselineskip=10pt
\centerline{\footnotesize\it Oxford, OX1 3PU, England}
\vspace*{10pt}
\centerline{\footnotesize 
A. M. STEANE}
\vspace*{0.015truein}
\centerline{\footnotesize\it Centre for Quantum Computation, 
Department of Atomic and Laser Physics, Clarendon Laboratory, Parks Road}
\baselineskip=10pt
\centerline{\footnotesize\it Oxford, OX1 3PU, England}
\vspace*{0.225truein}
\publisher{(received date)}{(revised date)}

\vspace*{0.21truein}

\abstracts{
We study the problem of designing electrode structures that allow
pairs of ions to be brought together and separated rapidly in
an array of linear Paul traps. We show that it is desirable for the
electrode structure to produce a d.c. octupole
moment with an a.c. radial quadrupole. For the case where electrical
breakdown limits the voltages that can be applied,
we show that the octupole is more demanding than the 
quadrupole when the characteristic distance scale
of the structure is larger than 1 to 10 microns (for typical materials).
We present a variety of approaches and optimizations of structures
consisting of one to three layers of electrodes. The three-layer
structures allow the fastest operation at given distance $\rho$ from
the trap centres to the nearest electrode surface, but 
when the total thickness $w$ of
the structure is constrained, leading to $w < \rho$, then
two-layer structures may be preferable.
}{}{}

\vspace*{10pt}

\vspace*{3pt}
\communicate{to be filled by the Editorial}

\vspace*{1pt}\textlineskip 


\section{Introduction}
\noindent

The application of ion traps to quantum computing \cite{98:WinelandB, 97:Steane, 02:Sasura} has
lead to interest in the construction of systems of multiple ion traps
situated close together. Experiments have shown that quantum
logic gates with good fidelity can be produced between a single pair
of ions in a single trap, and such methods can be extended to a few
ions in a single trap \cite{03:SchmidtKaler,03:Leibfried, 00:SackettExptEnt4,04:Riebe, 04:Roos}. However, to scale
this approach up to the manipulation of many qubits, it is not
feasible to try to manipulate large numbers of qubits in a single trap.
Instead, it is desirable to have an array of traps, with the possibility
to move quantum information between ions in separate
traps. A method to move the quantum information which has
advantages of relative simplicity and robustness is simply to move the
ions themselves \cite{02:KielpinskiArchitec,98:WinelandB,02:SteaneB}. 
To achieve this we need an array of ion traps, with
the possibility to move ions between traps. This concept has been
demonstrated in an impressive set of
experiments \cite{02:Rowe,04:Barrett,05:Hensinger} which demonstrate
its promise. The present work addresses the issue of how
to design a basic element of such an array: a region in which
a pair of ions can be brought together or separated rapidly,
by adjusting d.c. voltages on electrodes.
This includes the case of separating into two wells
a pair of ions which are initially in a single harmonic well. 

There are further constraints which make this a non-trivial problem:
\begin{itemize}
\item The electrode structure should
allow a route for ions to be moved into and out of the
region of the close pair of trapping centres.
\item The surfaces of the
electrodes should be kept as far as possible from the trap centres where
the ions sit. This is in order both to minimise heating of the motional state of
the trapped ions (see below) and to reduce the impact of surface irregularities
on the electric potential function experienced by the ions.
\item The trap confinement should be tight, i.e. the vibrational frequency
$\nu$ of the ions must exceed some desired minimum value. This is in order to maximise
the speed at which ions can be moved by displacing the trap centres.
\item The electric fields at the electrode surfaces must not exceed the electrical
breakdown limitations of the materials.
\item The electrode configuration must be capable of being fabricated accurately at
the required distance scale. This may require microfabrication techniques
which have their own intrinsic limitations.
\end{itemize}

For diagnostic purposes it may also be useful for the electrode 
structure to be open in order to allow optical access. 

The need to keep the electrode surfaces far away from the ions
is largely owing to the
observation of an anomalous heating rate in ion traps, which becomes
a severe problem with traps of small distance scale. The heating
process is not fully understood but appears to be associated with
impurities deposited on the electrodes, and studies suggest the heating rate
scales as $\rho^{-4}$ where $\rho$ is the distance from trap centre
to the nearest electrode surface \cite{00:TurchetteHeating,02:Rowe,04:Deslauriers}. 
Electrodes further from the trap
centre can also be bigger, which usually implies they are easier to make.

Typical distance scales are 1--10 microns for the separation of the trapped
ions, 1--100 microns for the distance to the electrodes
\cite{00:TurchetteHeating,02:Rowe,04:Barrett,04:Madsen}.

We begin the discussion by considering in section \ref{s:generic} the general
problem of a pair of 1-dimensional potential wells which can be moved together or pulled
apart while maintaining the tightness, i.e. the normal mode frequencies of a pair
of ions trapped in the wells. We argue that the best electrode structure is
one which produces an electric octupole potential, which is then `tweaked'
by the addition of quadrupole terms. In section \ref{s:scaling} we introduce
radial confinement by an a.c. quadrupole, and discuss in general terms the
expected scaling of the trap parameters with the size of the electrode
structure. We show that for structures whose dimensions are greater than
1 to 10 microns, the most demanding factor in the design is to obtain a large
octupole moment, assuming the applied voltages are limited by electrical
breakdown of the materials. We then proceed in section \ref{s:realise} to consider
the problem of obtaining an electric octupole with a minimum number
of electrodes. In section \ref{s:opt} we describe numerical calculations of
the electric potential for several types of electrode structure, obtaining
the values of two dimensionless factors that give the strength of the
three-dimensional d.c. octupole and two-dimensional a.c. quadrupole moment
at given breakdown field and trap scale.
Section \ref{s:mi} then briefly analyses the influence of manufacturing imprecision which
breaks the symmetries of an electrode structure and introduces stray electric fields. 

Section \ref{s:discuss} discusses
the main features of the numerical results. For example, we quantify
the relative merits of a 3-layer over a 2-layer or single-layer (planar)
design. We estimate the impact of requiring the complete structure to be thin
(to aid microfabrication). We summarize the optimizations
obtained by adjusting the positions 
and relative sizes of electrodes. Section \ref{s:conc} concludes.
The appendices give a discussion of axial micromotion and
some general information on charge distributions that
produce electric octupoles.

\section{Generic study of $V(x,y,z)$ for two traps close together} \label{s:generic}
\noindent

We consider first of all some general properties of the electric
potential for two Paul traps
located at $(x,y,z) = (0,0,\pm s)$. We assume the vibrational frequencies
$\omega_x, \omega_y, \omega_z$ of a single trapped ion are the
same for the two traps. The precise relative sizes of $\omega_x$, $\omega_y$
and $\omega_z$ are not crucial, but it will be useful if
$\omega_x, \omega_y \gg \omega_z$. The electrode
configurations to be discussed
in later sections of the paper will produce the confinement along $z$
primarily by d.c. voltages, and that along $x$ and $y$
primarily by an oscillating (r.f.) quadrupole field. However,
we do not need to assume this for the general discussion in this section
(for clarification of this point, see the appendix).
 
Let the distance from the origin to the nearest electrode surface be $\rho$.
Since we aim to achieve close traps and far electrodes, we assume
$s \ll \rho$. We can then usefully analyse the electric potential $V$ near the origin by
a Taylor or a multipole expansion. We further assume that the electrode construction is
close to symmetric so that odd order terms in the expansion almost
vanish. First let us consider $V$ along
the $z$ axis:
\beq
V(0,0,z) \simeq V_0 - E_0 z  +\alpha z^2 + \beta z^4 \label{Vz}
\eeq
where we have dropped the cubic term since we assume
that it is negligible. The signs of the coefficients
$E_0,\alpha,\beta$ have been chosen so that $E_0$ is the electric
field at the origin (assuming $dV/dx= dV/dy=0$ there), and to
create a double well potential we need $\alpha < 0$ and $\beta > 0$
(see figure 1).

\begin{figure} [htbp]
\centerline{\epsfig{file=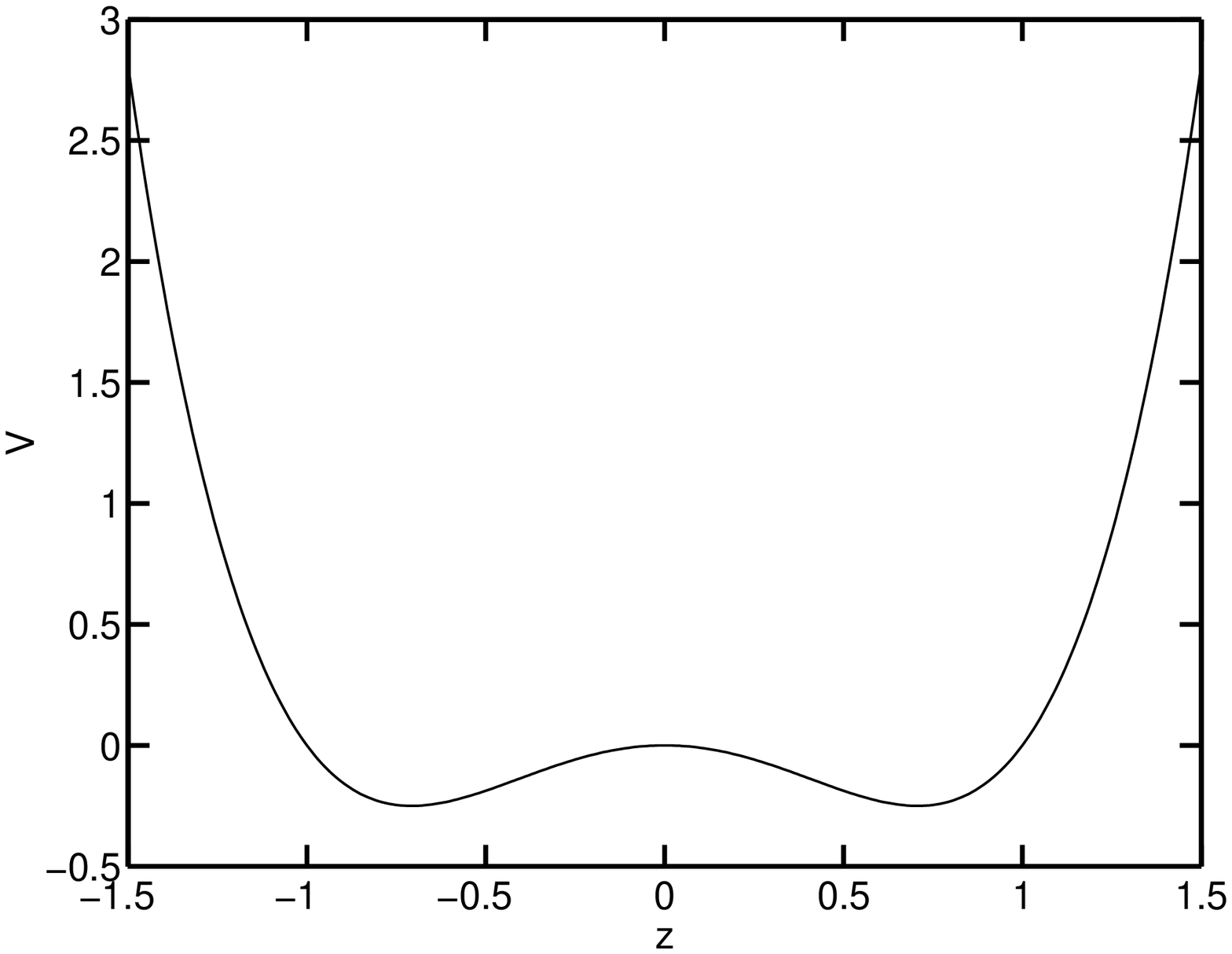, width=6.0cm}} 
\vspace*{13pt}
\fcaption{\label{f:1}Plot of $V(z)$ (equation (\ref{Vz})) for $V_0=0,\; E_0=0$,
$\alpha = -1,\;\beta=1$.}
\end{figure}

First suppose there is no linear term, $E_0=0$.
By solving $\partial V/\partial z=0$ we find for $\alpha < 0$
the potential wells are centred at
\beq
s = \sqrt{ \frac{|\alpha|}{2 \beta} \, }.    \label{s}
\eeq
Evaluating $\partial^2V/\partial z^2$ at either well centre gives
$\partial^2 V/\partial z^2 = 4
\alpha$. If the oscillating voltages are so arranged that the $z$
axis is an r.f. null, i.e. $V(0,0,z)$ is
constant in time or only has a relatively small oscillation, then
we may obtain $\omega_z$ for an ion trapped in either well
directly from $\partial^2 V/\partial z^2$, giving
\beq
-\alpha = \frac{m \omega_z^2}{4 q},  \label{alpha}
\eeq
where $m,q$ are the mass and charge of the ion.
For example, at $\omega_z/2 \pi =1$ MHz, $|\alpha| \simeq 4$ MV/m$^2$
for the charge to mass ratio of the
$^{43}$Ca$^+$ ion. The well depth $V(0)-V(s) = \alpha s^2/2$;
e.g. $0.2$~mV (2 Kelvin) at $s = 10\;\mu$m.

In summary, the quadratic term $\alpha$ in the
Taylor expansion sets the trap tightness (leading to $\omega_z$),
and the ratio of the quadratic to quartic terms sets the trap
separation.

We will arrange for the linear term $E_0$ to be small by using
electrodes symmetric under reflections in the $x-y$ plane, and
then nulling any remaining field if necessary by
compensation voltages. Let us
calculate next the value of $E_0$ which 
``tips the potential over''
just enough so that one of the traps cannot confine ions, i.e.
$\partial^2 V/\partial z^2$
goes to zero at a trap centre. This happens when
\beq
|E_0| \simeq \frac{4}{3} |\alpha| s.
\eeq
For example, for $^{43}$Ca$^+$ with
$\omega/2 \pi = 1$ MHz and $s=10\;\mu$m, $|\alpha| s \simeq
40$ V/m.

\subsection{Cancellation of quadratic term}
\noindent

Consider the Taylor expansion (\ref{Vz})
for a set of electrodes of distance scale $a$. In the absence
of special choices or symmetries which make one or more terms disappear,
we would obtain
\beq
|E| \simeq \frac{|V_0|}{a},\;\; |\alpha| \simeq \frac{|V_0|}{a^2},\;\;
|\beta| \simeq \frac{|V_0|}{a^4}       \label{scaling}
\eeq
(for example, consider the electric potential due to a point
charge, or a line charge, at distance $a$ from the point or line).
Therefore, without a special design, the order of magnitude of $s$
is expected to be $s \sim a$. To obtain $s \ll a$ we require
$\alpha / \beta \ll a^2$, and to obtain this without reducing $\omega_z$ 
we must increase $\beta$. This can be done
by constructing an electrode configuration in which the quadratic term
$\alpha$ is suppressed compared to $\beta$, and then
increasing the voltages on all electrodes together. Assuming the
increased voltages are attainable (and we will find they are), the available
$\beta$ at given $\alpha$ is limited by electrical breakdown associated
with high electric fields at the electrode surfaces.

Assume that the electrodes have reflection symmetry in $x$-$y$,
$x$-$z$ and $y$-$z$ planes, so that the electric field vanishes
at the origin. Then odd-order multipole moments of $V(x,y,z)$ vanish,
and so do mixed 2nd derivatives such
as $\partial^2 V / \partial x \partial y$ at the origin.
Under this assumption, we will next
show that the required electrode configuration is one
which produces, at lowest order in its multipole expansion around
the origin, an octupole moment.

An electrode set
that can produce $\alpha/\beta \ll a^2$ is close to the condition
$\alpha \rightarrow 0$. Since we require the possibility to adjust
$\alpha$ and $\beta$ (e.g. by adjusting voltages on the electrodes) 
the same electrode set can also attain $\alpha = 0$ (in practice,
it may or may not actually be used in that condition).
We therefore have to consider electric potential configurations
having $\partial^2 V/\partial z^2 = 0$ and $\partial^4 V/\partial z^4 \ne 0$.
From Laplace's equation, we then have
$\partial^2 V/\partial y^2 = - \partial^2 V/\partial x^2$, thus a 2D quadrupole
moment in the $x-y$ plane unless these derivatives are also zero.
Unless we take special measures to make
$|\partial^2 V / \partial x^2|$ small, we have $|\partial^2 V/\partial x^2|
\sim \beta a^2 \gg m \omega_z^2/q$ when $s \ll a$.
This implies the field is strongly
expelling along some direction in the $x-y$ plane.
To permit the ions to remain trapped, either the whole
field must be made to oscillate, or an oscillating quadrupole field
must be added whose strength is sufficient to overcome this expulsion. If
the whole field oscillates then in the pseudo-potential model, the effective
values of both $\alpha$ and $\beta$ are reduced, which is counter-productive.
If instead a separate oscillating quadrupole is added in the $x-y$ plane
(with small effect in the $z$-direction) then its strength must be large,
and this leads to electrical breakdown which will constrain the available
range of $\beta$. 

We conclude that the d.c. electrode structure must be designed to produce
small $|\partial^2 V / \partial x^2|$ (as well as small
$|\partial^2 V / \partial z^2|$)
and therefore we require an electrode structure at or close to
an electric hexapole or octupole configuration. 
A quantitative statement of this condition is given in sections
\ref{s:scaling} (eq. (\ref{nonoct})) and \ref{s:NIST}.
The octupole
has the advantage that reflection symmetries which cancel
odd-order multipole moments can be used to avoid producing an
unwanted electric field (i.e. a contribution to the electric potential
varying linearly with distance at the origin), and therefore we will
concentrate on that case. 

Confinement of the ions
in three dimensions is completed by adding to the d.c.
octupole field an oscillating 2D quadrupole field whose size is controlled separately.
Hence, after reinstating a non-zero value for $\alpha$, we obtain the general form
\beq
V(x,y,z,t) \!\! &\simeq& \!\! \alpha \left( z^2 - \frac{1}{2}(x^2 + y^2) \right) 
+ \beta V_4(x,y,z) \nonumber \\
&& + Q_{\rm ac} \cos(\Omega t) (x^2 - y^2)  \label{Vform}
\eeq
where $\alpha$, $\beta$ and $Q_{\rm ac}$ are time-independent, $V_4(x,y,z)$
is an octupole potential with $V_4(0,0,z) = z^4$, 
and $\Omega$ is the frequency of the a.c. quadrupole
providing confinement in the radial direction. Many ion trap
configurations in common use are described by (\ref{Vform}), but we have in
mind a case where $\beta/|\alpha|$ is as large as possible. We will see later
that none of the electrode sets to be discussed realise (\ref{Vform}) exactly,
because the time-dependent part depends slightly on $z$ as well as $x$ and $y$,
and because higher-order terms appear,
but it is useful to make clear what we are aiming at.

\subsection{Coulomb repulsion and normal mode frequencies} \label{s:normalmode}
\noindent

If there is one ion in each trap, then the distance between ions is not
$2 s$ owing to their mutual Coulomb repulsion. The equilibrium
positions are $z = \pm d/2$ where the separation $d$ is a real positive
solution of
\beq
\beta d^5 + 2 \alpha d^3 = \frac{q}{2 \pi \epsilon_0}.      \label{deq}
\eeq 
In order to bring out the general behaviour of solutions to this equation,
it useful to express it in the form
\beq
\left( \frac{d}{2s} \right)^5 + \frac{\alpha}{|\alpha|} \left( \frac{d}{2s} \right)^3 = \epsilon,
\eeq
where
\beq
\epsilon = \frac{q}{4 \pi \epsilon_0  |\alpha| (2s)^3}.
\eeq
Therefore the solutions are determined by the sign of $\alpha$ and a single
parameter $\epsilon$, which compares the Coulomb repulsion force to the
trapping force for ions situated at $\pm s$ in a harmonic well. 
When the $\beta$ term in (\ref{deq}) dominates, $\epsilon$ is large, and
$d \simeq (q/2 \pi \epsilon_0 \beta)^{1/5}$. 
When $\alpha$ is large and positive, i.e. two
ions in the same harmonic well, $d \simeq (q/4 \pi \epsilon_0 \alpha)^{1/3}$. 
When $\alpha < 0$ and $\epsilon \ll 1$, i.e. a pair of well-separated  traps, then
$d \simeq s(2 + \epsilon)$.

The system of two ions has two normal modes for motion along the $z$ direction:
the centre of mass mode
of frequency $\omega_1$ and the breathing (also called stretch) mode of frequency
$\omega_2$. These frequencies are given by
\beq
\omega_1^2 &=& \left( 2 \alpha   + 3 \beta d^2 \right)q/m,      \label{omega1} \\
\omega_2^2 &=& \omega_1^2 \left( 1+ \tilde{\epsilon} \right),
\eeq
where
\beq
\tilde{\epsilon} = \frac{q^2}{\pi \epsilon_0 m \omega_1^2 d^3}
\eeq
($m$ was defined previously, it is the mass of a single ion).

When the ions
are far apart, $\tilde{\epsilon} \rightarrow \epsilon \ll 1$ and the two
mode frequencies are both equal to the oscillation frequency of a single ion in
either trap. When $\alpha = 0$, $\tilde{\epsilon} = 2/3$ and when
$\alpha \gg 1$ (i.e. a single harmonic well), $\tilde{\epsilon} \rightarrow 2$.

\begin{figure} [htbp]
\centerline{\epsfig{file=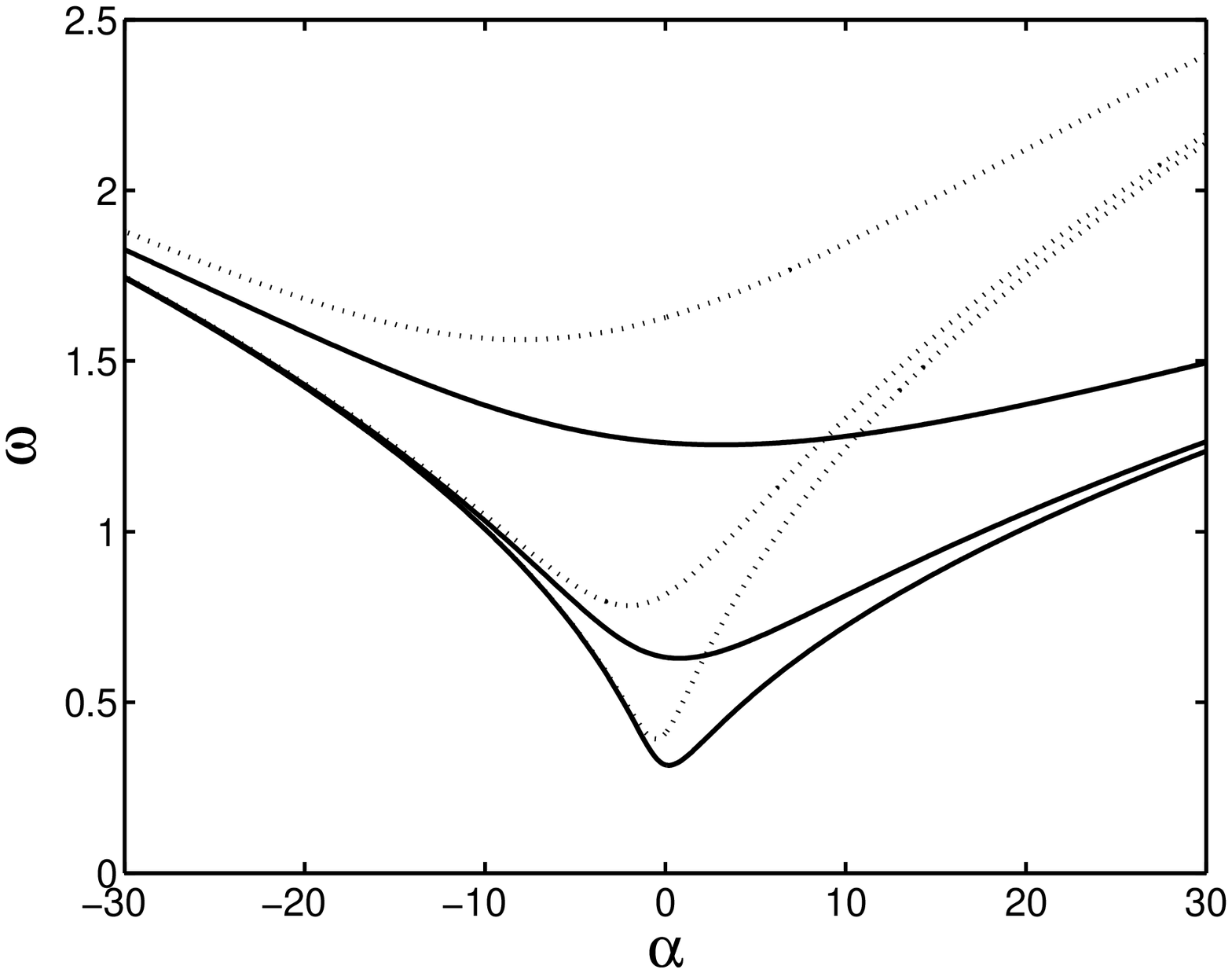, width=7.0cm}} 
\vspace*{13pt}
\fcaption{\label{fig:omegadependence}Plot of the
centre-of-mass frequency $\omega_1$ (solid curve) and stretch mode frequency
$\omega_2$ (dotted curve) vs. $\alpha$ for three different
values of $\beta$. The curves shown are for $\beta = 100, 10, 1$ in arbitrary units,
the higher value of $\beta$ giving the higher normal mode frequencies. 
The figure shows the generic behaviour, the units are arbitrary.}
\end{figure}

The reason to bring the ions close together is in order to perform
a two-ion quantum logic gate in which their Coulomb repulsion provides
the interaction energy between the qubits. The gate is fast
when the mode frequencies are high and 
well-separated \cite{03:Sasura, 00:Sorenson}. Therefore we
would like a high value of both $\omega_1$ and $\tilde{\epsilon}$.
Suppose the ions are initially in traps far apart, so that $\tilde{\epsilon} \ll 1$. Then
the vibrational frequencies are almost equal and are given by the value of $\alpha$
through eq. (\ref{alpha}). To bring the ions together, the voltages
on the electrodes may be changed so that $\beta$ gets larger while
$\alpha$ stays the same size. Eventually electrical breakdown
prevents further increase of $\beta$. If at this
point the values of $\omega_1$ and $\tilde{\epsilon}$ are high
enough to give an acceptable gate speed, and if it is
possible to apply the logic-gate laser pulses to ions in the separation zone, then
the gate is performed and the ions are subsequently separated. Otherwise, to bring
the ions into a common trap, $|\alpha|$ must be lowered and one accepts
an unavoidable reduction in mode frequencies while the potential
barrier between the twin traps is lowered, see figure \ref{fig:omegadependence}.
After $\alpha$ passes through
zero and becomes positive, the mode frequencies increase again and
$\tilde{\epsilon}$ rises towards its maximum value of 2. The ions
can then be moved as a pair to a convenient (e.g. less noisy)
region of the complete array, for the logic-gate laser pulses to be applied.

At the condition $\alpha = 0$, the full analysis is tractable and one finds
\beq
\omega_1 = \sqrt{\frac{3 q}{m}} \left( \frac{q}{2 \pi \epsilon_0} \right)^{1/5}  \, \beta^{3/10}
\label{om1}
\eeq
and $\tilde{\epsilon} = 2/3$ as mentioned above.

In the rest of this paper we will consider methods to produce this
condition, i.e. an octupole potential, with an added oscillating 
quadrupole potential in the $x-y$ plane, as in (\ref{Vform}). In practice
the electrode structures to be considered depart from (\ref{Vform}) primarily
in one respect, namely that the oscillating term includes a $z$-dependence,
i.e.
\beq
V(x,y,z,t) \!\! &\simeq& \!\! \alpha \left( z^2 - \frac{1}{2}(x^2 + y^2) \right) 
+ \beta V_4(x,y,z) \nonumber \\
&& + \cos(\Omega t) (\alpha_x x^2 - \alpha_y y^2 + \alpha_z z^2)  \label{Vform2}
\eeq
where we introduce the parameters $\alpha_x, \alpha_y, \alpha_z$ to characterise
the 3-dimensional a.c. quadrupole. If $\alpha_x \ne \alpha_y$ then necessarily
$\alpha_z \ne 0$, and in this case there is axial micromotion. It is shown in
the appendix that this leads to a Mathieu equation, which can be solved in
the standard way as a combination of slow secular motion and fast micromotion
\cite{{67:DehmeltRadiofreq}, {Bk:Ghosh}}. 
The secular motion can be modelled as motion in a pseudo-potential which
takes into account the influence of both the d.c. and the
a.c. (ponderomotive) terms. The above analysis then remains 
correct (within the pseudopotential approximation)
as long as we replace $\alpha$ by $\alpha'$, where $\alpha'$ is
the coefficient of the $z^2$ term in the total effective potential. 
For the structures discussed in this paper, it is found that the
electrode voltages required to produce  
$\alpha' = 0$ are very close to those required for $\alpha = 0$, hence 
changing from one condition to the other has little effect on $\beta$.
At $\alpha' = 0$ the secular frequency is completely determined by the
value of $\beta$ through (\ref{om1}).

\section{Radial r.f. confinement and scaling}  \label{s:scaling}
\noindent

Consider the potential (\ref{Vform}). A pair of trapped ions is confined
in the radial ($x$ and $y$) direction by the oscillating quadrupole field. 
To first approximation we may ignore the effect of the octupole
potential on the radial motion, then this motion is described by a Mathieu equation.
In the pseudo-potential model, for $\alpha=0$, the secular frequency of the centre-of-mass
vibrational motion in the radial direction is given by \cite{{03:Leibfriedreview},{67:DehmeltRadiofreq}, {Bk:Ghosh}}
\beq
\omega_r = q_r \Omega / 2 \sqrt{2},
\eeq
where  
\beq
q_r = \frac{4 q Q_{\rm ac}}{\Omega^2 m}
\eeq
is the Mathieu $q$-parameter. For stable motion, $q_r$ must not be large; values in
the range $0.1$ to $0.5$ are typical. This constrains the applied r.f. frequency $\Omega$
and hence we obtain
\beq
\omega_r = \left( \frac{q_r q}{2 m} \, Q_{\rm ac} \right)^{1/2}.  \label{omr}
\eeq

At the octupole condition, equations (\ref{om1}) and (\ref{omr}) give the centre of
mass vibrational frequencies for motion in the axial and radial directions respectively.
Since we want the ions to align themselves along the $z$ axis,
we require $\omega_r > \omega_1$ (and $\omega_r \gg \omega_1$ is desirable). This
sets a lower limit on the required value of $Q_{\rm ac}$ for given $\beta$:
\beq
Q_{\rm ac} = \left( \frac{\omega_r}{\omega_1} \right)^2
 \frac{6}{q_r} \left( \frac{q}{2 \pi \epsilon_0} \right)^{2/5} \, \beta^{3/5} .
\label{Qaclim}
\eeq

It is useful to examine the way these parameters scale with $\rho$, 
the distance from the origin to the nearest electrode surface, and
$E_{\rm max}$, the largest electric field at any electrode surface.
For a given electrode geometry $\rho$ sets the distance scale of the electrodes.
To keep $E_{\rm max}$ as small as possible, the electrode surfaces should have
as large a radius of curvature as possible, and therefore their radii
should be allowed to increase with $\rho$. In other words
the complete structure of the electrodes is assumed to scale with $\rho$. Nevertheless
for a given electrode set at a given value of $E_{\rm max}$, $\beta$ and $Q_{\rm ac}$
still have a range of possible values, because it is possible to
choose a range of values for the ratio between the sizes of the oscillating and
the d.c. voltages. Typically the variation is such that there is a competition between
high $\beta$ and high $Q_{\rm ac}$.

In order to obtain the main features by a relatively simple
analysis, we define two geometric factors $\gamma$ and $\mu$ such that
\begin{eqnarray}
\mbox{for } Q_{\rm ac} = 0,\;\;\; \beta = \frac{\gamma E_{\rm max}}{\rho^3}, \label{gam} \\
\mbox{for } \beta = 0, \;\;\; Q_{\rm ac} = \frac{\mu E_{\rm max}}{\rho}.  \label{mu}
\end{eqnarray}
Thus $\gamma$ and $\mu$ tell us about the maximum values of $\beta$ and $Q_{\rm ac}$
available for a given electrode geometry, but we keep in mind that it is not
possible to have both these maximum values at once.

We can now quantify the desirability of the octupole potential, as compared with
other possibilities. If the electrode structure is not designed to produce small or
zero d.c. quadrupole, then it will produce a d.c. quadrupole potential
in the $xy$ plane of strength $\alpha_{\rm r}^{\rm dc} \simeq \rho^2 \beta$. 
To achieve radial confinement, we require $\omega_r^2$ as given by (\ref{omr}) to
exceed $2 \alpha_{\rm r}^{\rm dc}q/m$. Hence the condition for
stable radial confinement is $Q_{\rm ac} > 4 \rho^2 \beta/q_r$ and therefore
\beq
\mu > \frac{4}{q_r} \gamma.   \label{nonoct}
\eeq
It is interesting that this condition is scale-independent. Since
it is satisfied for large $\mu$ or small $\gamma$, it shows
that suppressing the d.c. quadrupole in all directions is not necessary
for electrode structures which naturally produce a strong radial 
confinement or a weak octupole.
However, we are interested in making both parameters large, and it
will be shown that the structures offering high values of both parameters do
not satisfy (\ref{nonoct}), therefore the suppression of the d.c. quadrupole is useful.

\begin{figure} [htbp]
\centerline{\epsfig{file=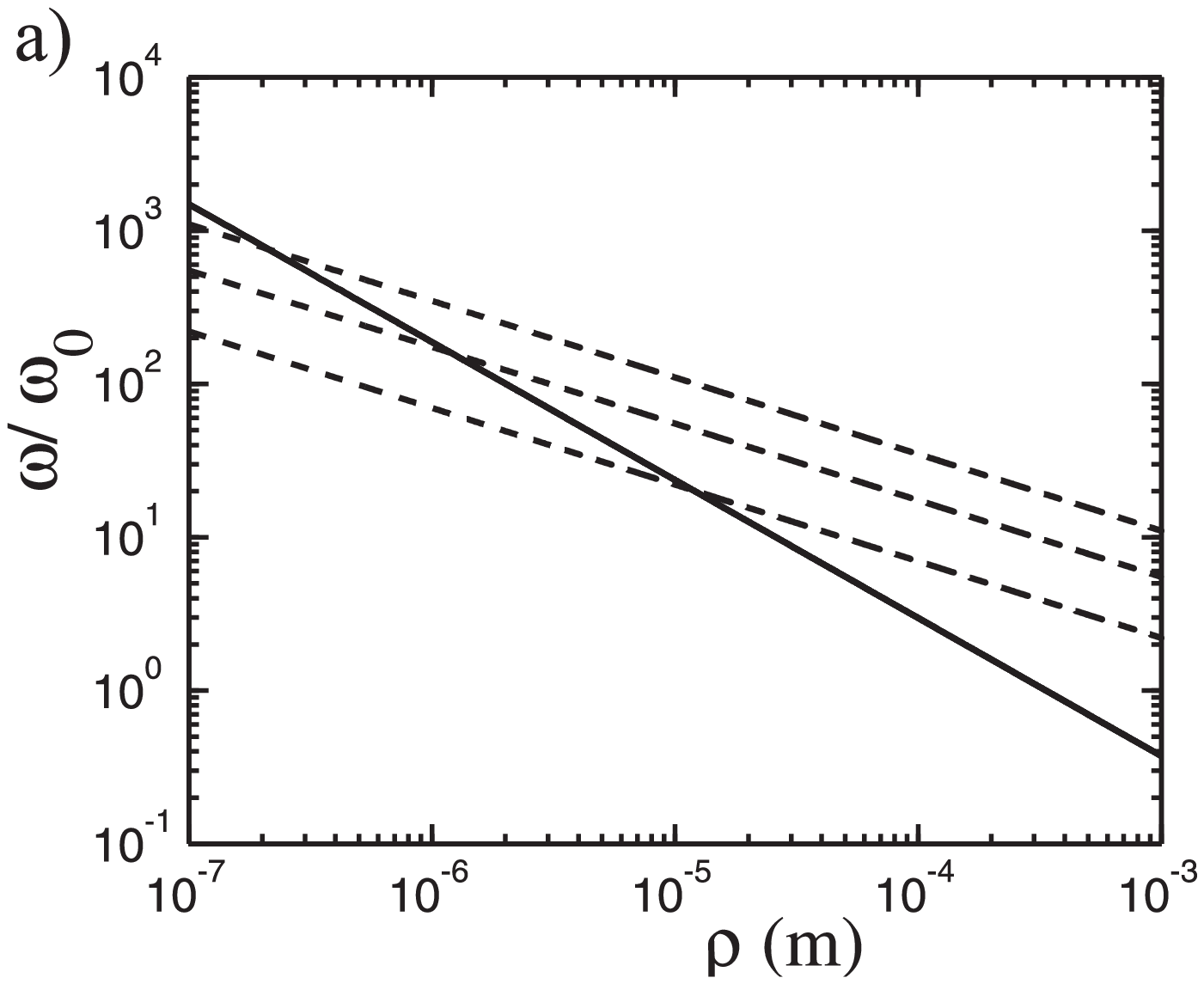, width=6.0cm} \epsfig{file=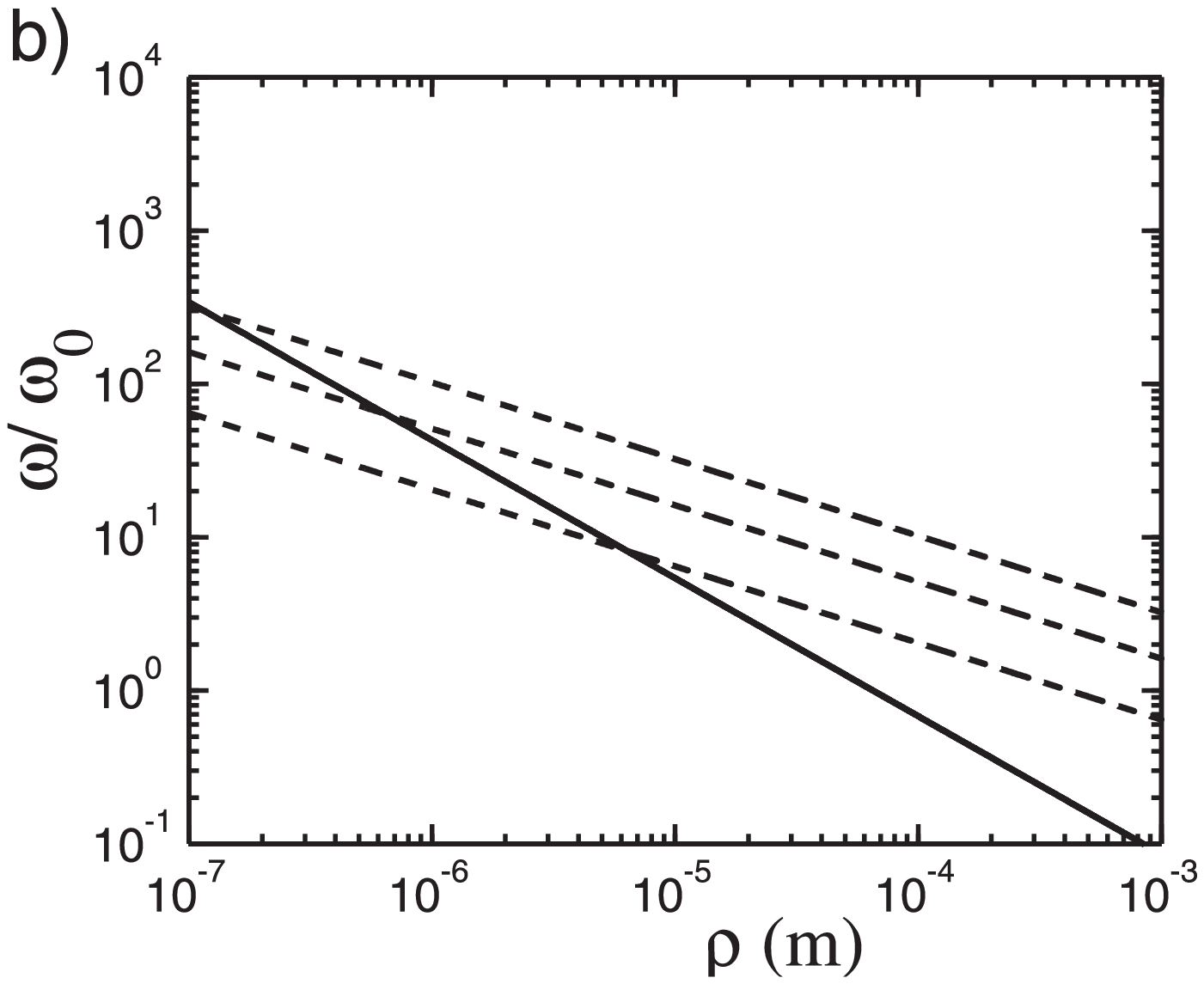, width=6.0cm}} 
\vspace*{13pt}
\fcaption{\label{fig:relrf} Centre-of-mass vibrational frequencies for axial motion of two 
ions in a d.c. octupole potential (solid line, $\omega_1$) and radial motion of two ions in a linear 
Paul trap (upper dashed line, $\omega_r$). The frequency unit $\omega_0$ is defined in the text.
(a): electrode arrangement 2.1, (b): electrode arrangement 5.1. The frequencies are obtained
using values of $\gamma$ and $\mu$ presented in section \ref{s:opt} and using
$E_{\rm max}=10^8$ V/m. The lower two dashed lines are $\omega_r / 2$ and $\omega_r/5$; these
permit the conditions $\omega_1 = \omega_r/2$ and $\omega_1 = \omega_r/5$ to be identified easily.}  
\end{figure}

Suppose the ions each have one electronic unit of charge and
let $A$ be the mass number,
then for a trap at the octupole condition ($\alpha=0$), eq. (\ref{om1}) gives
\beq
\frac{\omega_1}{2 \pi} \mbox{[MHz]} \simeq
\frac{840}{\sqrt{A}} \frac{ \left( \gamma E_{\rm max}  \right)^{3/10} }
{ \rho^{9/10} }.    \label{om1A}
\eeq
where $E_{\rm max}$ is in V$/\mu$m and $\rho$ is in $\mu$m.
Note that this mode frequency scales almost linearly with $1/\rho$, and is
relatively insensitive to $\gamma$ and $E_{\rm max}$. The radial frequency
(eq. (\ref{omr})), on the other hand, scales as
\beq
\omega_r \propto \frac{\mu^{1/2}}{\rho^{1/2}}.
\eeq
This difference in the scaling of frequency with trap size, for the radial
and axial motion, is illustrated in figure \ref{fig:relrf}.
The figure shows the frequencies $\omega_1$ (eq. (\ref{om1A}))
and $\omega_r$ (eq. (\ref{omr})) plotted as a function of $\rho$ for
two example electrode configurations discussed in section \ref{s:opt}.
In order that a single graph can show the behaviour for all 
ion species, the vibrational frequencies are shown in units of a basic
frequency $\omega_0$ which depends on the charge/mass ratio of the
ion. We define $\omega_0$ to
be the radial secular frequency (eq. (\ref{omr})) for an ion in a trap
having $Q_{\rm ac} = 10^8$ Vm$^{-2}$ and $q_r = 0.3$. $\omega_0/2 \pi$ takes the value
923 kHz for $^{43}$Ca$^+$, 574 kHz for $^{111}$Cd$^+$ and 2017 kHz for $^{9}$Be$^+$. 

Each point where the lines cross on figure \ref{fig:relrf} identifies
a distance scale $\rho_c$. It is found by
substituting (\ref{gam}) and (\ref{mu}) into (\ref{Qaclim}) and solving for
$\rho$:
\beq
 \rho_c \equiv 
 \frac{\gamma}{\mu} \left( \frac{1}{\gamma \mu} \right)^{1/4}
\left( \frac{1}{6 q_r} \right)^{5/4} 
\left( \frac{\omega_r}{\omega_1} \right)^{5/2}
L_0     \label{rhoc}
\eeq
where 
\beq
L_0 = 36 (3 q/ \pi \epsilon_0 E_{\rm max})^{1/2}.   \label{L0}
\eeq 
The distance scale $\rho_c$ answers the question, ``which motional frequency, the radial
or the axial, will be the more challenging to achieve when we design the electrode structure?"
It thus helps to establish priorities in the task. 
Below $\rho_c$ the challenge is to increase the radial frequency, while above $\rho_c$ the challenge
is to increase the axial frequency. This is because in order to maintain
a linear ion configuration, below $\rho_c$ the trap must be weakened in the axial direction,
by operating it with $\beta$ below the maximum available value, and therefore in this regime
$\mu$ gives the limit on the performance. 

We now show that the value of $\rho_c$ is of
order microns in practice.

In the expression (\ref{rhoc})
the factor $(1/6 q_r)^{5/4} \simeq 0.5$ to 2 for a choice of Mathieu $q_r$ parameter
in the range $0.1$ to $0.3$. The combination of geometric factors in 
(\ref{rhoc}) has a value in the range $0.1$ to 2 for the structures
described below (with the exception of those shown in figures 4 and 5 which
we reject as unsatisfactory owing to their poor radial confinement). Therefore
we find $\rho_c \simeq $(0.3 to 22)$L_0$ for $\omega_r = 2 \omega_1$. In practice
one would choose $\omega_r / \omega_1$ in the range 2 to 5 in order to keep the ions near
the axis without restricting $\omega_1$ too much. Increasing $\omega_r / \omega_1$
from 2 to 5 creates an order of magnitude increase in $\rho_c$. 
  
In practice we are concerned with singly charged ions, and therefore
$L_0$ is a distance scale set purely by the maximum electric field allowed in the structure.
For example when $E_{\rm max} = 10^9$ V/m (the order of magnitude when field emission
may occur \cite{04:Cruz, Bk:Gomer}), $L_0 \simeq 0.15\;\mu$m.
This field depends mainly on the materials, and partly on the geometry (for example a larger
field may be allowed across a vacuum gap than across an insulating material surface). However,
owing to the square root in (\ref{L0}) the dependence is not strong. 

Taking $q_r=0.3$ and combining these observations, it is seen that for a variety of structures, 
$\rho_c < 10 \;\mu$m for $\omega_r = 5 \omega_1$ at high electric field, 
and $\rho_c < 2 \; \mu$m for $\omega_r = 2 \omega_1$ and more modest fields.
This implies that for ion traps designed for quantum computing experiments, 
the more demanding problem is currently to obtain a large d.c. octupole moment,
but the radial confinement may become the more demanding problem in the
future as distance scales are reduced. The values of $\rho$ where the lines cross in 
figure \ref{fig:relrf} illustrate these points. 

Another important distance related to $L_0$ is the distance between the
ions when the electrodes are close to the octupole configuration. Using
(\ref{deq}) and (\ref{gam}) we obtain
\beq
d = \frac{\rho}{6} \left( \frac{L_0^2}{\gamma \rho^2} \right)^{1/5}.
\label{dmin}
\eeq
Therefore $d \ll \rho$ when $\rho \gg L_0 / \gamma^{1/2}$, 
which is usually the case. For example, putting $\gamma=0.02$ (c.f. table 2 below)
and $L_0 = 0.2\;\mu$m, $d \ll \rho$ when $\rho \gg 16$ nm. For electrodes
of micron scale or above, therefore, the available confinement is
such that when the electric potential is just at the point of producing a potential hill
between the ions, the distance between them is much smaller than the distance
to the electrodes. This implies that the Taylor expansion we have adopted is valid.

\subsection{Scaling of heating effects}  \label{s:heatscale}

The main reason why one would not make a quantum computer faster simply by fabricating
the traps at a smaller distance scale is the fact that heating problems will be exacerbated.
Heating rates observed in various studies scaled as $\rho^{-4}$ \cite{04:Madsen,00:TurchetteHeating}
and the scaling expected for heating caused by Johnson noise is $\rho^{-2}$. 
We will treat some general scaling law $\rho^{-k}$, and we
assume this heating depends on the electrode structure only through the distance scale
$\rho$. The time
to separate a pair of ions scales as $1/\omega_1$, therefore the number of phonons'
worth of heating per split time, $\Delta n$, is proportional to $1/(\omega_1 \rho^k)$. This gives
a good guide to the amount of cooling which will be needed to obtain a steady
average temperature in the computer. Using equation (\ref{om1A}) we obtain
\beq
\Delta n \propto \frac{ \rho^{-k+9/10} }{ \gamma^{3/10} }
\eeq
For a given electrode geometry, $\gamma$ is fixed, hence this scaling law governs the allowed
distance scale $\rho$. If we take $\Delta n$ as a fixed
quantity, then $\rho \propto \gamma^{3/(9-10k)}$. Substituting this behaviour into
equation (\ref{om1A}) we find the following scaling for the trap frequency in terms of $\gamma$:
\beq
\omega_1 \propto \gamma^{3k / (10k-9)}
\eeq
For example, for $k=2$ the scaling is $\omega_1 \propto \gamma^{0.545}$, for $k=4$ the scaling 
is $\omega_1 \propto \gamma^{0.387}$. When the radial rather than the axial frequency is
the harder to achieve, then a similar analysis may be used to relate $\omega_r$ to $\mu$.

\subsection{Study of a non-octupole trap structure in current use} \label{s:NIST}

An example of an ion trap in current use which has been used to separate ions is that 
operated at NIST Boulder for teleportation and other experiments \cite{04:Barrett}. 
It is a two-layer trap with 5 d.c. electrodes in two opposing quadrants, and the r.f. 
electrodes in the other quadrants. With $z$ along the trap axis, the ends of the electrodes
lie at $x = a$, $y = a$ where $a = 100\,\mu$m. It is not possible to 
produce an octupole using such an electrode configuration (this 
requires 7 d.c. electrodes, see section \ref{sec:twolayer}). 

We studied this structure by modelling it and solving Laplace's equation numerically.
The calculations were performed using  Simion \cite{pk:simion7} and CPO \cite{pk:CPO}.
Simion uses a finite difference method to find the potential 
of a regular 3D array of points. The array is
iteratively relaxed so as to satisfy Laplace's equation and the
boundary conditions, until the fractional changes per iteration
are all below $5 \times 10^{-3}$. However, from
simulations of analytically solvable cases, we found that
the error in electric fields close to spherical and cylindrical 
electrodes could be as high as $10\%$. CPO uses a surface charge method,
and we found it to be considerably faster and also much more precise.
All the numerical results presented in this paper were obtained using CPO.

To calculate a value for the parameter $\gamma = \rho^3 \beta / E_{\rm max}$, 
we numerically solve Laplace's
equation for the given electrode set-up, and then extract $\beta = (1/24) \partial^4
V / \partial z^4$ evaluated at the origin, and $E_{\rm max}$ by examining the
gradient of $V$ near the surfaces of all the electrodes.  

In order to find $\mu$, calculations were performed with 
the r.f. electrodes at a voltage of $10^6\,$V
which is large compared to the d.c. voltages of $\sim 1\,$V used for the d.c. electrodes.
The value of the radial confinement $\alpha_r \equiv Q_{ac}$ was extracted, 
along with the value of the maximum field $E_{\rm{max}}^\mu$
at the surface of the electrodes. The radial geometrical 
parameter $\mu$ was then calculated using equation (\ref{mu}). 

For the model of the NIST trap, first the r.f. electrodes were grounded and
the voltages on the d.c. electrodes were adjusted to 
satisfy $\partial^2 V/\partial x \partial y = \partial^2 V/\partial y \partial x = \partial^2 V/\partial z^2 = 0$.
We then found that $\partial^2 V/\partial y^2 = -\partial^2 V/\partial x^2 = 0.02$ V$a^{-2}$, 
$\beta = 0.01$ V$a^{-4}$ and $E_{\rm max} = 11.1$ V$a^{-1}$. 
The NIST trap has $\rho = 1.38 a$, hence we deduce $\gamma = 0.0024$. 
The same procedure was carried out 
with $\partial^2 V/\partial x^2 = \partial^2 V/\partial y^2 = \partial^2 V/\partial z^2 = 0$ but 
the remaining (undesirable) quadrupole terms were found to be 10 times larger. 
In each case we then raised the r.f. electrodes to a high voltage in order to find $\mu$, obtaining
the value 0.11 in both cases.

We observe that the d.c. quadrupole which exists in the $x-y$ plane when $\partial^2 V/\partial z^2 =0$
in this trap has a strength very close to $a^2 \beta$, i.e. it is just as large as we would expect
it to be from general scaling considerations when the structure is not designed to eliminate it. 
The values of $\gamma$ and $\mu$ satisfy condition (\ref{nonoct}) for $q_r > 0.006$, therefore
in this trap the radial confinement from the a.c. potential can overcome the d.c. anticonfinement before
the r.f. voltage is restricted by electrical breakdown. This can be regarded as owing to
a low value for $\gamma$. However, the presence of the d.c. quadrupole remains unwelcome, because
it results in a large increase in the required r.f. voltages. 

To estimate the maximum axial and radial frequencies which could be achieved at the point of ion separation, 
we use $q_r$ = 0.3 and assume that the maximum field allowed at the surface
of the electrodes is $10^8$ Vm$^{-1}$. Then the voltages on the d.c. electrodes are
67, -153, 509 V on the central, intermediate and outer electrodes respectively. For
$\omega_r \ge 3 \omega_z$ the r.f. voltage amplitude is $V_{rf} = 557\,$V. 
In the case of beryllium ions, these conditions lead to motional frequencies
$(\omega_x, \omega_y, \omega_z) = 2\pi \times (7, 30, 2)$ MHz. 
The large difference between $\omega_x$ and $\omega_y$ shows that most of the voltage on the r.f. electrodes 
is being used to overcome the d.c. anticonfinement along $x$. If the d.c. quadrupole were eliminated,
then one would obtain $\omega_x/2\pi, \omega_y/2\pi$ both equal to 20 MHz. In that case one
would have the option of reducing the r.f. voltage by an order of magnitude. 

\section{Realising octupoles} \label{s:realise}
\noindent

In this section we discuss general methods
to obtain an octupole moment at the origin of coordinates. This acts as a guide
in the design of the electrodes.

Three examples of octupole potentials are the axially symmetric octupole:
\beq
V = z^4 - 3 z^2\left( x^2 + y^2 \right) + \frac{3}{8}\left( x^2 + y^2 \right)^2 \label{Vaxoct}
\eeq
an octupole with cubic symmetry:
\beq
V= x^4 + y^4 + z^4 - 3 (x^2 y^2 + x^2 z^2 + y^2 z^2)  \label{Vcube}
\eeq
and the 2-dimensional octupole in the $x$-$z$ plane:
\beq
V = z^4 + x^4 - 6 z^2 x^2.      \label{V2Doct}
\eeq
The axially symmetric octupole can be produced by a set of 2 (+ve) end-cap and 3
ring (2 -ve, central one +ve) electrodes, all shaped to follow contours of $V$.
The 2D octupole can be
produced by a set of 8 (4 +ve, 4 -ve) rod electrodes parallel to the $y$ axis and
located at the corners of an octagon.
These two examples of pure octupoles give a useful pointer to the required
distribution of image charge.

The methods and image charge distributions listed below do not produce pure octupoles, but
have an octupole as the leading term in the multipole expansion of their potential near
the origin.

\subsection{Use of symmetry}  \label{s:methods}
\noindent

To produce an octupole at a point $\underline{r}$, 
19 constraints have to be satisfied. These are that the first, second 
and third derivatives of the potential at this point in three orthogonal 
directions should vanish. Many of these can be satisfied by the introduction 
of symmetry to the charge distribution producing the potential. This can take
the form of either rotation symmetry about an axis passing
through $\underline{r}$, or reflection symmetry in a plane 
containing the point $\underline{r}$.

A single reflection in a plane containing the 
point $\underline{r}$ produces zero odd-order moments in 
a direction perpendicular to the plane, and also forces 
the mixed second derivatives
such as $\partial^2 V/\partial x_i \partial x_j$ to vanish, where $x_i$ lies within and $x_j$ 
lies perpendicular to the plane of reflection. This leaves 12 constraints to be satisfied.

Two reflections in orthogonal planes containing $\underline{r}$ produces zero odd-order moments 
in the directions normal to the two planes, and all mixed second derivatives vanish.  This leaves 7 constraints to be satisfied.

Two-fold rotation symmetry about an axis passing through $\underline{r}$ causes the 
odd-order derivatives of the potential in all directions perpendicular to the axis to vanish, 
reducing the number of constraints to 9. 

Four-fold rotation symmetry about an axis passing through $\underline{r}$ makes the second 
derivatives in directions perpendicular to the axis equal, hence the number of constraints is reduced to 5.

Combinations of the above can be used to reduce the number of 
constraints further. Reflection symmetry in three orthogonal planes all 
containing $\underline{r}$ reduces the constraints to 3,
on the non-mixed second derivatives $\partial^2 V/\partial x_i^2$. 
Combining a two-fold rotation about an axis passing through $\underline{r}$ with 
a reflection in a plane normal to the axis and containing $\underline{r}$ reduces
the constraints to 4 (the non-mixed second derivatives and 
the derivative $\partial^2 V/\partial x_i \partial x_j$, where $x_i$ and $x_j$ are
orthogonal directions perpendicular to the rotation axis). A four-fold rotation
symmetry about an axis passing through $\underline{r}$ combined with a reflection
in a plane normal to the axis and containing $\underline{r}$ leaves 2 constraints
(on the three non-mixed second derivatives, two of which are equal).  

Finally, Laplace's equation reduces the degrees of freedom of the non-mixed 
second derivatives $\partial^2 V/\partial x_i^2$ by one.

\subsection{Image charge constructions} \label{s:image}
\noindent

(A). An octupole with cubic symmetry is produced by any arrangement of charge
having cubic symmetry. Proof: the symmetries under reflections in three orthogonal
planes imply the odd-order multipole moments vanish, and so do mixed
second derivatives of $V$ such as $\partial^2 V/\partial x \partial y$;
the symmetries under rotation
through 90$^{\circ}$ imply the coefficients of $x^2$, $y^2$ and $z^2$ in a Taylor
expansion of the potential at the centre of symmetry are all the same, and therefore
by Laplace's equation they all vanish.

(B). An octupole with cylindrical symmetry is produced by any arrangement of charge having
cylindrical symmetry, reflection symmetry in $z=0$,
and placed such that $\partial^2 V/\partial z^2=0$ (where $z$ is
the symmetry axis). Proof: the second derivatives w.r.t. $x$ and $y$ must be equal by
symmetry and therefore by Laplace's equation they also vanish.

(C). An octupole is produced by any arrangement of charge having
four-fold rotational symmetry in the $x-z$ plane, reflection symmetry in
$y=0$, and arranged so that
$\partial^2 V/\partial y^2=0$. The reasoning is the same as in the previous case.
The resulting octupole potential is
a combination of the axially symmetric case (\ref{Vaxoct}) and
the 2-dimensional case (\ref{V2Doct}) oriented such that $y$ is the common axis.

(D). An octupole can be produced, starting with any arrangement of charge $\rho(x,y,z)$,
by the following recipe. First ensure reflection symmetries in order to cancel
odd multipole moments, by forming
\beq
\bar{\rho} \equiv \sum_{i=0}^1 \sum_{j=0}^1 \sum_{k=0}^1 \rho((-1)^i x, (-1)^j y, (-1)^k z).
\label{rhobar}
\eeq
Next, form
\beq
\bar{\rho}' \equiv \bar{\rho}(x,y,z) - \bar{\rho}(x/f, y/f, z/f).   \label{rhotilde}
\eeq
where $f \ne 1$ is a numerical factor.
$\bar{\rho}'$ produces an octupole at the centre of symmetry. This can be proved by evaluation
of the second derivatives of $V({\bf r}) \propto \int \bar{\rho}'({\bf r}') /
|{\bf r} - {\bf r'}| d\tau$ at the origin. A proof giving further physical
insight is as follows. $\bar{\rho}(x/f, y/f, z/f)$
produces the same functional form of $V(x,y,z)$ as $\bar{\rho}(x,y,z)$ but on a
distance scale expanded by a factor $f$.
Modelling $\bar{\rho}$ as a set of
point charges, each charge in the rescaled distribution
is now further from the origin by a factor $f$, and
is increased in size by a factor $f^3$. Since the second derivatives of $V$
due to a point charge go as (distance)$^{-3}$, it follows that the second derivatives
of the new potential are equal and opposite to those of the original one, QED. The fourth
derivatives  scale as (distance)$^{-5}$, so
\beq
\frac{\partial^4 \bar{V}'}{\partial x_i^4} = \left(1 - \frac{1}{f^2}\right)
\frac{\partial^4 \bar{V}}{\partial x_i^4}.
\eeq
where $\bar{V}'$ ($\bar{V}$) is the potential due to $\bar{\rho}'$ ($\bar{\rho}$)
respectively.

A natural extension of this method can produce higher order multipole configurations.

(E). An octupole can be formed by taking almost any charge distribution
$\rho(x,y,z)$, and displacing and reflecting it in three dimensions.
A general displacement to $\rho(x-x_0, y-y_0, z-z_0)$
has three free parameters $(x_0, y_0, z_0)$. One can be absorbed into an overall
scale factor, leaving two. This is just enough free parameters to allow the
two constraints
\beq
\pd{V}{x}(0,0,0) &=& 0   \nonumber \\
\pd{V}{y}(0,0,0) &=& 0   \label{constraint}
\eeq
to be satisfied, where $V$ is the potential due to $\rho(x-x_0, y-y_0, z-z_0)$,
except in rare cases when the form of $\rho$ leads to
no solution. The nulling of $\partial^2 V/\partial z^2$
follows from
Laplace's equation. Finally, introduce reflection symmetry as in (\ref{rhobar}),
which causes odd-order moments and mixed second derivatives ($\partial^2 V
/\partial x \partial y$ etc.) to vanish.

Another way to understand this construction
is to argue that most charge distributions produce a potential with zero second
derivatives with respect to fixed orthogonal axes at some point in space;
it suffices to place the origin at such a point
and then introduce reflection symmetry. For example, if $\rho$ corresponds to
a single point charge, this construction leads to point charges on the corners
of a cube. More generally if $\rho(x,y,z)$ lies in the $x$-$z$ plane, this construction
leads to a distribution of charge in two parallel planes at $y = \pm y_0$.

(F). All the above constructions involve constraints on the locations of the image charges
in addition to the basic assumption of three-fold reflection symmetry. Our next
construction does not. An octupole can always be formed
(with rare exceptions arising
in pathalogical cases) by a charge distribution
$\bar{\rho}$ with reflection symmetry as in (\ref{rhobar}), based on
\beq
\rho(x,y,z) = \rho_0 + f_1 \rho_1 + f_2 \rho_2  \label{rho012}
\eeq
where $\rho_i(x,y,z)$ ($i=0,1,2$) are three different charge distributions located
anywhere in the positive octant, and $f_1$ and $f_2$ are parameters whose
values are given by solving the simultaneous equations
\beq
\left.
\begin{array}{ccc}
  \pd{V_0}{x} + f_1 \pd{V_1}{x} + f_2 \pd{V_2}{x} &=& 0 \\
  \pd{V_0}{z} + f_1 \pd{V_1}{z} + f_2 \pd{V_2}{z} &=& 0
\end{array}
\right\}          \label{f1f2}
\eeq
(evaluated at the origin),
in which $V_i$ are the potentials due to $\rho_i$. This construction uses
the same property invoked in the previous one, namely
that after cancellation of odd-order moments there are only two
constraints (\ref{constraint}) we need to satisfy. This
implies that we only need two free parameters, and these are provided by
$f_1$ and $f_2$.

(G). When we abandon a reflection symmetry, the octupole can be regained
by inserting further charges. For example, suppose we only have reflection
symmetry in the $x$--$y$ and $y$--$z$ planes. This is the case, for example,
when all the charges are located in or on a single
substrate, with the octupole centred above the top surface of the substrate.
Then in addition to (\ref{constraint}) we have 4 further constraints,
on $\partial V/ \partial y,\, \partial^3 V / \partial y^3,\, 
\partial^3 V / \partial x^2 \partial y,\, \partial^3 V / \partial z^2 \partial y$. One or more of these 
can be satisfied by a careful placement of a single
charge distribution. Alternatively, charges 
are placed at any convenient position, and 
then their magnitudes are adjusted, similar to construction (F). 
To obtain the desired behaviour of the trapping centres, we
don't need to insist on an octupole, however. Of these further
constraints, only $\partial V/\partial y = 0$ is strictly necessary.

\subsection{Application to electrode design}
\noindent

The distributions of charge discussed above are to be realised on the surfaces
of a set of conducting electrodes at fixed voltages (or else they
are image charge distributions which lead to the same equipotential
surfaces). We wish to avoid complicated
electrode shapes so we restrict ourselves to electrodes approximating to
simple combinations of lines, rings, or sheets.

In order to satisfy $N$ constraints on derivatives of $V(x,y,z)$ 
merely by adjusting voltages on
electrodes, it is necessary to have $N+2$ electrodes, since one
defines a voltage zero, and the next one sets an overall scale factor.
In order to produce radial confinement, a set of r.f. electrodes is introduced
around (or in between) the d.c. ones which create the octupole. These r.f.
electrodes also serve to define the d.c. zero. Thus, in construction (F)
we require one r.f. voltage at d.c. ground, and 3 adjustable d.c. voltages;
in construction (G) we require one r.f. voltage at d.c. ground,
and 7 adjustable d.c. voltages.

Most of the constructions presented in the previous section can be ruled out,
as follows.

The case of cubic symmetry (A) involves 8 electrodes, but to preserve precise
symmetry these would have to lie along the edges or the extended major diagonals of
a cube, which is difficult to fabricate, and a linear r.f. electrode set
would also break the symmetry.

Case (B) can be realised with just two ring-shaped electrodes.
This involves the fewest electrodes. For a Penning trap one
would choose the magnetic field direction as the axis of symmetry
of the rings. For a Paul trap the need to provide an additional r.f. quadrupole
producing radial confinement in the $x-y$ plane suggests that there
will be further electrodes parallel to the $z$ axis. This implies
the fabrication is more straightforward if the rings are placed
around another axis, which we take to be the $y$ axis, so that
all the electrodes lie in a set of parallel planes.
The r.f. electrodes then destroy the circular symmetry. To
regain a simple design, the rings can be replaced with rectangular
loops, c.f. construction (C).

The other cases can all be fabricated in layers.
The method of (\ref{rhotilde}) (case (D)) involves up to 16 electrodes in
total or 8 if they all lie in a plane; the method by displacement
(case (E))
involves 8 electrodes in two parallel planes, the method
of (\ref{rho012}) (case (F)) involves up to 24 electrodes in general,
or up to 12 if they all
lie in a plane, which we take to be the $x-z$ plane.
This can be reduced to 10 if one pair lies on
the $x$ axis. We avoid placing electrodes along the $z$ axis,
because we want to be able to bring ions into
and out of the structure by moving them along this axis.

The minimal number of electrodes for a perfectly constructed system
is not the only consideration, however.
Manufacturing imprecision, patch potentials, and the
desire to adjust the trap separation all imply that further electrodes
will be needed to trim the system. The electrode structures which rely on
precise construction to produce an octupole (cases (A)-(E)) require more 
of these extra electrodes than those which 
depend on variable voltages (cases (F) and (G)).

Screening effects arise when we consider
conducting electrodes rather than merely charge distributions. The innermost
electrodes tend to act like a Faraday cage to modify and screen the effects of
the outer ones. Let $P$ be the centre of the region in which we require
a certain potential form (here, $P$ is at the origin).
A rough rule of thumb is that if there is not an unobscured line of sight from
$P$ to a given electrode, then the influence
of that electrode on the potential at or near $P$ is greatly modified by the
obstructing electrode. However, the outer charge distribution in (\ref{rhotilde})
is, by construction, exactly on an extended line from the origin to the inner
charge distribution. Therefore an electrode set built with the aim of
realising the recipe (D) suffers from screening effects and
it is difficult to obtain a surface charge distribution of the form
(\ref{rhotilde}).

In the remainder of this paper we will concentrate on the case
(F) of 10 electrodes all in the $x$-$z$ plane, of which one pair
lies along the $x$ axis, the case (E) of 8 or more electrodes
in two planes at $y = \pm y_0$, and the case (G) of an octupole
situated above all the electrodes. For the sake of completeness, the
appendix lists some other examples of octupoles
based on simple distributions of point, line and ring charges.

\section{Optimization of some example structures}  \label{s:opt}
\noindent

In this section we study various electrode arrangements
and obtain the geometric factors $\gamma$ (eq. (\ref{gam})) and $\mu$ (eq. (\ref{mu})) in each case.
We optimize the most promising designs by adjusting electrode positions
and relative sizes in order to maximise both factors.

In all the configurations to be discussed, we choose the $z$ axis
along the line where the twin traps are situated, the $x-z$ plane is parallel to the planes
containing sets of electrodes, and hence these planes
are separated along the $y$-direction. 

The geometric factors $\gamma$ and $\mu$ were obtained by the
numerical method described in section \ref{s:NIST}.
To obtain $\mu$ the r.f. electrodes were assigned a large voltage
and the second derivative of $V(x,y,z)$ 
was calculated along the principal axes of the quadrupole
perpendicular to the $z$ axis. In most cases these were the $x$ and
$y$ directions (giving $\mu_x$ and $\mu_y$). Where this is not the
case the radial geometrical parameters are denoted by $\mu_{x'}$
and $\mu_{y'}$. Where $\mu_x$ and $\mu_y$ differ, a component of
the oscillating quadrupole potential lies along $z$, hence trapped ions 
would experience axial micromotion and the associated pseudopotential
would contribute to their confinement along the $z$ direction, 
as discussed at the end of section \ref{s:normalmode}.

Since the numerical solution of Laplace's equation can be slow, we first gained some general
insights by analytic methods. We roughly modelled the electrodes as a
set of line charges, and wrote down the complete potential function. This permits
$\gamma$ to be found as a function of various parameters, such as the line charges
and their positions, and it can be maximised analytically. Results from
such analyses gave a starting point for the electrode structures used in 
the numerical studies.

In order to know what value of $\gamma$ and $\mu$ one might hope
for, we numerically calculated these parameters for two simple cases.
These can be regarded as a `standard' against which the results for our various
electrode structures can be measured.
As a standard for $\mu$, we used four cylindrical electrodes of radius 0.41254$a$ with
axes aligned along the $z$ axis, and centred at the corners of a square of
side $2a$. This arrangement gave $\mu = 0.20$. Electrodes of
radius $0.3a$ centred at the same positions gave $\mu = 0.15$.  

To obtain a standard value for $\gamma$, an electrode structure 
loosely based on the equipotentials of the axially symmetric 
octupole in equation \ref{Vaxoct} was used. This consisted of three toroidal
electrodes centred on the $z$ axis at $z = 0$ and $z = \pm 0.643a$, with two
spherical ``end-cap'' electrodes of radius $0.35a$ at $z = \pm a$. The
cross-section radii of the torii tubes and the radii of the centres of
the tubes were $0.2a$ and $a$ respectively for the $z = 0$ torus
and $0.2a$ and $0.766a$ respectively for the $z = \pm 0.643a$ torii. This
configuration gave $\gamma = 0.133$.

\subsection{8 d.c. electrodes in 2 planes, 2 r.f. electrodes}
\label{sec:8dc2rf}
\noindent

\begin{figure} [htbp]
\centerline{\epsfig{file=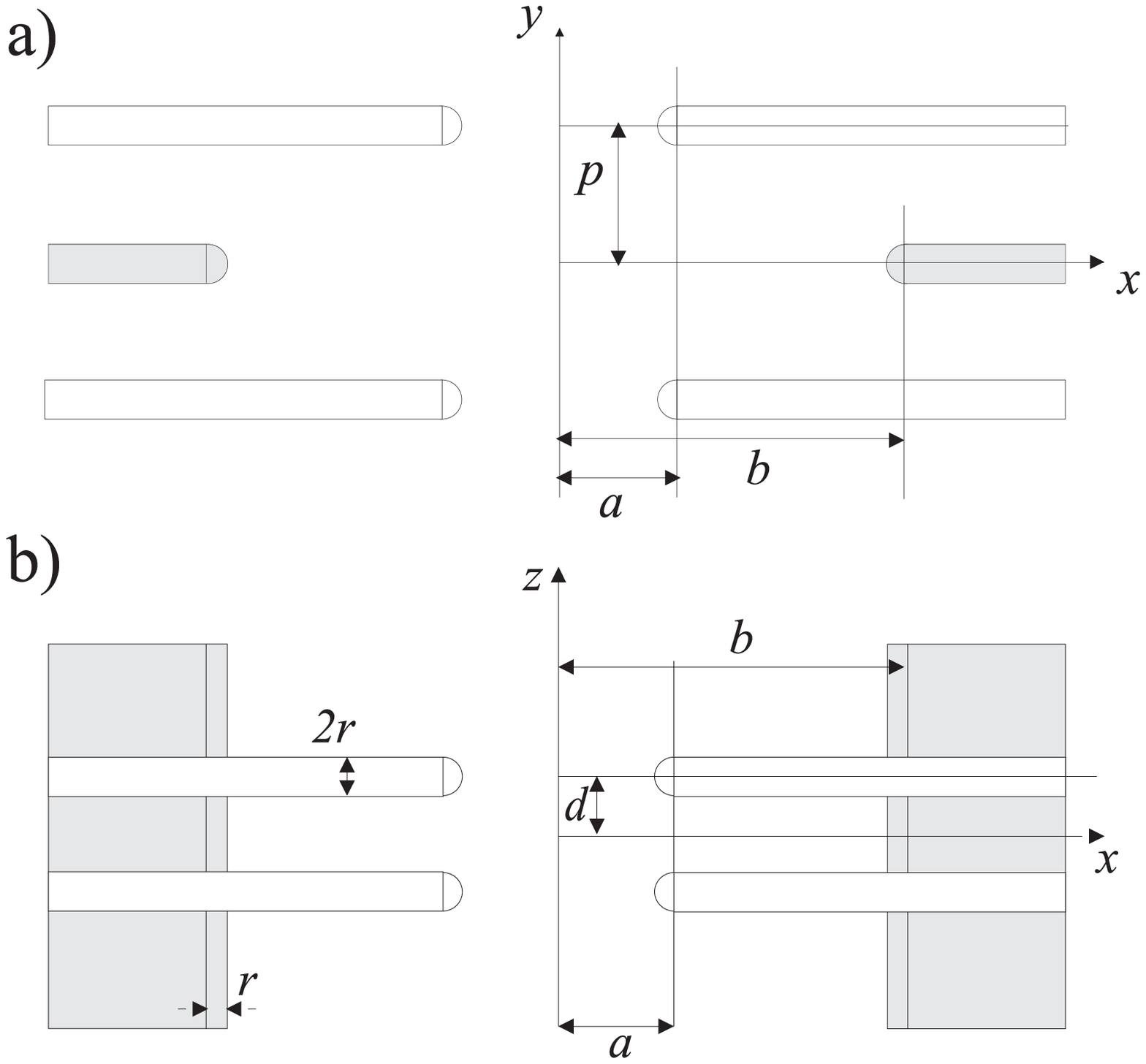, width=6.0cm}} 
\vspace*{13pt}
\fcaption{\label{fig:sandwich}Views of the electrode structure with 4 d.c. electrodes in each of two
planes, and 2 r.f. electrodes. a) cross-section in the $x-y$ plane showing
the sandwich structure.  The r.f. electrodes are filled grey. b) View 
looking down on the $x-z$ plane from above.}
\end{figure}

Let us first consider structures consisting of rod d.c. electrodes placed at a distance $p$ above 
and below the plane containing the r.f. electrodes. This ``sandwich'' arrangement is 
shown in figure \ref{fig:sandwich} a). It is based on construction (E) of section
\ref{s:image}. All the d.c. electrodes are at the same voltage.

\begin{figure} [htbp]
\centerline{\epsfig{file=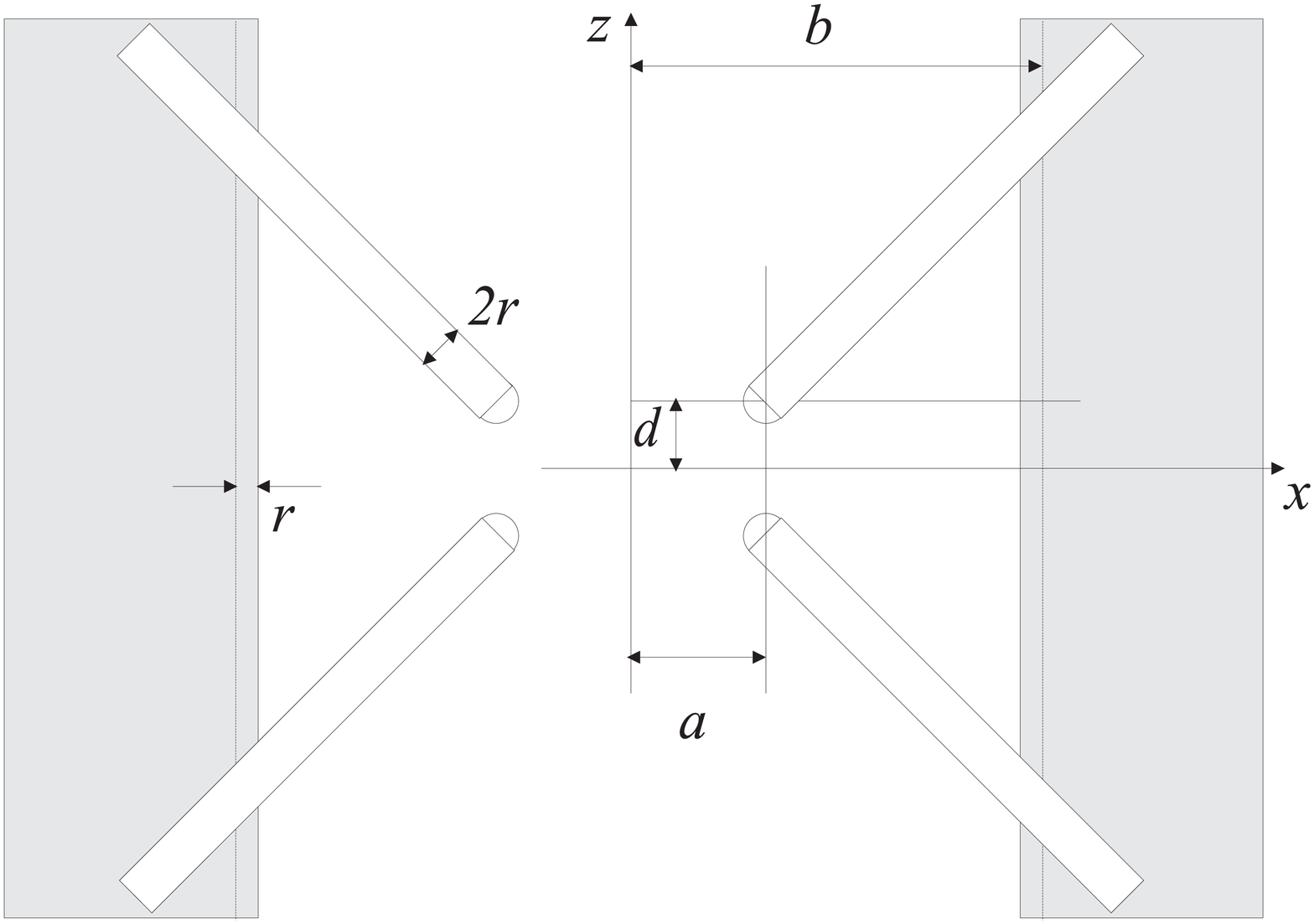, width=6.0cm}} 
\vspace*{13pt}
\fcaption{\label{fig:diag}A view of the electrode structure with 45 degree rod electrodes having 
hemispherical ends centred at ($\pm a, \pm p, \pm d$). The r.f. electrodes have 
thickness $2r$, and lie in the $y = 0$ plane. The end of these 
electrodes are hemicylinders of radius $r$ and centered on $x = \pm b, y = 0$.  
The grey fill represents the r.f. electrodes.}
\end{figure}

Plan views of the two d.c. electrode structures tried are shown 
in figures \ref{fig:sandwich}b and \ref{fig:diag}. 
We let $a$ set the overall distance scale, and then there are
4 parameters ($b$, $d$, $p$, $r$) and 2 constraints. The octupole 
configuration was found by fixing $r$ and $b$, and then adjusting $p$ and $d$.

The equipotential along
the r.f. electrodes tends to prevent an octupole being produced at the origin.
We found that it was necessary to restrict the influence of this equipotential
by keeping $p$ small. Since $p$ must be larger than the electrode diameter
this means that $r$ must not be too large. 

Results of the calculations for both structures are presented in table \ref{tab:8dc2rf}. 
The two cases produce similar values of $\gamma$ and $\mu$, those (labelled 1.1)
for the structure shown in fig. \ref{fig:sandwich}b being somewhat higher than those
(labelled 1.3) for
the structure shown in fig. \ref{fig:diag}. It was found that the r.f. electrodes
had to be placed well back (a large value for $b$) in order to get a high
value for $\gamma$.

Results 1.1 and 1.2 show that the value of $\mu$ decreases 
as $b$ is reduced.  This is because $p$ and $d$ also decrease,
with the result that the influence of the   
r.f. electrodes is screened by the d.c. electrodes.

The geometry shown in figure \ref{fig:sandwich} provides greater screening of the
r.f. electrode by the d.c. electrodes, compared to that shown in figure \ref{fig:diag}, and this
is probably the main reason why it provides the higher $\gamma$.

\vspace*{4pt}   
\begin{table}[hb]
\tcaption{Values of the distance parameters $r,b,p,d$ which give
an octupole for three example electrode 
structures, with the distance to an electrode surface $\rho$ and the
calculated geometric factors $\gamma$ and $\mu$ in each case. Results 
1.1 and 1.2 are for electrode configurations of the form shown in
figure \ref{fig:sandwich}. Result 1.3 is for an electrode 
configuration like that shown in figure $\ref{fig:diag}$. }
\centerline{\footnotesize\smalllineskip
\begin{tabular}{|l|l|l|l|l|l||c|c|c|} \hline
Result & $r$ & $b$ & $p$ & $d$ & $\rho$ & $\gamma$ & $\mu_x$ & $\mu_y$\\
\hline units & ($a$) & ($a$)& ($a$)& ($a$) & ($a$)& $10^{-3}$ & $10^{-3}$ & $10^{-3}$ \\
\hline
1.1 & 0.30 & 6.30 & 1.94 & 2.01 & 2.67 & 26 & 6.40 & 6.45 \\
1.2 & 0.37 & 3.34 & 1.16 & 1.28 & 1.63 & 18 & 3.44 & 3.77 \\
\hline
1.3 & 0.26 & 4.95 & 1.20 & 0.88 & 1.53 & 15 & 4.80 & 3.14 \\
\hline
\end{tabular}}
\label{tab:8dc2rf}
\end{table}

\subsection{10 d.c. electrodes in a plane, 4 r.f. electrodes}
\label{sec:10dc4rf}
\noindent

\begin{figure} [htbp]
\centerline{\epsfig{file=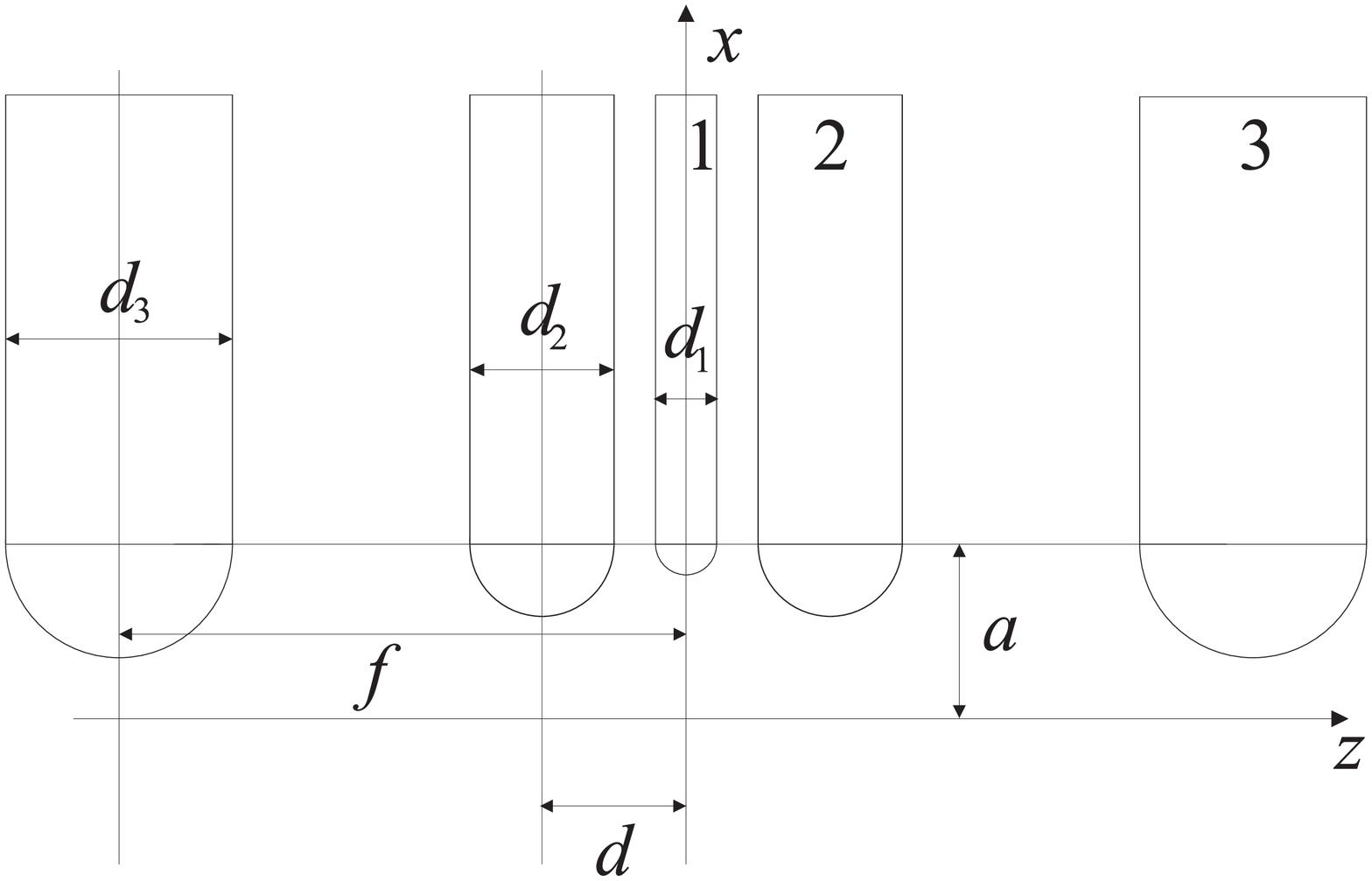, width=6.0cm}} 
\vspace*{13pt}
\fcaption{\label{fig:5Electrodes}10 d.c. electrodes, 2 r.f. electrodes: A cross section 
view of the d.c. electrodes in the $x-z$ plane with $y = 0$ for positive $x$.  
The trap centre is at the intersection of the x and z axes. The electrode structure is 
repeated for negative $x$ by reflection in the $z-y$ plane.}
\end{figure}

This structure consists of 10 d.c. electrodes in the $y = 0$ plane,
and 4 r.f. electrodes in two planes parallel to this.
It has reflection symmetry in $x - y$, $x-z$ and and $y - z$ planes. 
The d.c. electrodes are cylindrical, with hemispherical ends centred
at $x = \pm a$. Figure \ref{fig:5Electrodes} shows a cross 
section in the $x-z$ plane for positive $x$.

Each r.f. electrode
is effectively a semi-infinite plane, with its edge parallel
to the $z$ axis.  The $y$ position of the r.f. electrodes is $\pm p$, 
and the $x$-position is $\pm (a + t)$, hence $t$ gives the retraction of the r.f. electrodes. 
Where $p = 1.03$ the r.f. electrode had thickness $a/4$, otherwise the thickness was negligible. 
For the finite thickness cases, the edge of the electrode is hemicylindrical, and
$p$ and $t$ give the position of
the axis of the hemicylinder. An electrode structure with $p = 1.03a$ and $t = a$ is illustrated in figure 6.

\begin{figure} [htbp]
\centerline{\epsfig{file=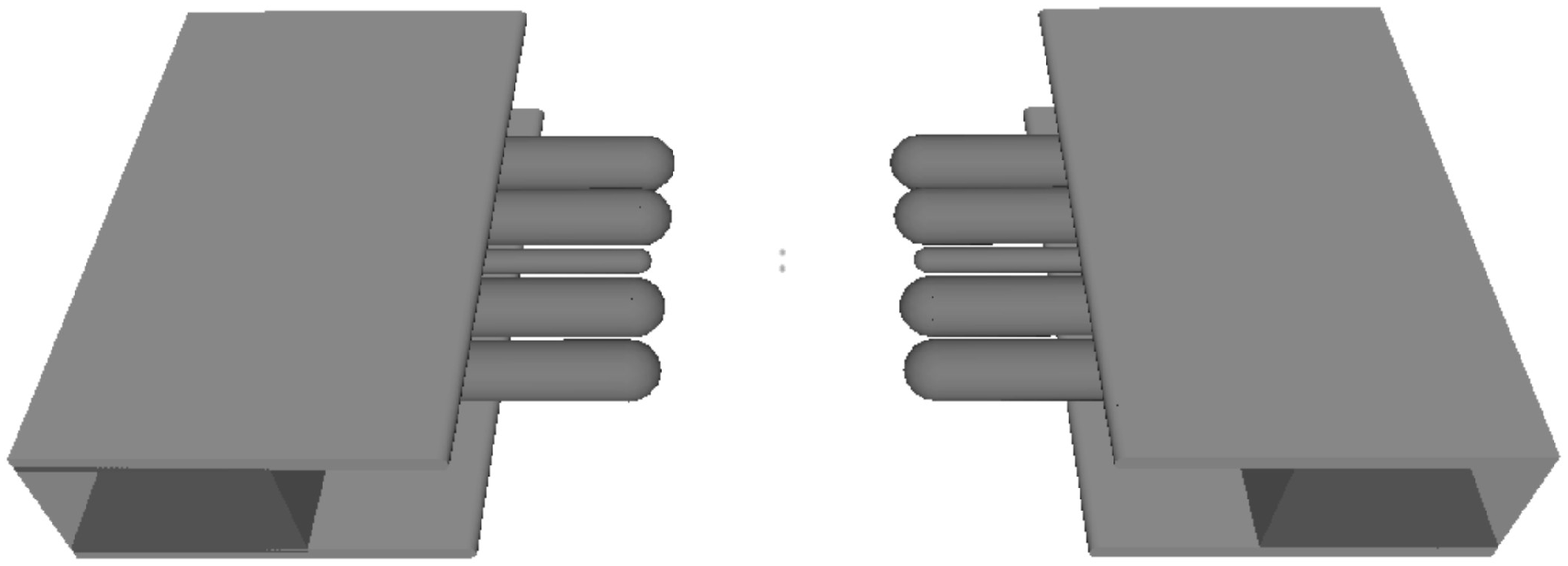, width=6.0cm}} 
\vspace*{13pt}
\fcaption{\label{fig:3D5Electrodes4}A perspective view of the electrodes for r.f. electrodes positioned at $p = 1.03a$ and $t = a$. The 10 cylindrical 
electrodes have d.c. voltages applied to them. r.f. voltages are applied to the  planar 
electrodes. Also shown are insulating spacers which hold the electrodes in place.}
\end{figure}

The voltage on electrode 3 (V$_3$) was set to 1.4V in all the calculations in this
section. With a fixed geometry our free parameters are the voltages on 
electrodes 1 and 2 (V$_1$ and V$_2$ respectively). These were adjusted
to fulfill the octupole condition, using a two-dimensional Newton-Raphson 
method in order to find the solution with few trials. Then
the values of $\beta$, $E_{\rm max}$ 
and $\gamma$ were calculated. Adjustments were then made to $f$, $d$ and $d_3$, and 
the calculation repeated. A selection of positions and their corresponding values 
of $\gamma$ are shown in table \ref{tab:10dc4rfoct}.  The nearest electrode
surface distance $\rho = 0.825 a$ in all cases.

The initial choice of electrode diameters was $d_1 = 0.35 a$, $d_2 = 0.825 a$, $d_3 = 1.3 a$, 
based on the indications from our initial line charge calculations. Using these diameters, CPO 
calculations were carried out with no r.f. electrodes for several values of $f$ and $d$. 
The optimal values were found to be $f = 3.25 a$ and $d = 0.825 a$.

\begin{figure} [htbp]
\centerline{\epsfig{file=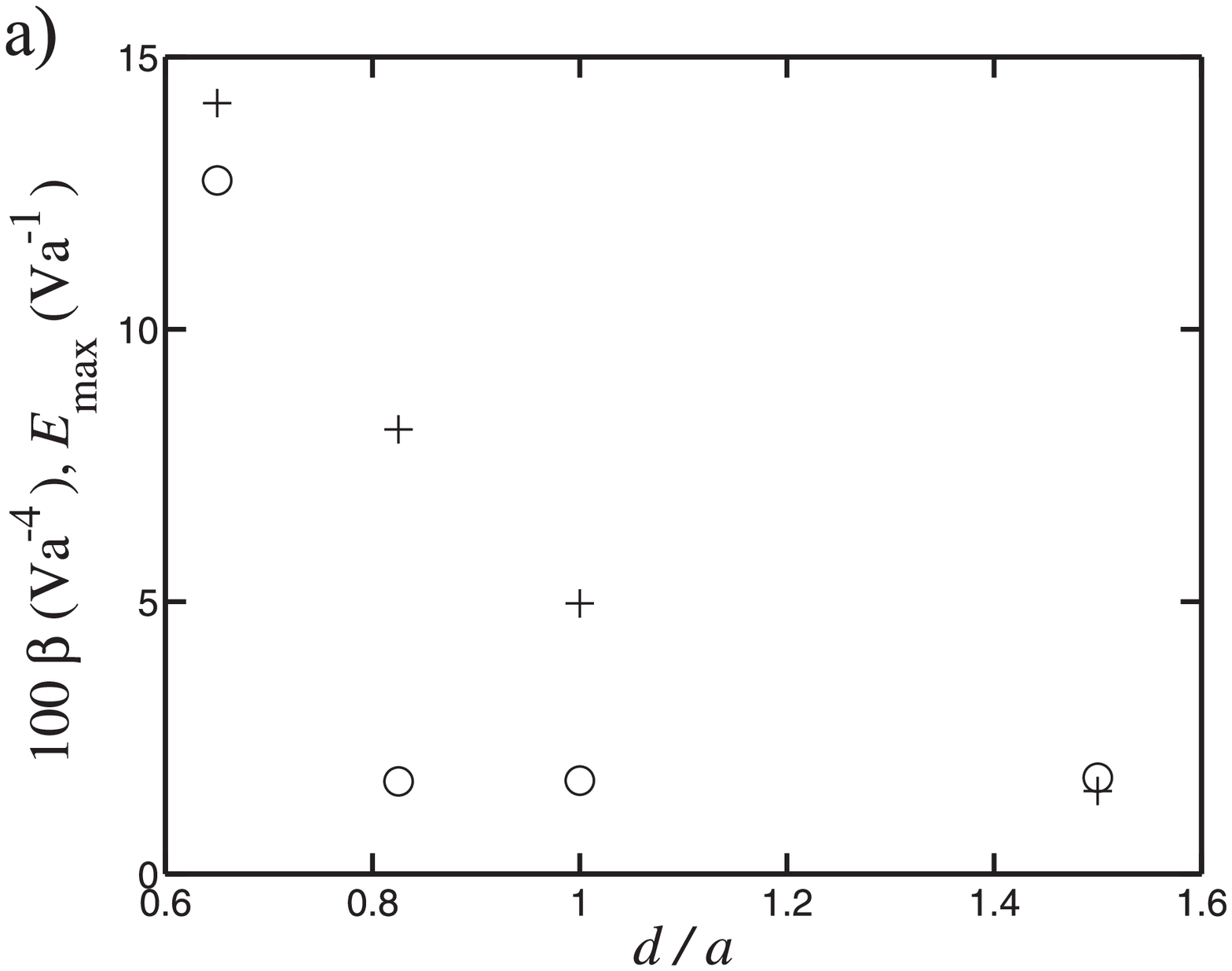, width=6.0cm}, \epsfig{file=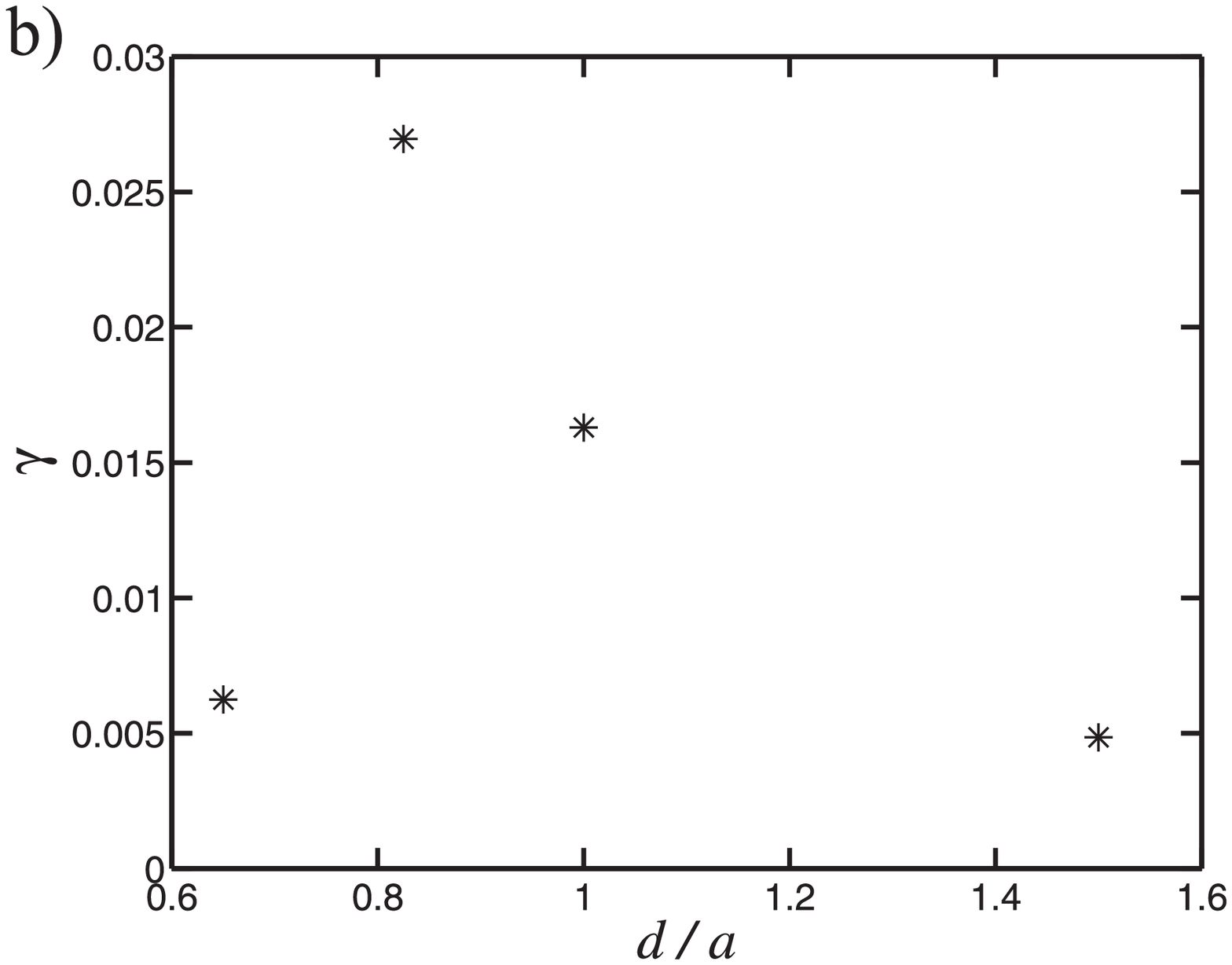, width=6.0cm}} 
\vspace*{13pt}
\fcaption{\label{fig:changed}The effect on $\beta$, $E_{\rm max}$ and $\gamma$ of adjusting $d$ for result 2.1 in table \ref{tab:10dc4rfoct}. 
$100 \beta$ is represented by $+$, and $E_{\rm max}$ by unfilled circles. If the largest field 
lies between the first and second electrodes, then reducing $d$ increases both $\beta$ 
and $E_{\rm max}$, and $\gamma$ drops. Where this is not the case, the increase in $\beta$ 
causes an increase in $\gamma$.}
\end{figure}

We then introduced the r.f. electrodes. With the r.f. electrodes far 
away from the d.c. electrodes ($p = 2a$ and $4a$) the 
maximum field is found between the d.c. electrodes, hence varying $f$ makes little 
difference to the value of $\gamma$. This is because an increase (decrease)
in $\beta$ is countered by an increase (decrease) in $E_{\rm max}$.  Cases 2.1 
and 2.2 in table \ref{tab:10dc4rfoct} illustrate this.  For large $d$, the 
maximum field is found between electrodes 2 and 3, hence as $d$ is
reduced, $E_{\rm max}$ decreases. For small $d$, the maximum field is
found between electrodes 1 and 2, and then as $d$ is reduced, 
$E_{\rm max}$ increases. This
behaviour is shown in figure \ref{fig:changed}.

\vspace*{4pt}   
\begin{table}[hb]
\tcaption{Results from CPO calculations. Results 2.1-2.6 are for the structures described 
in section \ref{sec:10dc4rf} (c.f. figures \ref{fig:5Electrodes}, \ref{fig:3D5Electrodes4}),
results 3.1-3.3 are for the inverted structures described in section \ref{sec:20dc2rf}.
All the cases listed have $d = 0.825 a$, $d_1 = 0.35 a$, $d_2 = 0.825 a$, $V_{\rm rf} = 0$
and $V_3 = 1.4$V. The table illustrates effects of changing the values of $f$, $d_3$, $p$ and $t$. 
}
\centerline{\footnotesize\smalllineskip
\begin{tabular}{|l|l|l|l|l|l|l||c|c|c||c|c|} \hline
\emph{\textup{Result}} & $f$ & $d_3$& $p$ & $t$ & \emph{$V_1$} & \emph{$V_2$} & $10^4 \beta$ & $E_{\rm max}$  &\emph{$\gamma$} & $\mu_x$ & $\mu_y$ \\
\hline  & (a) & (a) & (a) & (a) &  (V) & (V) & (V$/a^{4}$)&(V$/a$)& $10^{-3}$ & & \\ 
\hline 
 2.1 & 3.25  & 0.65  & 2 & 1 & 0.4605 & 0.2156  & 816 & 1.7 & 27 & 0.121 & 0.148 \\
 2.2 & 2.513 & 0.65  & 2 & 1 & 0.6657 & 0.2501  & 1421 & 2.6 & 31 & 0.111 & 0.147 \\
 2.3 & 3.25  & 0.413 & 4 & 3 & 0.6392 & 0.5065 & 500 & 1.9 & 14 & 0.096 & 0.110 \\
 2.4 & 3.25  & 0.65  & 1.03 & 0 & 0.0973 & -0.0407 & 519 & 9 & 3.3 & 0.053 & 0.068 \\
 2.5 & 3.25  & 0.65  & 1.03 & 1 & 0.3085 & 0.0650 & 764 & 9 & 4.8 & 0.030 & 0.037 \\
 2.6 & 1.9   & 0.413 & 1.03 & 1 & 0.5218 & -0.0055 & 1819 & 5 & 20 & 0.053 & 0.069 \\
 \hline
 3.1 & 3.25  & 0.65  & 2 & 1 & 6.183 & -1.207 & 506 & 55 & 5.0 & 0.150 & 0.152 \\
 3.2 & 3.25  & 0.65  & 1 & 1 & 0.3936 & -0.2378 & 1387 & 15 & 6.2 & 0.046 & 0.043 \\ 
 3.3 & 3.25  & 0.65  & 2 & 3 & 4.969 & -0.3090 & 86 & 32 & 10 & 0.027 & 0.030 \\
\hline
\end{tabular}}
\label{tab:10dc4rfoct}
\end{table}

\begin{figure} [htbp]
\centerline{\epsfig{file=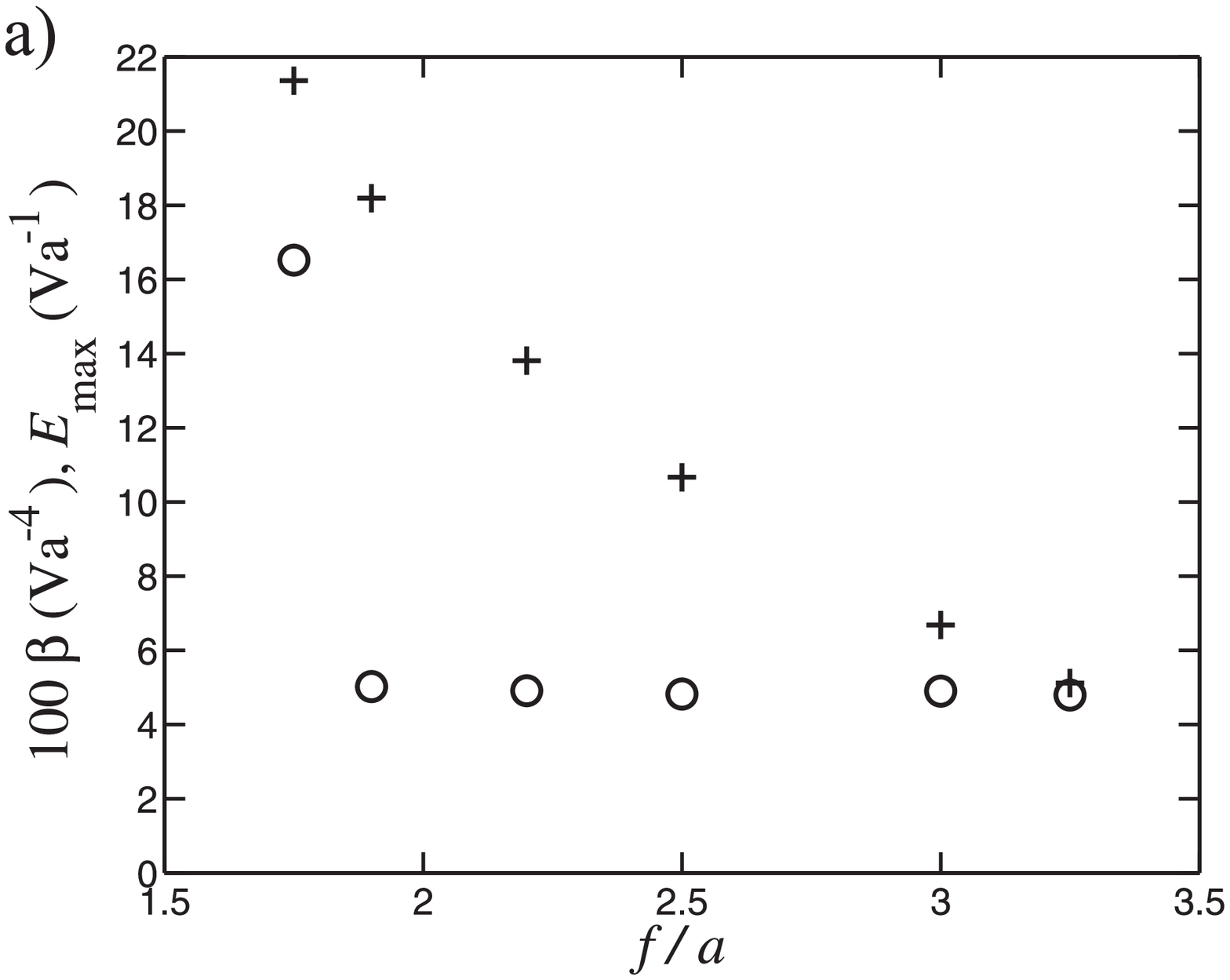, width=6.0cm}, \epsfig{file=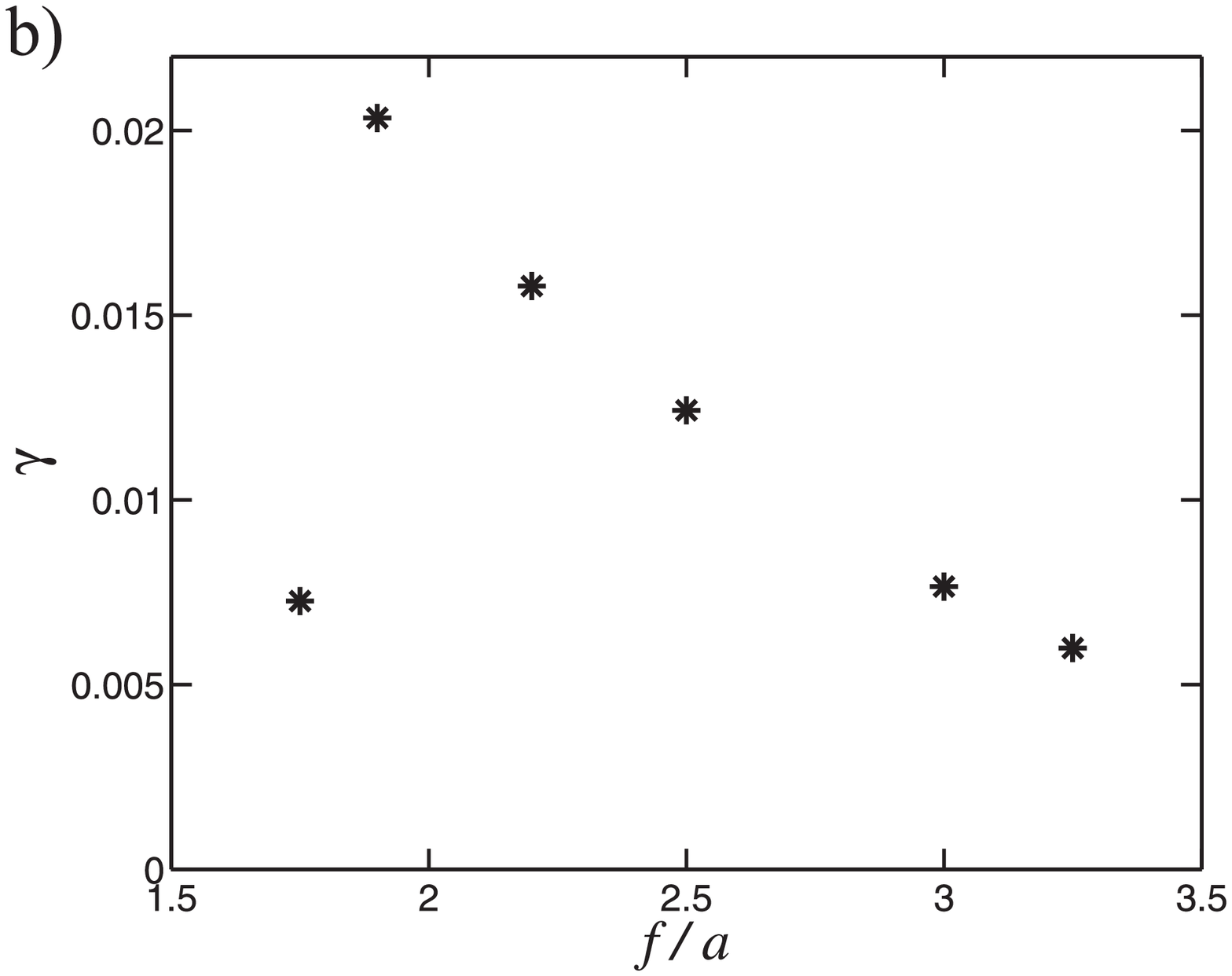, width=6.0cm}} 
\vspace*{13pt}
\fcaption{\label{fig:changef}The effect on $\beta$, $E_{\rm max}$ and $\gamma$ of adjusting $f$, with $d = 0.825 a$, $d_1 = 0.35 a$, $d_2 = 0.825 a$, 
$d_3 = 0.825 a$, r.f. electrodes at $p = 1.03a$, $t = 0$. $100 \beta$ is 
represented by $+$, and $E_{\rm max}$ by circles. 
Reducing $f$ has the effect of increasing $\beta$. At larger $f$, the maximum field is found between the r.f. electrodes and the 
d.c. electrodes, hence we can increase $\gamma$ by decreasing $f$. 
After reaching the point where the largest field is found between the d.c. electrodes,  
$\gamma$ is much reduced for further reduction of $f$.}
\end{figure}

When the r.f. electrodes are close to the d.c. electrodes ($p = 1.03a$), the maximum field is found between the d.c. electrode 3 and
the r.f. electrodes. This field can be reduced by reducing $d_3$.
$\beta$ increases as $f$ is reduced, hence $\gamma$ can also be increased
by reducing $f$ until the maximum field is found between the d.c. electrodes.
Figure \ref{fig:changef} shows this behaviour for structures with $p = 1.03a$ and $t = 0$,
and the net effect is illustrated by results 2.5, 2.6 in table \ref{tab:10dc4rfoct}.

We also examined the effect of making electrode 3 extend further along the $z$
direction. It was replaced by a planar electrode of thickness $d_3$ with
edges rounded off by hemicylinders, located such that the edge nearest the
origin was situated at the same place as the cylindrical electrode in
the other results. This had little effect on $\gamma$ or $\mu$. 

Finally, we examined an example in which the d.c. electrodes are
extended along the $z$ direction into planar shapes. This will 
reduce $\gamma$, but it also reduces the $z$-dependence
of the a.c. part of the potential, therefore giving less
micromotion along $z$. Each electrode had two flat surfaces parallel 
to the $x-z$ plane, and sides in the shape of hemicylinders having the
same radius of curvature as before. The electrode centres were placed 
as in result 2.6 and they were extended along the $z$ direction
until the gaps between them were $0.1\, a$. This arrangement
gave $\gamma = 2 \times 10^{-3}$, i.e. a factor 10 reduction compared
to result 2.6.

\subsection{20 d.c. electrodes in two planes, 2 r.f. electrodes}
\label{sec:20dc2rf}
\noindent

This arrangement is like the previous one, but with the d.c. and r.f. electrode positions swapped. 
It consists of 2 planar r.f. electrodes of thickness $0.5a$ lying in the $y = 0$ plane. These are bounded by hemicylindrical edges  centred at $x = \pm(a + t)$. The d.c. electrodes lie in two 
planes at $y = \pm p$, each containing two sets of five d.c. electrodes
(with $f$, $d$ and electrode diameters the same as result 2.1 in table \ref{tab:10dc4rfoct}). 
Results for these arrangements are shown in table  \ref{tab:10dc4rfoct} as results 3.1-3.3. The values of $\gamma$ and $\mu$ are similar to those in section \ref{sec:10dc4rf}. Some improvement in $\gamma$ would be expected through optimisation of the relative positions of the d.c. electrodes. This has not been carried out in obtaining results 3.1-3.3.

\subsection{Low Aspect Ratio Electrode Configurations}
\noindent

\begin{figure} [htbp]
\centerline{\epsfig{file=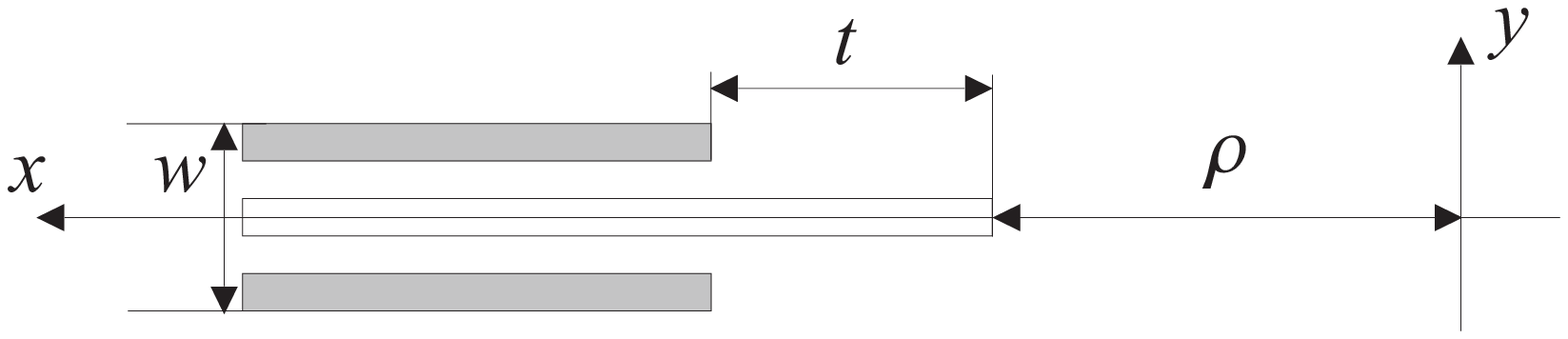, width=6.0cm}} 
\vspace*{13pt}
\fcaption{\label{fig:lowasdia}A cross-section view of a low aspect ratio electrode structure for positive $x$ showing the 
definition of the parameters $w$ and $t$. The r.f. electrodes are represented by grey fill. 
The electrodes for negative $x$ are obtained by a reflection in the $x = 0$ plane.}
\end{figure}

Let $w$ be the total thickness, i.e. depth in $y$, of the electrode structure
(c.f. figure \ref{fig:lowasdia}). We define the aspect ratio $g=w/\rho$.
The three-layer trap designs considered in sections \ref{sec:8dc2rf}, \ref{sec:10dc4rf} 
and \ref{sec:20dc2rf} have $g > 1$.
Though microfabrication techniques are available  which could be used to construct
traps with $\rho \le 100$ $\mu$m and $g > 1$ \cite{04:Reid}, many methods are 
limited to a maximum structure thickness of 
around $20\mu$m \cite{04:Madsen, 04:Blain}.
This means that traps may have a low aspect ratio, $g < 1$.

\begin{figure} [htbp]
\centerline{\epsfig{file=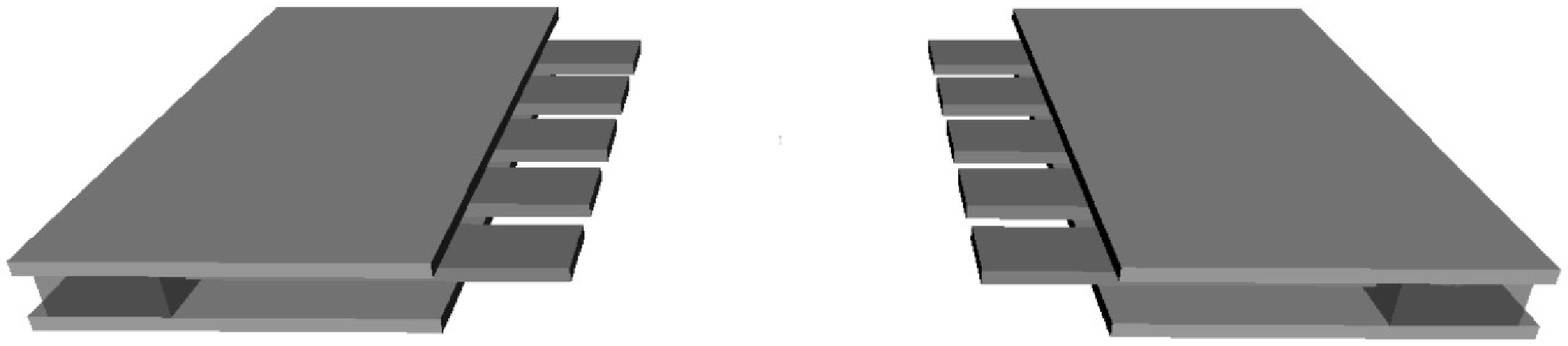, width=6.0cm}} 
\vspace*{13pt}
\fcaption{\label{fig:lowas}Perspective view of a low aspect ratio electrode structure with $g = 0.4$ and $t/\rho = 0.6$.}
\end{figure}

We studied structures similar to those of section
\ref{sec:10dc4rf}, but now with $g < 1$. To speed the calculation process
we used cuboid shapes for the electrodes (see figures \ref{fig:lowasdia} and
\ref{fig:lowas}). 
The d.c. electrodes had width in $z$ of $d_1 = d_2 = d_3 = w$, and were separated
by gaps of $w /2$ (hence $d = 3w/2$, $f = 3w$). The thickness of the electrodes in $y$ 
was $w/5$. With a cuboid shape
the electric field diverges near the electrode edges. 
In order to calculate $\gamma$ and $\mu$ we need to make some
reasonable assumption to handle this. Our aim was to use the calculation to
indicate approximately what would happen if in fact the electrode
edges were smoothed. To this end we calculated electric field values
at $(x,y,z) = (\rho, 0, 0),\, (\rho+w/5, 0, d_1/2), (\rho+t+w/5, w/10, 0)$
and at similar places around the other electrodes, and took $E_{\rm max}$
to be the largest of these.

\begin{figure} [htbp]
\centerline{\epsfig{file=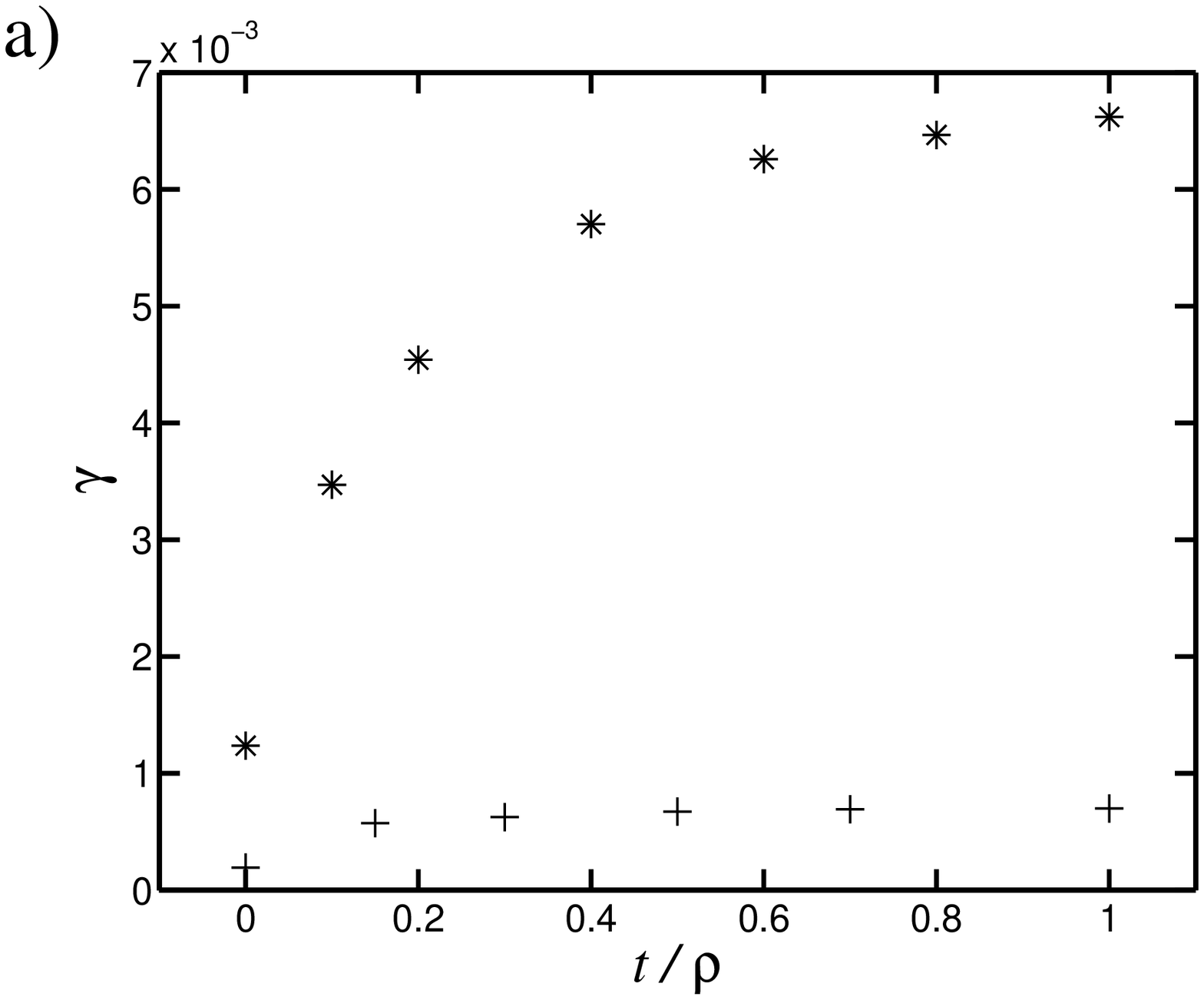, width=6.0cm}, \epsfig{file=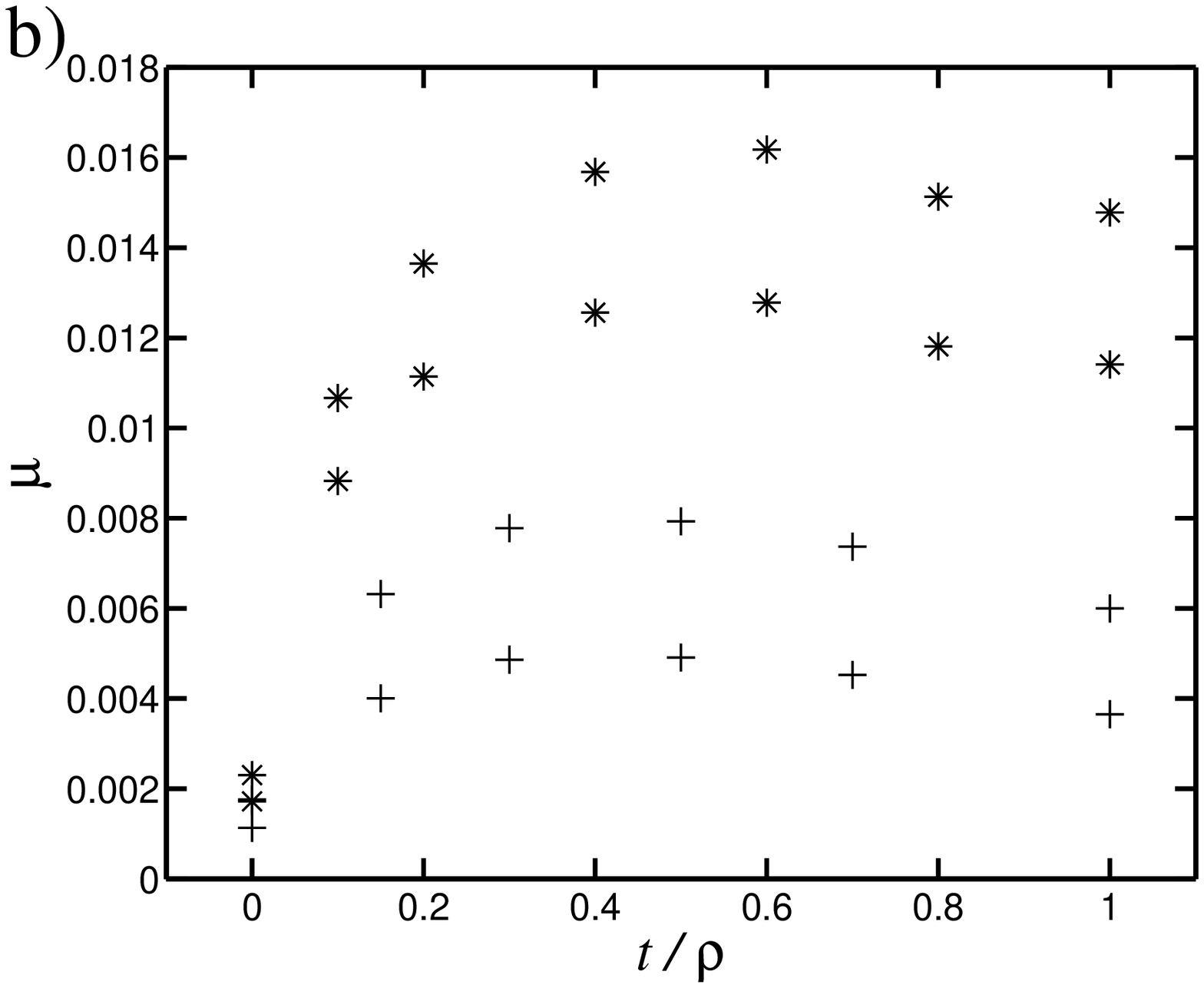, width=6.0cm}} 
\vspace*{13pt}
\fcaption{\label{fig:mulowas}Values of a) $\gamma$, b) $\mu$ for low aspect-ratio traps, 
as a function of the distance the r.f. electrodes are retracted with
respect to the d.c. electrodes. Values for $g = 0.4$ structures are marked by $*$, 
and values for $g = 0.2$ are marked with a $+$. For a given $g$, the lower values
of $\mu$ are $\mu_y$, the upper values are $\mu_x$.}
\end{figure}

Results for structures with $g=0.4$ and $g=0.2$ are shown in figure
\ref{fig:mulowas}. We found that retracting the r.f. electrodes (i.e.
using $t > 0$) gave a useful increase in both $\gamma$ and $\mu$.
As the aspect ratio decreases, $\gamma$ falls rapidly (and $\mu$ less
rapidly).

\subsection{Two-Layer Electrode Structures.}
\label{sec:twolayer}
\noindent

The most natural configuration for producing a radial r.f. quadrupole in the $x-y$ plane 
consists of electrodes placed at four corners of a square. 
Equal r.f. voltages are applied to diagonally opposite electrodes. To produce an 
octupole by such an arrangement, 
one or more of these electrodes must be segmented. Choosing the structure to 
have 2-fold rotational symmetry about the $z$-axis, and reflection 
symmetry in the plane $z = 0$, there remain 3 constraints. We
chose to satisfy the constraints by adjusting voltages rather than by
accurately placing electrodes, and therefore two diagonally opposite electrodes
must be split into seven segments. 

\begin{figure} [htbp]
\centerline{\epsfig{file=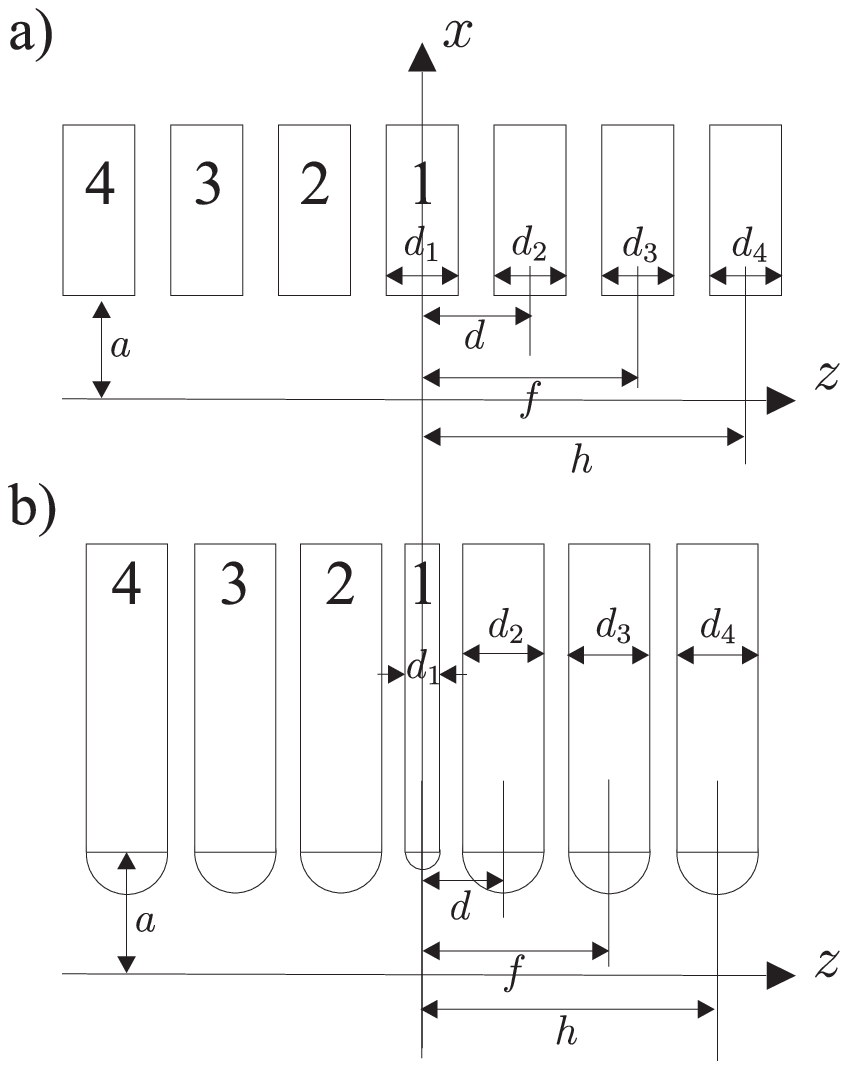, width=6.0cm}} 
\vspace*{13pt}
\fcaption{\label{fig:twolayseg}The segmentation of the d.c. electrodes used in the two-layer 
electrode structures for a) cuboid electrodes and b) cylindrical electrodes. For cases where 
cuboidal electrodes were used, the electrode widths $d_1$ to $d_4$ were equal, 
and the separation between electrodes was $d_1 /2$. 
For structures with cylindrical electrodes, $d_2 = d_3 = d_4 = 2.36 d_1$.}
\end{figure}

\begin{figure} [htbp]
\centerline{\epsfig{file=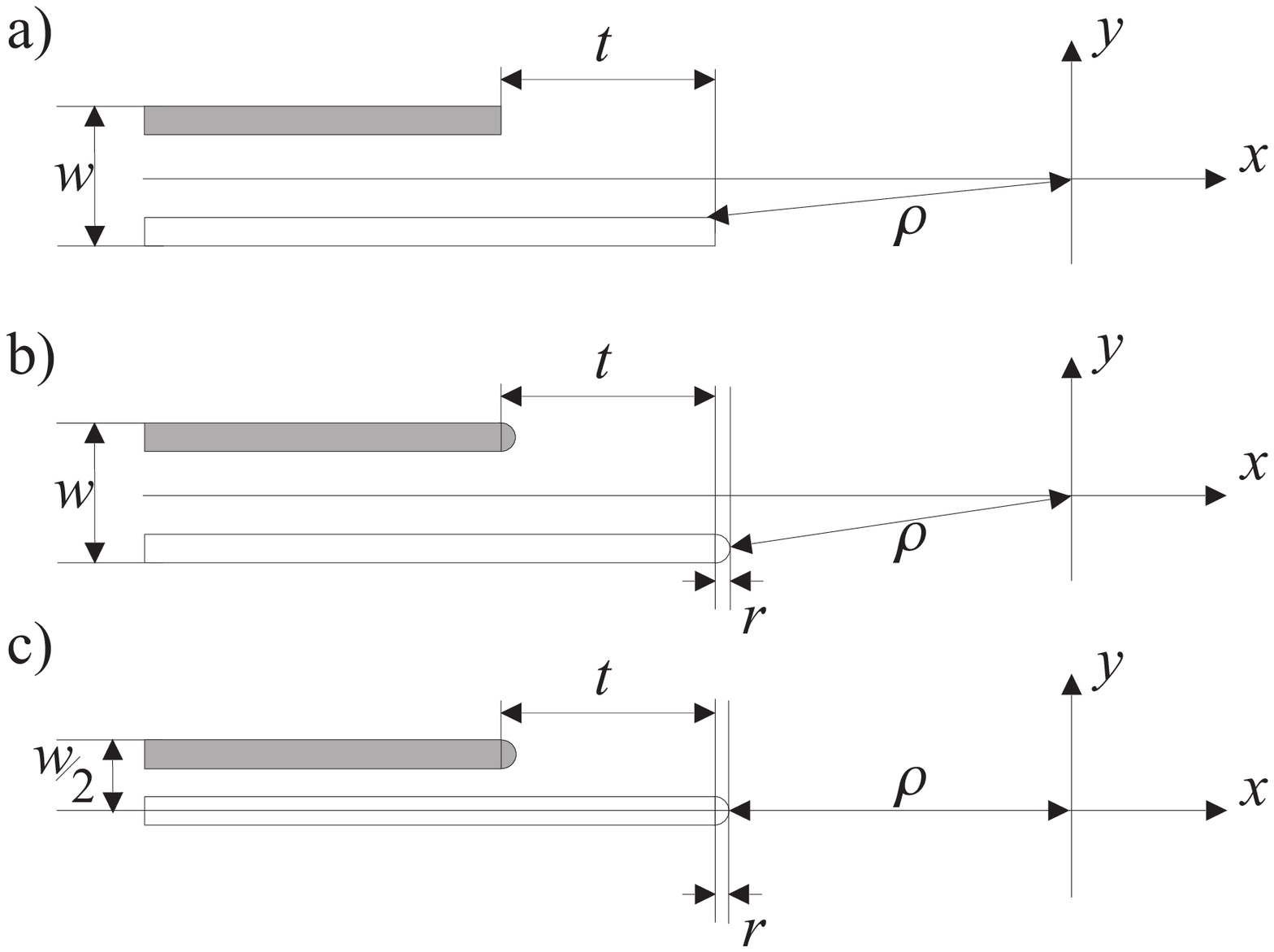, width=6.0cm}} 
\vspace*{13pt}
\fcaption{\label{fig:twolayer}Profiles of the two-layer electrode structures for negative $x$ viewed
in the $x-y$ plane. The r.f. electrodes are represented by grey fill. a) is the
structure used to obtain results 4.1 to 4.12 in table \ref{tab:twolayer}, 
b) was used for results 4.13 to 4.15, and c) for results 4.16 and 4.17. 
The aspect-ratio of the trap was defined by $g = w/\rho$. 
The structure is repeated for positive $x$ by a rotation of 180$^\circ$ about the $z$ axis.}
\end{figure}

The d.c. electrode segmentations used in the 
results presented below are shown in figure \ref{fig:twolayseg}. 
The cuboid electrode structures consist of electrodes of equal width 
($d_1 = d_2 = d_3 = d_4$), equally spaced by ${d_1} /2$ (hence $d = 3/2 d_1$, 
$f = 3d$ and $h = 9/2d$).  
The cylindrical electrode structures were made up of cylinders parallel to the $x$
axis with hemispherical ends. The diameters of the cylinders 
were $d_2 = d_3 = d_4 = 2.36 d_1$. The axes of the cylinders were positioned 
at $d = 2.36 d_1$, $f = 5.43 d_1$ and $h = 8.43 d_1$. The hemispherical ends 
of the cylinders were centred on $x = a$.   

The r.f. electrodes consist of extended planes of thickness $w_{\rm rf}$. For
structures with cuboid electrodes, the end of the r.f. electrodes were placed
at $x = \pm(a + t)$. For structures with cylindrical d.c. electrodes, the r.f. 
electrodes had hemicylindrical ends of radius $w_{\rm rf} /2$ with axes parallel
to the $z$ axis at $x = \pm(a + t)$. 

The voltage on d.c. electrode 4 was set to 1.4V, and the r.f. electrode voltage 
was set to 0V. Numerical calculations were performed for a range of values of 
$d_1$, $t$ and $g$, as defined in figures \ref{fig:twolayseg} 
and \ref{fig:twolayer}. Example results of these calculations are
presented in tables \ref{tab:twolayer} and \ref{tab:twolayervolts}. 

\vspace*{4pt}   
\begin{table}[hb]
\tcaption{Results from numerical calculations for the two-layer electrode structures. The value of $E_{\rm{max}}^\mu$ was 
calculated using an r.f. electrode voltage of $10^{6}$V. Results 4.1 to 4.12 use 
cuboid electrodes. Results 4.13 to 4.17 
use cylindrical d.c. electrodes in the arrangement shown in figure \ref{fig:twolayseg}.
4.13 to 4.15 are for case (b) in figure  \ref{fig:twolayer}, 4.16 and 4.17 for
for case (c) in figure \ref{fig:twolayer}.}
\centerline{\footnotesize\smalllineskip
\begin{tabular}{|l|l|l|l|l||c|c|c||c|c|c|c|} 
\hline Result & $g$ & $w_{\rm rf}$ & $d_1$& $t$ & $10^4 \beta$ & $E_{\rm{max}}$ & $\gamma$ & $E_{\rm{max}}^\mu$ & $\mu_{x'}$ & $\mu_{y'}$ \\
\hline units &  & ($\rho$) & ($\rho$) &  ($\rho$) & (V$a^{-4}$) & (V$a^{-1}$) & ($10^{-3}$) & ($10^6$V$a^{-1}$) & & \\
\hline
4.1 & 2  & 0.16 & 0.39 & 0 & -710 & 21 & -7.0 & 1.77 & 0.147  & 0.154 \\
4.2 & 2  & 0.16 & 0.39 & 0.39 &  -728 & 23 & -7.4 & 1.70 & 0.114 & 0.124 \\
4.3 & 2  & 0.16 & 0.39 & 0.78 & -872 & 25 & -7.5 & 1.53 & 0.097 & 0.109 \\
4.4 & 2  & 0.16 & 0.78 & 0 &  -131 & 7.8 & -1.30 & 1.32 & 0.199 & 0.199 \\
\hline
4.5 & 1.1  & 0.19 & 0.94 & 0 &  -98 & 9 & -1.3 & 2.25 & 0.122 & 0.158 \\
4.6 & 1.1 & 0.19 & 0.47 & 0 &  -1304 & 20 & -7.7 & 2.42 & 0.122 & 0.141 \\
\hline
4.7 & 0.4 & 0.08 & 0.40 & 0 &  -1563 & 34 & -4.6 & 4.37 & 0.041 & 0.042 \\
4.8 & 0.4 & 0.08 & 0.40 & 0.5 &  -1848 & 33 & -5.5 & 4.59 & 0.028 & 0.031 \\
\hline
4.9 & 0.24 & 0.08 & 0.20 & 0 & -1094 & 233 & -0.47 & 12.6 & 0.007 & 0.011 \\
4.10 & 0.24 & 0.08 & 0.40 & 0 &  -1250 & 38 & -3.3 & 12.7 & 0.011 & 0.012 \\
4.11 & 0.24 & 0.08 & 0.40 & 0.5 &  -1636 & 35 & -4.7 & 12.8 & 0.0099 & 0.011 \\ 
4.12 & 0.24 & 0.08 & 0.40 & 1   &  -1649 & 34 & -5.0 & 12.7 & 0.0061 & 0.0068 \\
\hline
\hline 
 4.13 & 1.28 & 0.24 & 0.17 & 0 &  -399 & 229 & -1.6 & 19.1 & 0.097 & 0.100 \\
 4.14 & 2.56 & 0.42 & 0.28 & 0.8 & 12724 & 257 & 8.5 & 16.8 & 0.120 & 0.112 \\
 4.15 & 2.56 & 0.42 & 0.4 & 0.4 &  -482 & 9.2 & -9.0 & 16.2 & 0.167 & 0.136 \\
\hline
 4.16 & 2.56 & 0.61 & 0.42 & 1.21 & -970 & 10 & -5.4 & 36.5 & 0.030 & 0.041 \\
 4.17 & 2.56 & 0.61 & 0.42 & 0.4 &  -885 & 10 & -5.0 & 33.3 & 0.048 & 0.067 \\
\hline
\end{tabular}}
\label{tab:twolayer}
\end{table}

\vspace*{4pt}   
\begin{table}[hb]
\tcaption{Voltages on the d.c. electrodes required to produce the results presented in table \ref{tab:twolayer}}
\centerline{\footnotesize\smalllineskip
\begin{tabular}{|l|l|l|l|} 
\hline Result &  $V_1$ & $V_2$ & $V_3$ \\
\hline units & (V) & (V) & (V) \\
\hline
4.1 & -1.8451 & 1.4055 & -3.2924  \\
4.2 & -1.6997 & 2.2963 & -3.1155  \\
4.3 & -1.8592 & 2.4805 & -3.3316 \\
4.4  & -0.0885 & 0.3648 & -1.6934  \\
\hline
4.5  & -0.0290 & 0.2095 & -1.6090 \\
4.6 & -0.7951 & 1.4055 & -2.5750  \\
\hline
4.7  & -1.4109 & 2.0817 & -3.1413 \\
4.8  & -1.4960 & 2.1987 & -3.0247 \\
\hline
4.9 & -11.582 & 10.442 & -6.3685  \\
4.10 & -1.3316 & 1.9730 & -3.0171  \\
4.11 & -1.2679 & 1.8959 & -2.7263  \\ 
4.12 & -1.1612 & 1.9083 & -2.5576 \\
\hline
\hline 
 4.13 &  -22.030 & 4.7513 & -4.5768  \\
 4.14 &  25.107 & -5.2939 & 18.252  \\
 4.15 &  -0.3793 & 0.1728 & -1.3831 \\
\hline
 4.16 & -0.1479 & 0.1103 & -0.8449  \\
 4.17 &  -0.1427 & 0.0917 & -0.7954  \\
\hline
\end{tabular}}
\label{tab:twolayervolts}
\end{table}

\begin{figure} [htbp]
\centerline{\epsfig{file=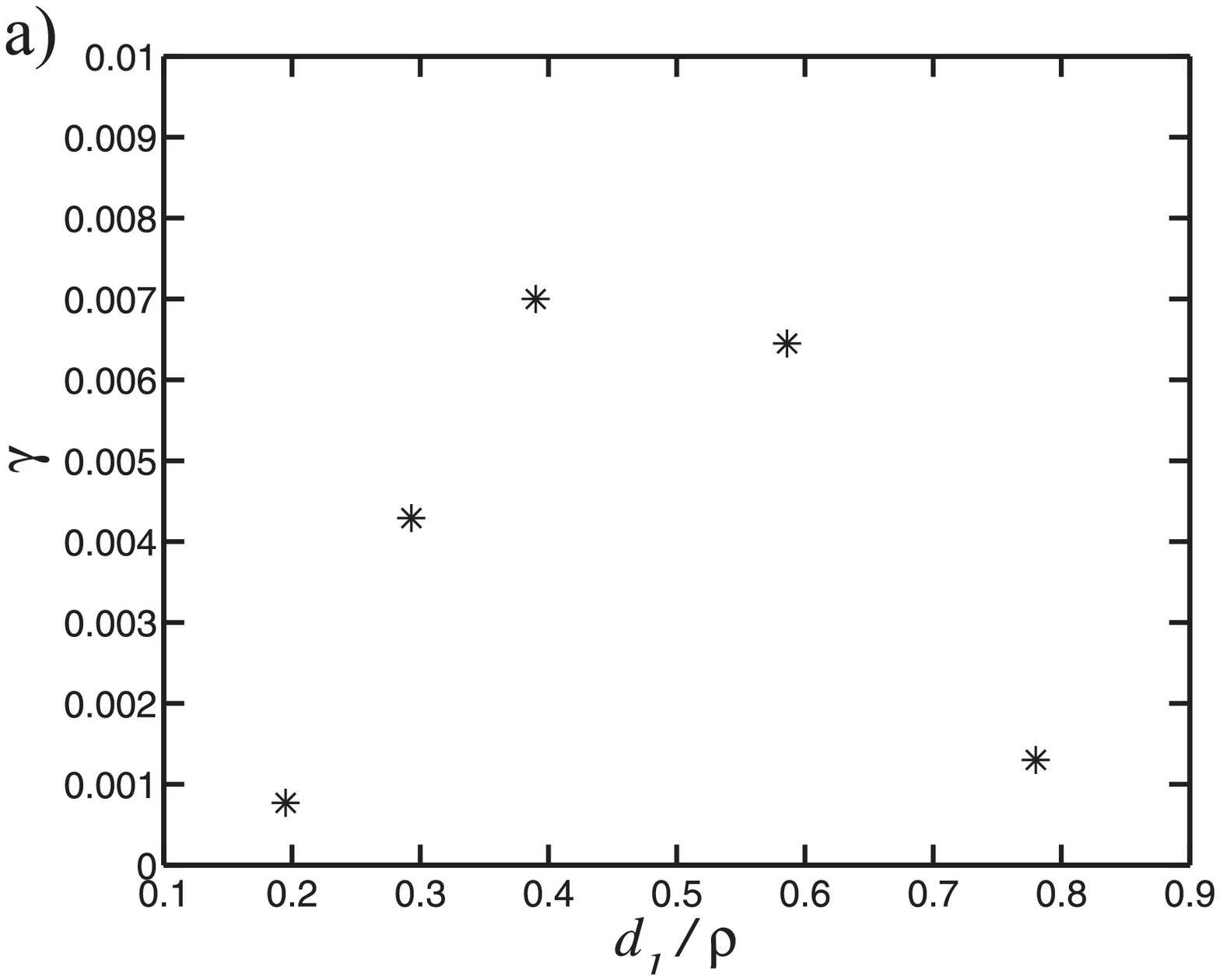, width=6.0cm}, \epsfig{file=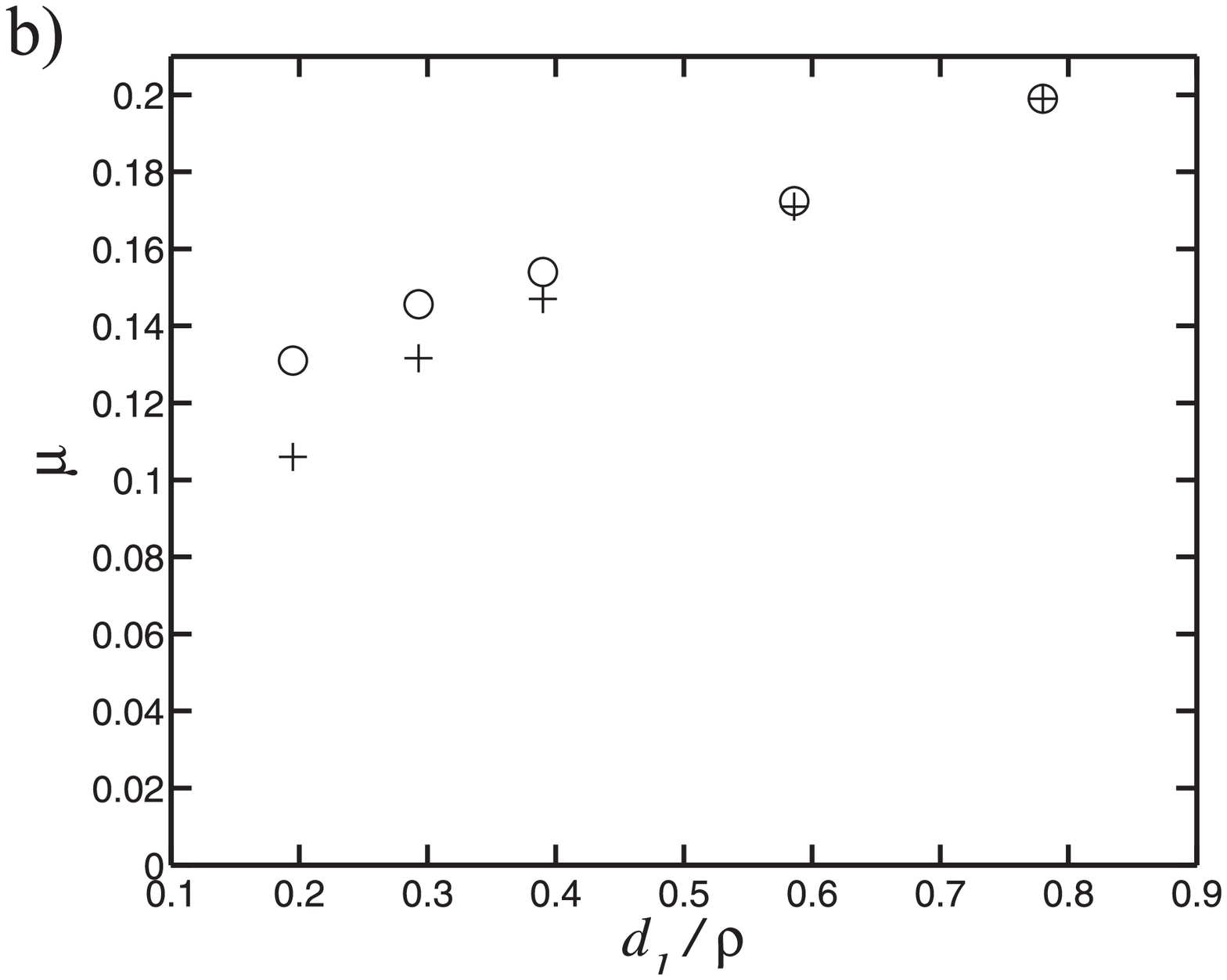, width=6.0cm}} 
\vspace*{13pt}
\fcaption{\label{fig:changewe}The effect on a) $\gamma$ and b) $\mu$ of adjusting the width of the 
central electrode for a two-layer electrode structure having $g=2$ and $t=0$.
$\mu_{x'}$ is represented by unfilled circles and $\mu_{y'}$ by +.}
\end{figure}

We found that the width $d_1$ of the central electrode had a strong effect
on $\gamma$ and a smaller effect on $\mu$ (see figure \ref{fig:changewe}
and c.f results $\{4.5,4.6\}$, $\{4.9,4.10\}$).
The optimum value of $d_1$ for producing a d.c. octupole was found to be 
around $0.4 \rho$, for all values of $g$.
This is comparable with the value used for the three
layer electrode structures in section \ref{sec:10dc4rf}, which was obtained by 
examining the equipotentials of line-charge calculations. 
The monotonic increase in $\mu$ with $d_1$ 
may be ascribed to an increased influence of all of the d.c. electrodes. 

\begin{figure} [htbp]
\centerline{\epsfig{file=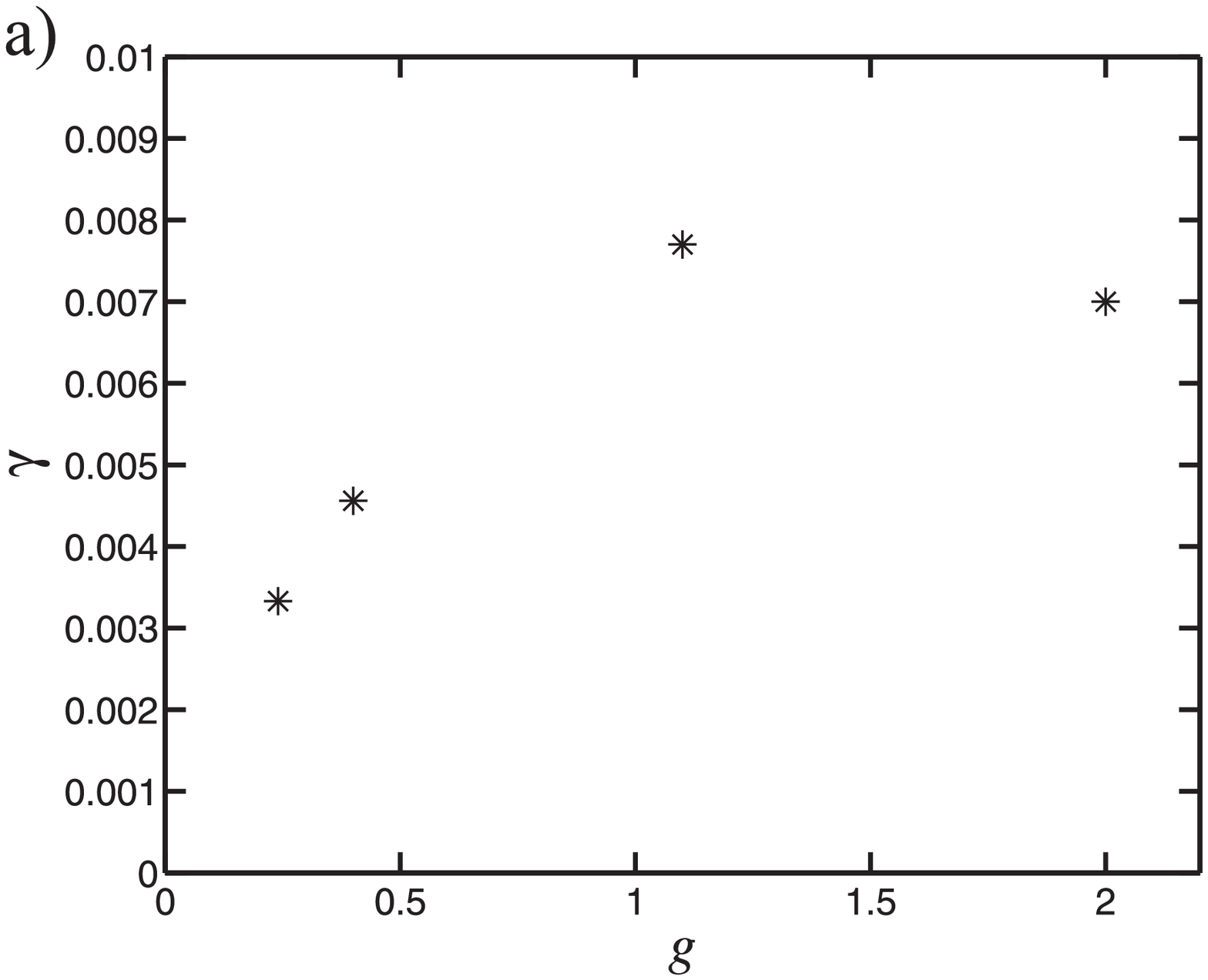, width=6.0cm}, \epsfig{file=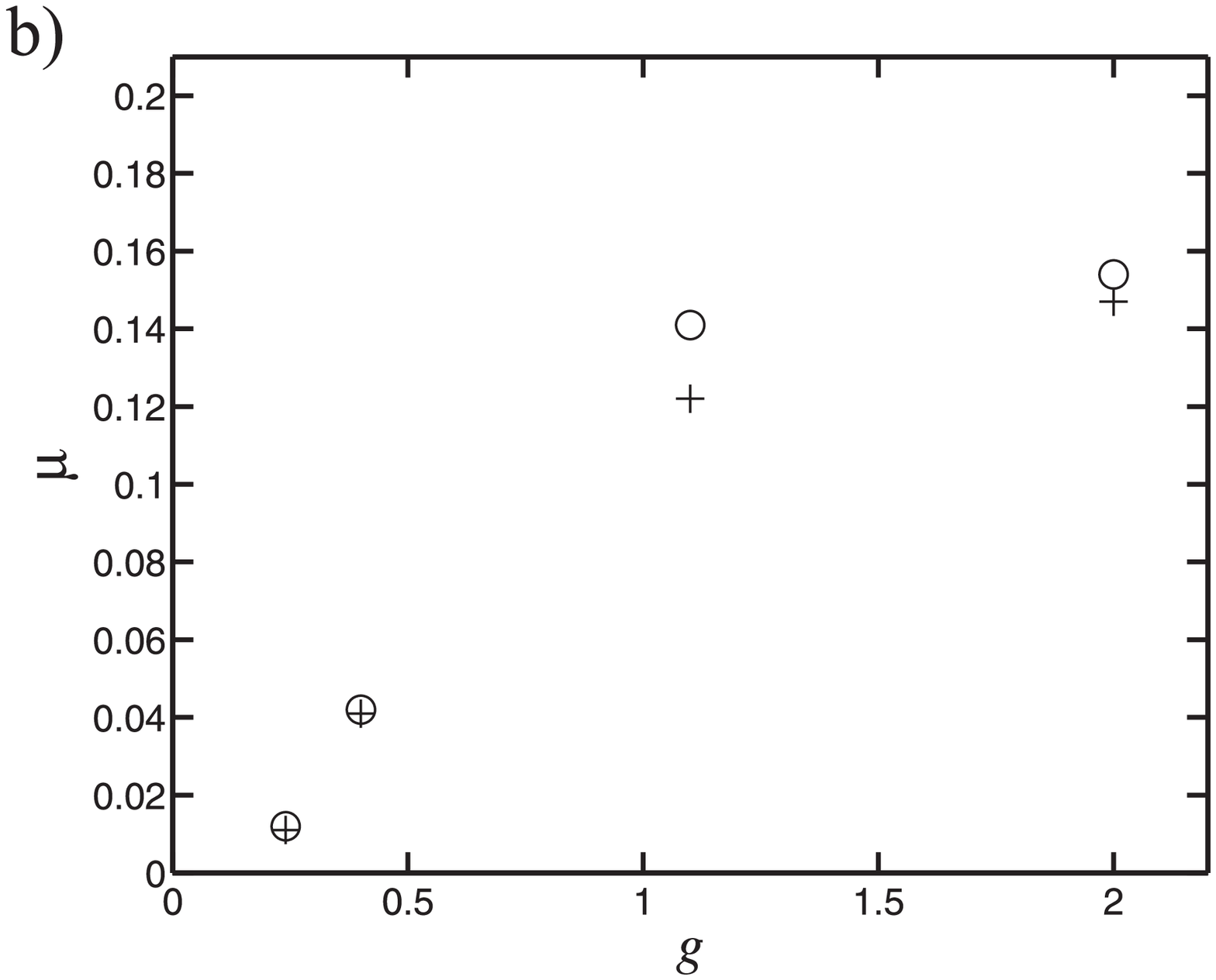, width=6.0cm}} 
\vspace*{13pt}
\fcaption{\label{fig:changeg}The effect on a) $\gamma$ and b) $\mu$ of adjusting the aspect 
ratio $g = w/\rho$ for a two layer electrode structure. In b), the values of $\mu$ plotted are 
those along the $x'$ (represented by $+$) and $y'$ (represented by $\circ$) axes of the quadrupole. 
The values of $d_1$ for these cases are similar 
but not identical (0.39, 0.47, 0.4 and 0.4 $\rho$ from left to right on the plots).}
\end{figure}

Figure \ref{fig:changeg} shows the effect on $\gamma$ and $\mu$ of changing the
aspect ratio of the trap. For $g < 1$,
both parameters increase with $g$. At higher $g$, $\mu$ continues to increase with $g$ 
while $\gamma$ decreases, presumably due to the reduced shielding of the 
r.f. electrodes by the d.c. electrodes.  

We found that, for $g=2$, retraction of the r.f. electrodes had no significant
effect on $\gamma$, but reduced $\mu$ somewhat (results $\{$4.1--4.3$\}$).
When $g=0.4$ retraction allows a small increase in $\gamma$ with a 
drop in $\mu$ (results 4.7,4.8). When $g=0.24$, increasing $t$ from 0 to $\rho/2$
increased $\gamma$ by 40\% with little effect on $\mu$ (results 4.10,4.11).

We examined the structure shown in figure \ref{fig:twolayer}c,
with the d.c. electrodes placed in the $y=0$ plane, with a view to finding out whether
it would produce a higher $\gamma$. However, comparison of results 4.17 and 4.15
suggest that it has the reverse effect. Since $\mu$ decreases also there
is nothing to be gained by adopting this structure.

\subsection{Planar or near-planar arrangments}
\noindent 

\begin{figure} [htbp]
\centerline{\epsfig{file=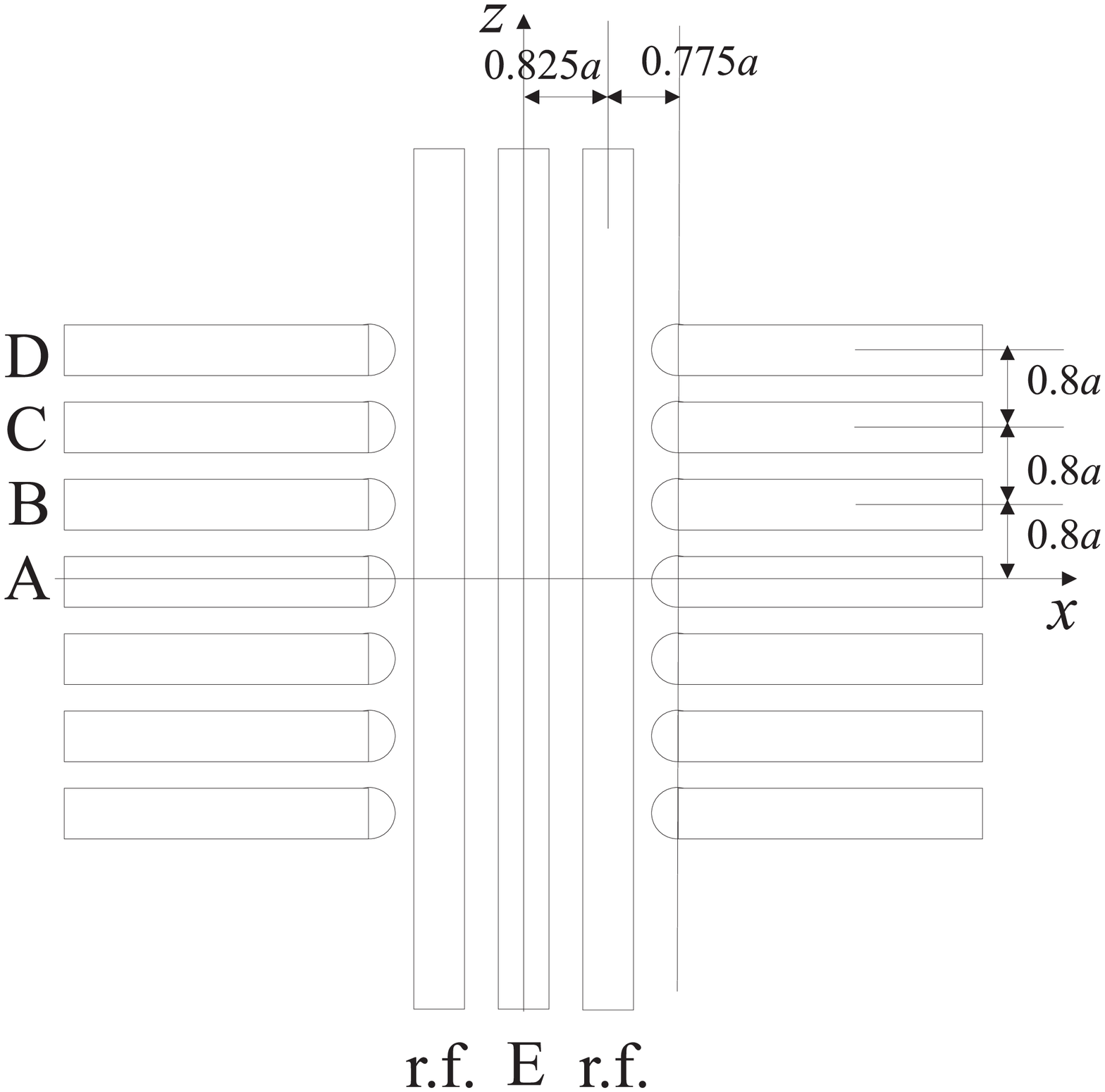, width=6.0cm}} 
\vspace*{13pt}
\fcaption{\label{fig:svnelec}The planar electrode arrangement (result 5.1). The 
cylindrical electrodes all lie in the same plane and have the same 
radius of 0.3$a$. The ends of the electrodes parallel to the $x$ axis are hemispherical.}
\end{figure}

\begin{figure} [htbp]
\centerline{\epsfig{file=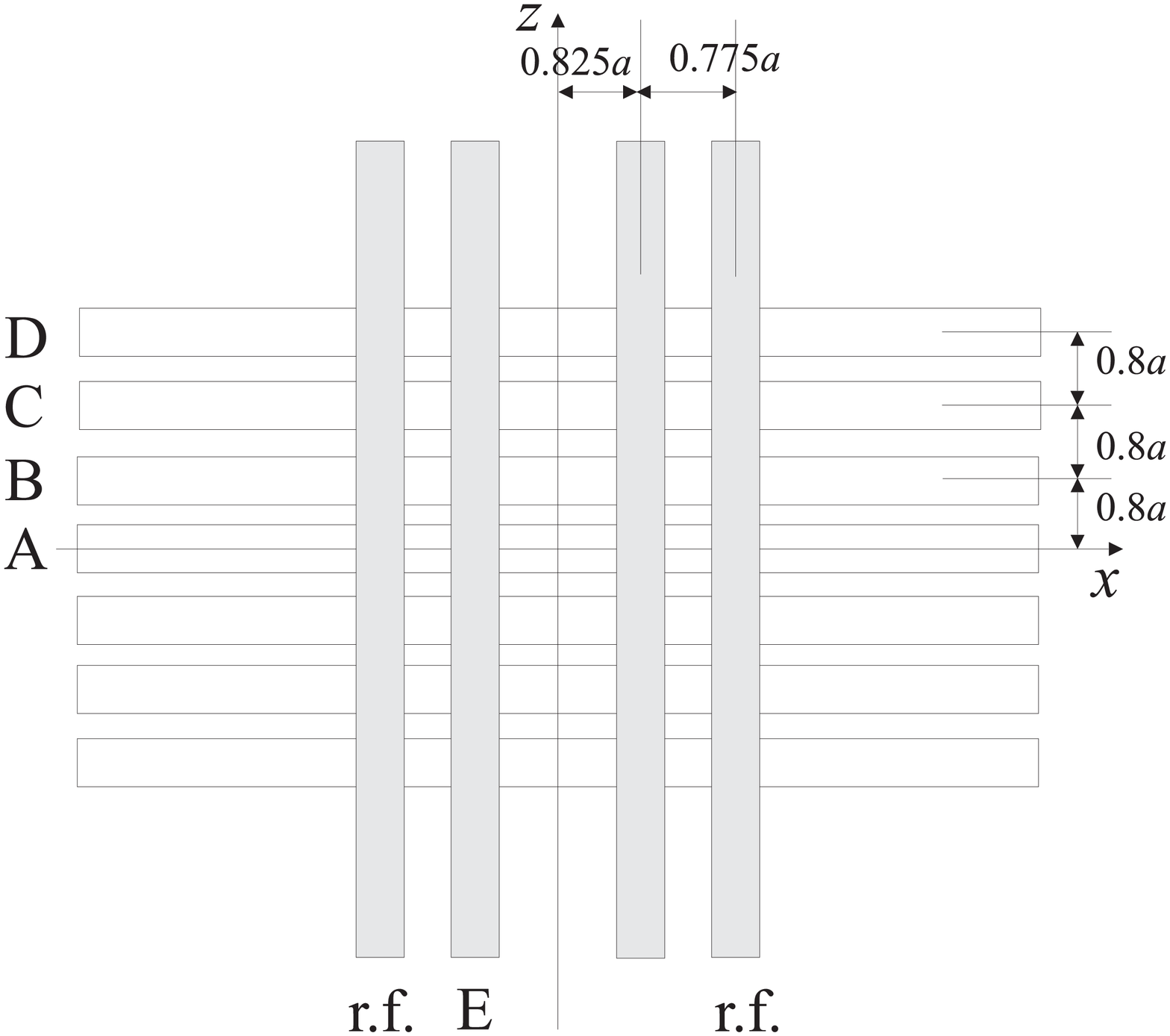, width=6.0cm}} 
\vspace*{13pt}
\fcaption{\label{fig:railway}Railway track electrode arrangement (result 5.2). The 
electrodes parallel to the $z$ axis are situated above those parallel to the $x$ 
axis by a distance of $a$. All electrodes are cylindrical, of radius $0.3 a$. }
\end{figure}

We considered cases where all the electrodes of the twin-trap system lie in a single
plane (figure \ref{fig:svnelec}), and a simple `railway track' arrangement in which
rod-shaped r.f. electrodes
run orthogonal to a set of simple rod-shaped d.c. electrodes (figure
\ref{fig:railway}). The simplicity of
such designs gives obvious advantages from the point of view of microfabrication,
but we find that the penalty in terms of reduced factors $\gamma$ and $\mu$ is severe. 
These results are presented in table \ref{tab:planar}.

The voltages on these structures are to be adjusted to produce a trapping region
centred outside the electrode structure. Therefore they lack one reflection symmetry,
and there are 6 constraints if we require an octupole (construction (G) of
section \ref{s:image}). To satisfy all these constraints by adjusting voltages would
require 7 d.c. voltages (in addition to the d.c. ground set by the r.f. electrodes)
and therefore a large number of electrodes. In order to reduce the number
of electrodes, we chose to examine structures such that the 
hexupole moment may be small but non-zero. We adjusted the 5 voltages labelled
A--E in figures \ref{fig:svnelec} and \ref{fig:railway} so as to cancel
the quadrupole terms and $\partial V/ \partial y$ and 
$\partial^3 V / \partial y^3$, but we did not constrain 
$\partial^3 V / \partial x^2 \partial y$ or $\partial^3 V / \partial z^2 \partial y$.

To find the condition where the d.c. hexupole/octupole is centred at the same height
as the r.f. quadrupole, we proceeded as follows. First the r.f. electrodes
were set to zero, and the d.c. voltages were adjusted to produce a hexupole/octupole
at a few different heights. Next, the r.f. electrodes were allowed to go
to $\pm 1$ and the quadrupole position identified. We could then make a good
first guess of the right d.c. voltages to get the desired coincidence, and the
solution was found iteratively.
\vspace*{4pt}   
\begin{table}[hb]
\tcaption{The two cases studied which produced almost-octupoles 
above the plane of the electrodes. Result 5.1 uses the single plane electrode 
geometry shown in figure \ref{fig:svnelec}. 
Result 5.2 uses the ``railway track" geometry shown in figure \ref{fig:railway}.}
\centerline{\footnotesize\smalllineskip
\begin{tabular}{|l|l|l|l|l||l|l|l|l||l|l|} \hline
Result &  $V_A$ & $V_B$ & $V_C$ &  $V_E$ &$10^4 \beta$ & $E_{\rm max}$& $\rho$ & $\gamma$ & $\mu_x$ & $\mu_y$ \\
\hline units &  (V) & (V) & (V)  & (mV) &(V$/a^4$)&(V$/a$)& ($a$) & ($10^{-3}$) & ($10^{-3}$) & ($10^{-3}$) \\
\hline 5.1 &  -4.257 & 4.560 & -3.527 & -2.197  & -0.0155 & 44.74 & 0.829 & 0.198 & 10.4 & 10.5 \\
5.2 &  1.089 & 1.212 & -4.526 & -201 & 0.000247  & 20.47 & 2.5 & 0.189 & 3.5 & 3.9 \\
\hline
\end{tabular}}
\label{tab:planar}
\end{table}

\section{Manufacturing imprecision and stray electric fields} \label{s:mi}
\noindent

Throughout the above we relied on symmetries in the electrode structure in order to
achieve the desired electric potential function. It is necessary to check whether
there might be a high sensitivity to manufacturing imprecision and patch potentials
which will in practice
break such symmetries. To this end we studied an example case: the electrode
structure leading to result 2.2 in table \ref{tab:10dc4rfoct}, c.f. figure \ref{fig:5Electrodes}.

In order to illustrate manufacturing imprecision,
the electrode which has its spherical end centred at $(a,0,d)$ was displaced by $0.05 \rho$ in the
$z$ direction. The d.c. electrode voltages were then adjusted to 
satisfy $\partial^2 V/\partial x^2 = \partial^2 V/\partial y^2 =\partial^2 V/\partial z^2 = 0$ at 
the centre of the trap, and the electric field $E_z(z = 0)$ along the $z$ axis at the 
centre of the trap was calculated. The voltage on the displaced electrode was then 
adjusted to null this axial field, and the resulting potential shape $V(x,y,z)$ examined.
This was repeated for displacements of the electrode
in the $x$ and $y$ directions. The results are tabulated in table \ref{tab:mi}.
The voltages are given such that for $\rho=10\;\mu$m and $E_{\rm max} = 100$ V$\mu$m$^{-1}$,
the axial centre of mass
frequency is $\omega_1 = 2 \pi \times 22$ MHz, and the radial
centre of mass secular frequencies are  $\omega_x = 2 \pi \times 41$ MHz and $\omega_y
= 2 \pi \times 52$ MHz, for calcium ions.  

\vspace*{4pt}   
\begin{table}[hb]
\tcaption{Example of the effect of manufacturing imprecision.
The first three rows show 
the voltages required for $\partial^2 V/\partial x^2 
= \partial^2 V/\partial y^2 =\partial^2 V/\partial z^2 = 0$ at the centre of the trap. 
$E_z(z = 0)$ is the axial electric field at $z = 0$ when this condition is fulfilled. 
$z_c$ is the displacement
of the centre of mass of an ion pair in a $(1/2)m\omega_1^2 z^2$ potential well
in the presence of electric field $E_z$.
$V_2^{\rm null}$ is the voltage 
on the displaced electrode which is required to null the electric field. 
In all cases the r.f. electrode voltage amplitude is $\sim 240$ V, so that $\omega_{r} \simeq 2 \omega_1$.}
\centerline{\footnotesize\smalllineskip
\begin{tabular}{|l|l|l|l||l|c|l|} \hline
 direction & $V_1$  & $V_2$ & $V_3$ & $E_z({z = 0})$ & $z_c = qE_z/m \omega_1^2$ & $V_2^{\rm null}$\\
\hline &(V) & (V) & (V)  & (V$\mu{\rm m}^{-1}$) & ($\mu$m) & (V)\\
\hline $x$ & 281.7 & 97.6 & 593.0& -0.17 & 21 & 85.8 \\
$y$ & 281.6 & 100.7 & 593.0&  -0.058     & 7.3 & 97.4 \\
$z$ & 277.3 & 100.0 & 593.0 & -0.073     & 9.2 & 95.1 \\
\hline
\end{tabular}}
\label{tab:mi}
\end{table}

In this study we did not introduce any further electrodes, whereas in 
practice further electrodes would be present if the structure is part of a larger array,
and such further electrodes could be used to cancel stray electric fields along
$z$ such as the one obtained here. However by using the given set of electrodes we
obtain an estimate of the precision required for the d.c. voltages, and an
upper limit on the size of unwanted d.c. quadrupole terms which are introduced.

The values of $z_c$ in table \ref{tab:mi} show that when the voltages are adjusted to
cancel the quadrupole, then the stray electric field (owing to manufacturing imprecision)
is large enough to cause a displacement of order $\rho$
of the ions, which would cause a problem. Comparing the values of
$V_2$ in the 3rd and last columns of the table, however, it is seen that only
a $\sim 5$\% adjustment in $V_2$ is needed to cancel this field. Since we are dealing
with a 5\% misplacement of the electrode, we infer that the required values of the
electrode voltages are not especially sensitive to electrode misplacements.
Equally, the figures show that the behaviour is sensitive to voltage inaccuracies,
because without this same 5\% voltage adjustment there would exist a severe problem.

To characterise the latter, argue as follows.
In the presence of a field $E_0$ the centre of mass
of the ions is displaced by approximately $z_c = qE_0 / m \omega_1^2$, and at the
octupole condition this evaluates to $z_c = E_0 / 3 d^2 \beta$ (using eq. (\ref{om1})).
This is a problem when the displacement is such that the potential hill introduced
when $\alpha$ becomes negative does not appear in between the ion pair, so fails
to separate the ions. Therefore the displacement is unacceptable if 
it exceeds approximately $d/2$, so we require
\beq
E_0 < d^3 \beta.
\eeq
In the example under discussion, $\beta \simeq 2.5 \times 10^{21}$ Vm$^{-4}$ and $d \simeq 1\;\mu$m,
so the requirement is $E_0 < 2500$ V/m. It is in practice easy to 
cancel fields of this magnitude, but in view of the fact that patch potentials and 
manufacturing imprecision can result in fields much larger than this, it
may be necessary to characterise the electrode structure in situ in a vacuum
system by transporting ions through it. 

Next, consider the d.c. quadrupole terms.
When the electrode structure lacks symmetry owing to its manufacturing imprecision,
we can no longer assume mixed derivatives such as $\partial^2 V/\partial x \partial y$, 
$\partial^2 V / \partial y \partial x$ are zero, so
we expect to get a d.c. quadrupole in the $xy$ plane when the electrodes are adjusted
to satisfy the constraints $\partial V / \partial z=0$, $\partial^2 V / \partial z^2=0$.
In order to estimate the typical size of d.c. quadrupole
caused by manfacturing imprecision, we left the electrode voltages at the
values indicated in table \ref{tab:mi}, without attempting any further nulling
of quadratic terms, and examined the quadratic terms in the potential near the origin.
These terms were of order $10^9$~Vm$^{-2}$. Under the conditions
of the test, the radial confinement from the a.c. quadrupole is such
as to achieve radial secular frequency around $45$ MHz. We find that the influence of
the d.c. quadrupole is small: it merely changes the radial secular frequencies to 
$\sim (45^2 \pm 10^2)^{1/2} \simeq 45 \pm 1$ MHz.

\section{Discussion}  \label{s:discuss}
\noindent

\begin{figure} [htbp]
\centerline{\epsfig{file=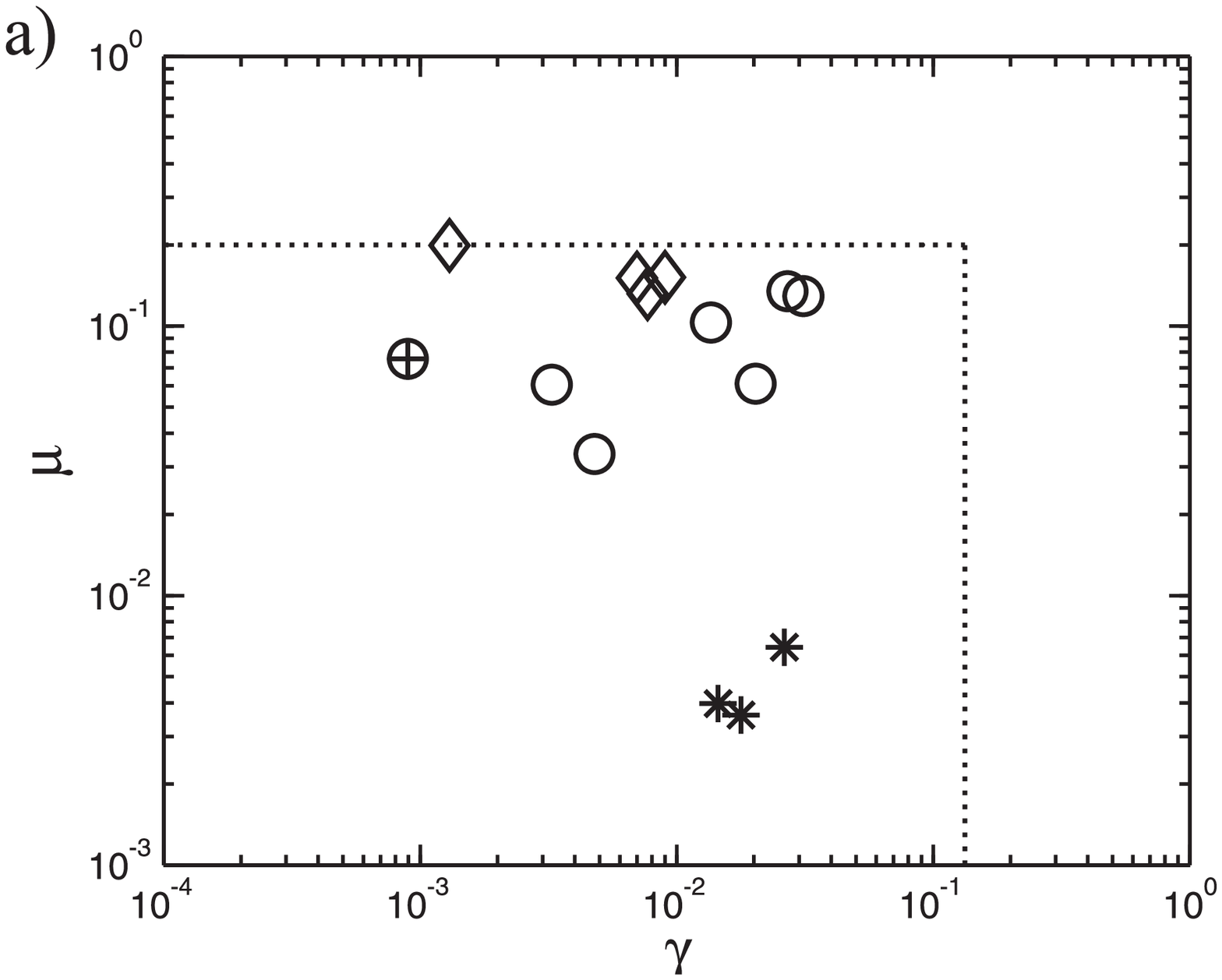, width=6.0cm}, \epsfig{file=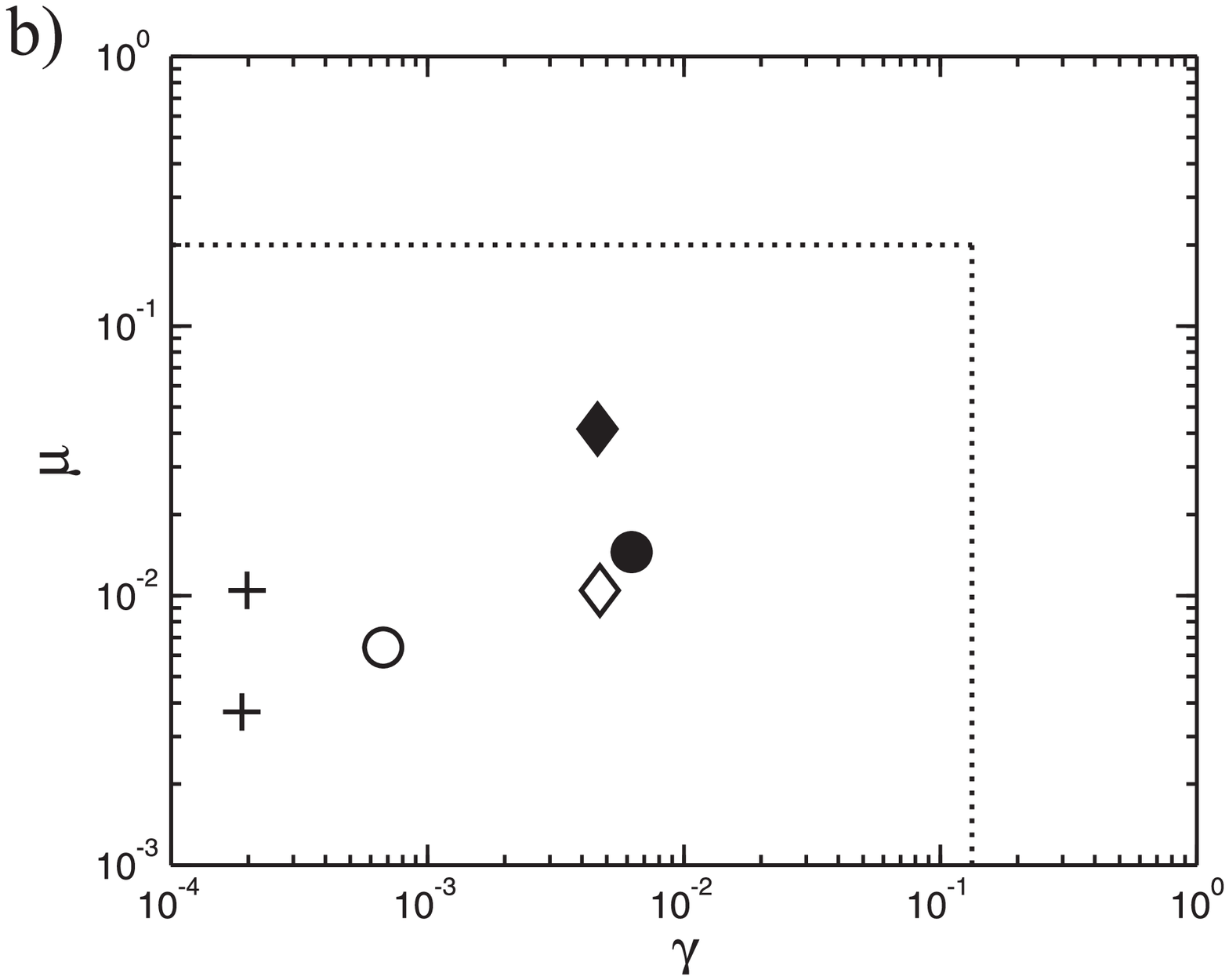, width=6.0cm}} 
\vspace*{13pt}
\fcaption{\label{f:gammamu}Values of $\gamma$ and $\mu$ for the structures considered in
the text. (a): high aspect ratio, $g \ge 1$; 
$*$: 1.1-1.3, $\circ$: 2.1--2.6, $\oplus$: 3.1, 
$\diamond$: 4.1, 4.2, 4.7, 4.14.
(b): low aspect ratio, $g \le 0.75$; $\circ$: 3-layer, $\diamond$: 2-layer,  
$+$: `planar' 5.1,5.2 ($g \simeq 0.65$). The filled symbols represent results for $g \simeq 0.4$,
the open symbols represent results for $g \simeq 0.2$.  The dotted lines give the values
which were obtained for the two `standard' cases described in
section \ref{s:opt}, for comparison.}
\end{figure}

A summary of the values of $\gamma$ and $\mu$ obtained for the various structures
we have considered is given in figure \ref{f:gammamu}. We find that, with the exception
of results $1.1$-$1.3$, $\gamma$ is always smaller than $\mu$.
This may be because we have
not allowed the r.f. electrode to be divided. It presents a flat potential
surface along the $z$ direction which tends to suppress the variation in potential
which is needed for large $\gamma$. It may be possible to increase $\gamma$
without introducing multiple r.f. electrodes, by tailoring the shape of
the edge of the r.f. electrodes to follow an equipotential produced by the d.c.
octupole. 

One might expect a competition between $\mu$ and $\gamma$ to be observed in
the results, but there is little indication of this. The best
structures give high $\gamma$ and high $\mu$ simultaneously.

Many of the optimizations studied, by adjusting the widths and placements
of electrodes, gave useful improvements in $\gamma$ and/or $\mu$. This
implies that a careful optimization of any chosen design is worthwhile.

Of the three-layer electrode structures we examined, those of the general
form shown in figure \ref{fig:3D5Electrodes4} performed best. The highest
$\gamma$ value ($\gamma \simeq 0.03$, result 2.2), was obtained  
with $\mu$ close to the highest value, using a 3-layer structure of
aspect ratio $g \simeq 3$. 

At lower aspect ratio, $g < 0.5$, the two-layer structures performed better
than three-layer ones of the same $g$
(figure \ref{f:gammamu}b). The larger value for $\gamma$ may be understood
from the fact that when $g$ is small, $\gamma$ varies rapidly with $w$
because thicker electrodes can extend their influence on $V(x,y,z)$
over larger distances. For a given total depth $w$ of the whole structure, if
there are $m$ layers and $m-1$ gaps all of the same thickness, then the thickness
of each layer is $w/(2m-1)$. Therefore a two-layer structure of given $g$
may be expected to give a similar value for $\gamma$ as a three-layer
structure with $g$ larger by a factor $5/3$. This may be seen by comparing 
the values for a three-layer structure at $g=0.4$ in 
figure \ref{fig:mulowas} with those for a two-layer structure at $g=0.24$, e.g. 
result 4.11 (and c.f. figure \ref{f:gammamu}b). 
The two-layer structures also give larger values for $\mu$ at given $g$.
This is in part for a similar reason, and also because the electrodes
are more conveniently placed to realise a radial quadrupole.
The two-layer structures of large aspect ratio satisfy the condition
(\ref{nonoct}) at $q_r=0.3$ which means that, owing to their good radial confinement, it is
not necessary to cancel d.c. quadrupole terms in the potential, but doing so
remains useful in order to reduce the required r.f. voltages.

At $g=0.24$ the highest $\gamma$ value
($\gamma \simeq 0.005$, result 4.11) was smaller than the best
overall by a factor 6, and $\mu$ was reduced
by an order of magnitude. 

The planar or near-planar arrangements, cases 5.1 and 5.2, gave
a further order of magnitude reduction in $\gamma$ compared to the
two-layer low-aspect-ratio structure. Owing to this small $\gamma$ value,
these structures satisfy the condition (\ref{nonoct}) so the
cancellation of d.c. quadrupole terms is not necessary.

\section{Conclusions} \label{s:conc}
\noindent

We have studied the general problem of separating or bringing together pairs
of trapped ions, while keeping the frequencies of the normal
modes of oscillation of the ions as high as possible. A simple picture
of this is to say the central electrode introduces a potential hill which
pushes the ions apart. However we have found that the distance between the ions 
can be small compared to the distance scale of the electrodes, even
when the potential well is about to divide into two local minima. This means
that the potential shape in the region between the ions is really a joint
property of the whole set of electrodes. There are various contributions 
to the potential that have a quadratic dependence on position near the
origin, and when the ions are being split or recombined these are
balanced, so as to leave a quartic dependence. 

The problem under consideration reduces to two main issues: radial
confinement, and axial confinement at the point where it is weakest,
which is to say when the axial potential is quartic. We found that
axial confinement is the harder to achieve in large structures, and
radial confinement in small ones. The size scale $\rho_c$ when
the radial confinement becomes the limiting problem is of order
microns. It is set largely by a length scale $L_0$ (equation (\ref{L0})) 
which is determined purely by materials considerations, i.e.
by the maximum electric field allowed on electrode surfaces.
Since $L_0$ varies as $E_{\rm max}^{1/2}$ it will not
change greatly for different materials or fabrication methods.
Ion trap micro-fabrication methods being attempted at present are in the
regime $\rho \gg \rho_c$ \footnote{Methods have been suggested to approach the nanometre
scale, such as the use of carbon nanotubes as electrodes. The use
of very tight traps would require novel non-optical methods to achieve
quantum logic gates, because the Lamb-Dicke parameter would be very small.}.

We have studied a variety of electrode structures and characterised them
mainly through the geometrical factors $\mu$ and $\gamma$ (equations
(\ref{mu}) and (\ref{gam})) which characterise the strength of r.f. quadrupole
and d.c. octupole moments. Many of the structures exhibit
$z$-dependence in the oscillating part of the potential,
and therefore micro-motion along the $z$ axis. This motion is found
to have a small Matthieu $q$-parameter and is not expected to be
a problem\footnote{We have carried out example numerical integrations of the
equations of motion for a pair of ions undergoing controlled
separation in order to gain evidence of this.}. We assumed reflection symmetries
in the structures, and for one example case we examined the effects
of manufacturing imprecision which broke the symmetry. This did not
suggest any major problem with the design, in that the unwanted
electric field owing to manufacturing error
could be cancelled by small adjustments to the electrode voltages,
and the unwanted quadrupole terms were small.  

We found that for large aspect ratio, 3-layer designs gave
the highest values of $\gamma$ and simultaneously almost the highest $\mu$. At
low aspect ratio, 2-layer designs performed best at a given aspect ratio. 
Current efforts to fabricate trap arrays
are focussed on distances of order $\rho \simeq 100\;\mu$m 
and this suggests a low aspect ratio $g \simeq 0.2$ will be necessary
if lithographic methods are used. In the future, however, there
may be interest in $\rho \simeq 10\;\mu$m and then $g \simeq 2$
will be available. 

It is not surprising that the 3-layer designs offer tighter traps than
the planar designs, but it is noteworthy that when considering the problem
of separating ions, this is especially true. The factor $\simeq 150$ decrease
in $\gamma$ (comparing case 2.2 in table \ref{tab:10dc4rfoct} with 
case 5.1 in table \ref{tab:planar}) means
that the planar design would need to be fabricated at a distance scale
a factor $150^{1/3} \simeq 5$ times smaller in order to obtain the
same motional frequency $\omega_1$ (equation (\ref{om1A})). If we assume heating
rates scaling as $\rho^{-4}$, the heating rate in the latter case would be increased
compared to the former by a factor $\simeq 800$. If instead one compares
structures having the same number of phonons of heating per ion separation
time (c.f. section \ref{s:heatscale}), then the planar design leads to
a motional frequency a factor $150^{0.387} \simeq 7$ lower. 

In view of the greater ease of fabrication of planar structures, a sevenfold
reduction of motional frequencies might be regarded as acceptable. However, for
a quantum computer in which moving ions around and separating and combining
strings of ions are essential ingredients to most logical operations, the
separation/combination time may be the main limitation on the overall
logic speed. To establish a definite design preference between 
a small planar structure and a larger layered
structure, a greater understanding of the heating mechanism will be needed.

\nonumsection{Acknowledgements}
\noindent
We thank D.N.Stacey, M.G.Blain and J.Fleming for helpful discussions. We thank D.Leibfried for information on the NIST trap dimensions.

This work was supported by the EPSRC, the Research Training and Development and Human Potential Programs of the European Union,
the National Security Agency (NSA) and
Advanced Research and Development Activity (ARDA)
(P-43513-PH-QCO-02107-1).

\nonumsection{References}
\noindent

\section*{Appendix 1: Axial micromotion}

Consider the axial motion of an ion in the potential given in equation (\ref{Vform2}): 
\beq
V(0,0,z) = \alpha z^2 + \beta z^4 + (\alpha_{z} z^2 + \beta_r z^4)\cos(\Omega t)
\eeq
where $\Omega$ is the frequency of the voltage applied to the r.f. electrodes, and we added
an octupole term to the oscillating part in case that is important (it will turn out that it
is not). For small departures from equilibrium, the equation of motion for ion $i$
of an $n$-ion chain (assuming the ions are confined strongly in the radial direction) is 
\beq
m \frac{d^2 z_i}{d t^2} = -q (2 \alpha z_i + 4 \beta z_i^3 + 
(2 \alpha_z z_i + 4 \beta_r z_i^3)\cos(\Omega t)) + \sum_j k_{ij} (z_i - z_j) 
\eeq
where $z_i$ is the position of ion $i$, and $k_{ij}$ is the spring constant 
arising from the Coulomb term in the expansion of the ion-ion repulsion (this
expansion has only been taken to first order in the ion separation). 

When we sum the equations for each of two ions, and the ion-ion repulsion term drops out, leaving
\beq
\label{diffdsq}
m \left(\frac{d^2 z_1}{d t^2} + \frac{d^2 z_2}{d t^2}\right) = -q ( 2 \alpha + 2 \alpha_z\cos(\Omega t)) (z_1 + z_2) + (4 \beta + 4 \beta_r \cos(\Omega t))(z_1^3 + z_2^3) .
\eeq 
Now substitute $z_c = (z_1 + z_2)/2$, and $z_1 - z_2 = d + \zeta$  
where $d$ is the equilibrium separation of the ions and $\zeta$ represents 
excursions of the ions from their equilibrium separation. Equation (\ref{diffdsq}) becomes
\beq
\label{diffcom}
m \frac{d^2 z_c}{d t^2} = -2 q ( \alpha + \alpha_z\cos(\Omega t)) z_c  - q (\beta + \beta_r \cos(\Omega t))(3 (d + \zeta)^2  + 4 z_c^2)z_c .
\eeq
In the limit $\zeta, z_c \ll d$ we neglect the terms of 
order $\zeta d$, $\zeta^2$ and $z_c^2$ in the second term on the right hand side. The 
resulting equation can be transformed to the standard form of the Matthieu equation
\beq
\label{eq:Matt}
\frac{d^2 z_c}{d \xi^2} + (a_z - 2q_z \cos(2 \xi))z_c = 0
\eeq
using the substitutions 
\beq
\xi = \frac{\Omega t}{2},\ a_z = \frac{4 q}{m \Omega^2}(2 \alpha + 3 \beta d^2),\ q_z = -\frac{4 q \alpha_z}{m \Omega^2} , 
\eeq
where we assume the d.c. octupole is such that $\beta d^2 \sim \alpha$, but the a.c. octupole term can be neglected since $\beta_r \sim \alpha_z/a^2 \ll  \alpha_z/d^2$.  
The secular frequency associated with the solutions of equation (\ref{eq:Matt}) is given by
\beq
\omega_c^2 & = & \left(\frac{\Omega}{2}\right)^2 (a_z + q_z^2/2 ) \\
& = & \frac{q}{m} \left(2 \alpha + 3 \beta d^2 + \frac{|q_z \alpha_z|}{2}\right). \label{omegac}
\eeq
This equation should be compared with (\ref{omega1}). We can obtain (\ref{omegac}) from
(\ref{omega1}) by the substitution 
\beq
\alpha \rightarrow \alpha' = \alpha +  \frac{1}{4}|q_z \alpha_z|
\eeq
The same result is obtained by treating the effect of the oscillating part of $V(0,0,z)$ using a static
pseuodopotential.

In the electrode structures discussed in this paper, the oscillating part of the
potential has a quadrupole form with axial coefficient always less than half
the radial coefficients, $|\alpha_z| < (1/2) |\alpha_r|$. This means that
if we choose the radial Mathieu $q$-parameter $q_r < 0.3$, then $|q_z| < 0.15$. 
In the most extreme case we find that to obtain $\alpha'=0$ we 
require $\alpha \sim -0.04 |\alpha_r|$.
We find that this requires only a small adjustment to the voltages, compared
to the values required for $\alpha=0$.

\section*{Appendix 2: Basic octupole constructions}
\noindent

We list some combinations of point and line charges that produce octupoles
using a minimal number of separate points or lines.

\subsection{Systems of point charges}
\noindent

8 point charges $q$ at the corners of a regular cube of side $2 a$
produce an octupole having cubic symmetry (\ref{Vcube})
with $\beta = 7 q/81 \sqrt{3} \pi \epsilon_0 a^5$.

If we search for a system of 8 point charges all lying in a single plane $y=0$, the
only solution is the one produced by the construction (\ref{rhobar}), (\ref{rhotilde}).
This is shown in figure \ref{fig:linech}a.
In the positive quadrant there is a charge $q$ at $(x,z) = (a,d)$ and
a charge $-f^3 q$ at $(fa, fd)$. For example, $a=d$ gives
$\beta = (-13 q / 128 \sqrt{2} \pi \epsilon_0 a^5)(1-1/f^2)$;
the coefficients of $x^4, y^4$ and $z^4$ are then in the ratios $13:-12:13$, and the
potential is a combination of axial and 2D octupoles in the ratio $-24:35$.
When $d=a/\sqrt{2}$ then
each set of 4 charges on its own
produces a potential with $\partial^2 V/\partial z^2(0,0,0) = 0$, and the two sets
as described produce equal and opposite $\partial^2 V/\partial x^2(0,0,0)$. This gives
$\beta = (-14 \sqrt{2/3 \,} q / 81 \pi \epsilon_0 a^5)(1-1/f^2)$;
the coefficients of $x^4, y^4$ and $z^4$ are in the ratios $26:-54:56$. Similar remarks
apply with $x$ and $z$ swapped when $d=\sqrt{2} a$.

\subsection{Systems of ring and line charges}
\noindent

The potential produced by a semi-infinite
straight line charge is $V = (\lambda/2 \pi \epsilon_0) \ln(r+z)$,
in a coordinate system such that the line charge lies on the negative
$z$ axis and ends at the origin ($r=(x^2 + y^2 + z^2)^{1/2}$).

\begin{figure} [htbp]
\centerline{\epsfig{file=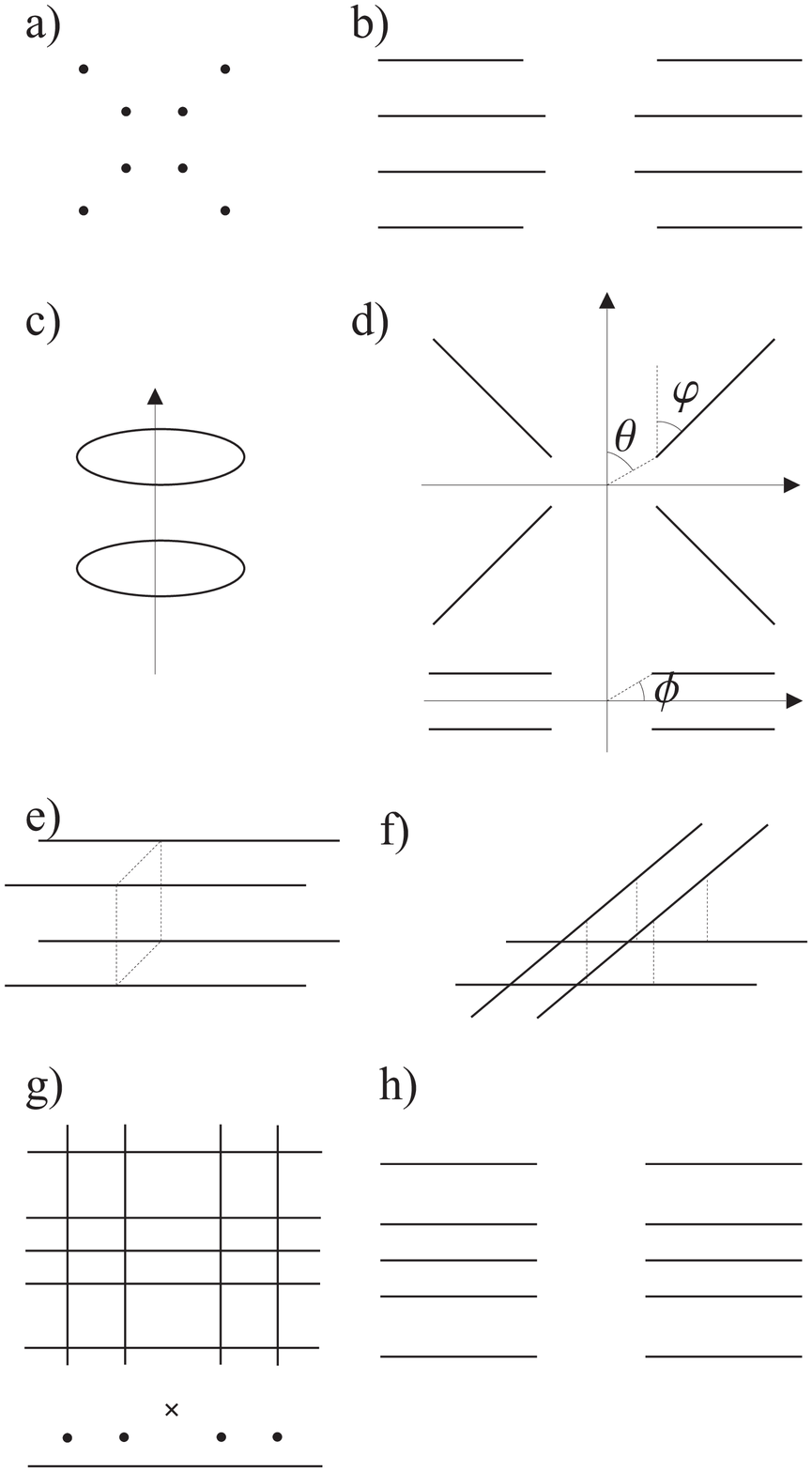, width=6.0cm}} 
\vspace*{13pt}
\fcaption{\label{fig:linech}Examples of octupole contructions using line charges.}
\end{figure}

An axially symmetric octupole (\ref{Vaxoct}) can be created by a single pair of identical
ring charges on a common axis, whose separation is $\sqrt{2}$ times
larger than their radius (shown in figure \ref{fig:linech}c). For charge per unit length $\lambda$ and radius $a$,
$\beta = -14 \sqrt{2/3} \lambda /81 \epsilon_0 a^4$. This charge distribution can also be interpreted as a rotation about
the $z$ axis of point charges at the `special' location $x = z/\sqrt{2}$ which
we identified in the preceeding section.

Addition of a further pair of rings allows a solution at any values of the
ring separations, by adjusting the free parameter of the relative charge
per unit length.

A set of 8 identical line charges, ending at the corners of a regular
cube of side $2a$, and extending outwards along the extended major diagonals, produces
an octupole having cubic symmetry (\ref{Vcube})
with $\beta = -56 \lambda/ 2592 \pi \epsilon_0 a^4$, where
$\lambda$ is the charge per unit length.

Next let us consider two sets of 4 line charges lying in two parallel planes
(construction (E) of section \ref{s:methods}).
To reduce the number of parameters, we consider semi-infinite lines.
We take the origin at the centre of symmetry, and orient axes so
that the two planes are parallel to the $x-z$ plane.
Apart from an overall scale factor, there are then three parameters
which describe the layout: the angular coordinates of the end of one of the lines,
which may be specified for example by
spherical polar angles $\theta$, $\phi$, and the angle $\varphi$
between the line charge and the $z$-direction. This is shown in figure \ref{fig:linech}d.
Since we have three parameters ($\theta, \phi, \varphi$) and two constraints (\ref{constraint}),
one parameter may be chosen arbitrarily and the others adjusted to find
a solution. For example in the construction (E)
we pick $\varphi$ and then adjust $\theta$ and $\phi$. We will list
some example solutions. For convenience, we specify $\theta,\phi$ by giving the
rectangular coordinates $(x,y,z)$ of the end of the line lying in the positive octant,
with an arbitrary scale factor. Example solutions are $(x,y,z,\varphi) =$
$(1,1,0,0)$, $(1,1.434,0.578,\pi/6)$, $(1,1.799,1,\pi/4)$, $(1,2.482,1.731,\pi/3)$,
$(0,1,1,\pi/2)$.

A set of 4 parallel infinite line charges passing through the corners of a square
in a plane perpendicular to them (figure \ref{fig:linech}e) produces a 
2D octupole (\ref{V2Doct})
with $\beta = \lambda/8 \pi \epsilon_0 a^4$, where the square has
side $2 a$ (see figure \ref{fig:linech}e).

A set of 4 infinite line charges passing along the sides of a cube as
in figure \ref{fig:linech}f produces an octupole having $\beta = -3 \lambda/4 \pi \epsilon_0 a^4$ , at the centre of
the cube, where the cube has side $2 a$.

Next, consider sets of semi-infinite line charges all lying in a single plane.
Let the plane be the $x-z$ plane and
assume symmetry under reflections in the $x$ and $z$ axes.
First suppose there is only a single
line charge in the positive quadrant (so 4 line charges in total). The
geometry has two parameters: the angle $\theta$ between the $z$ axis
and a vector from the origin to the end of the line, and the angle $\varphi$
between the line itself and the $z$-direction. This suggests it might be possible
to obtain $\partial^2 V/\partial x^2 = \partial^2 V/\partial z^2 = 0$ by a
good choice of $\theta$ and $\varphi$, but the pair of simultaneous equations
has no solution. 
The construction (D) (eq. (\ref{rhotilde})) leads to an infinite
set of solutions with 8 electrodes in two groups of four.
The two groups have the same values of $\theta$ and $\varphi$;
the charge per unit length
on the outer group is greater than that on the inner group
by a factor $f^2$. There are further solutions when the two
groups have different angles. As an example case, shown in figure \ref{fig:linech}b, 
$\theta = \pi/4$, $\varphi=\pi$ (line charges parallel to the
$x$ axis and finishing on the line $x=z$) gives
$\beta=-(15 \lambda (16 + 11\sqrt{2})/ 16 \pi \epsilon_0 (1 + \sqrt{2})^4 a^4) (1-1/f^2)$. 


\begin{thebibliography}{10}

\bibitem{98:WinelandB}
D.~J. Wineland, C.~Monroe, W.~M. Itano, D.~Leibfried, B.~E. King, and D.~M.
  Meekhof.
\newblock Experimental issues in coherent quantum-state manipulation of trapped
  atomic ions.
\newblock {\em J. Res. Natl. Inst. Stand. Technol.}, 103:259--328, 1998.

\bibitem{97:Steane}
A.~M.~Steane
\newblock The ion trap quantum information processor.
\newblock {\em App. Phys.}, B64, 623--642, (1997)

\bibitem{02:Sasura}
M.~Sasura, V.~Buzek
\newblock Cold Trapped Ions as Quantum Information Processors
\newblock {\em J. Mod. Opt.}, 49, 1593--1647 (2002)

\bibitem{03:SchmidtKaler}
F.~Schmidt-Kaler, H.~Häffner, M.~Riebe, S.~Gulde, G.~P.~T.~Lancaster, T.~Deuschle, C.~Becher, C.~F.~Roos, J.~Eschner and R.~Blatt.
\newblock Realization of the Cirac-Zoller controlled-NOT quantum gate.
\newblock{\em Nature}, 422:408--411, March 2003

\bibitem{03:Leibfried}
D.~Leibfried, B.~DeMarco, V.~Meyer, D.~Lucas, M.~Barrett, J.~Britton, W.~M.~Itano, B.~Jelenkovic, C.~Langer, T.~Rosenband and D.~J.~Wineland.
\newblock{Experimental demonstration of a robust, high-fidelity geometric two ion-qubit phase gate}
\newblock{\em Nature}, 422:412--415, March 2003

\bibitem{00:SackettExptEnt4}
C.~A. Sackett, D.~Kielpinski, B.~E. King, C.~Langer, V.~Meyer, C.~J. Myatt,
  M.~Rowe, Q.~A. Turchette, W.~M. Itano, D.~J. Wineland, and C.~Monroe.
\newblock Experimental entanglement of four particles.
\newblock {\em Nature}, 404:256--258, March 2000.

\bibitem{04:Riebe}
M.~Riebe, H.~Häffner, C.~F. Roos, W.~Hänsel, J.~Benhelm, G.~P.~T. Lancaster,
  T.~W. Körber, C.~Becher, F.~Schmidt-Kaler, D.~F.~V. James, and R.~Blatt.
\newblock Deterministic quantum teleportation with atoms.
\newblock {\em Nature}, 429:734--737, 2004.

\bibitem{04:Roos}
C.~F. Roos, M.~Riebe, H.~Haffner, W.~Hansel, J.~Benhelm, G.~P.~T. Lancaster,
  C.~Becher, F.~Schmidt-Kaler, and R.~Blatt.
\newblock Control and measurement of three-qubit entangled states.
\newblock {\em Science}, 304(1478), 2004.

\bibitem{02:KielpinskiArchitec}
D.~Kielpinski, C.Monroe, and D.~Wineland.
\newblock Architecture for a large-scale ion-trap quantum computer.
\newblock {\em Nature}, 417:709--711, June 2002.

\bibitem{02:SteaneB}
A.~M. Steane.
\newblock Quantum computer architecture for fast entropy extraction.
\newblock {\em Quant. Inf. and Comp.}, 2:297--306, 2002.
\newblock quant-ph/0203047.

\bibitem{02:Rowe}
M.~A. Rowe, A.~Ben-Kish, B.~DeMarco, D.~Leibfried, V.~Meyer, J.~Beall,
  J.~Britton, J.~Hughes, W.~M. Itano, B.~Jelenkovi\'{c}, C.~Langer,
  T.~Rosenband, and D.~J. Wineland.
\newblock Transport of quantum states and separation of ions in a dual rf ion
  trap.
\newblock {\em Quantum Information and Computation}, 2(4):257, 2002.

\bibitem{04:Barrett}
M.~D. Barrett, J.~Chiaverini, T.~Schaetz, J.~Britton, W.~M. Itano J.~D.
  Jost, E.~Knill, C.~Langer, D.~Leibfried, R.~Ozeri, and D.~J. Wineland.
\newblock Deterministic quantum teleportation of atomic qubits.
\newblock {\em Nature}, 429:737--739, 2004.

\bibitem{05:Hensinger}
W.~K.~Hensinger, S.~Olmschenk, D.~Stick, D.~Hucul, M.~Yeo, M.~Acton, 
L.~Deslauriers, C.~Monroe, and J.~Rabchuk. 
\newblock T-junction ion trap array for two-dimensional shuttling, storage and manipulation.
\newblock quant-ph/0508097

\bibitem{00:TurchetteHeating}
Q.~A. Turchette, D.~Kielpinski, B.~E. King, D.~Leibfried, D.~M. Meekhof, C.~J.
  Myatt, M.~A. Rowe, C.~A. Sackett, C.~S. Wood, W.~M. Itano, C.~Monroe, and
  D.~J. Wineland.
\newblock Heating of trapped ions from the quantum ground state.
\newblock {\em Phys. Rev.}, A61:063418, 2000.

\bibitem{04:Deslauriers}
L.~Deslauriers, P.~C. Haljan, P.~J. Lee, K.-A. Brickman, B.~B. Blinov, M.~J.
  Madsen, and C.~Monroe.
\newblock Zero-point cooling and low heating of trapped cd ions.
\newblock {\em Phys. Rev. A}, 2004.
\newblock quant-ph/0404142.

\bibitem{04:Madsen}
M.~J. Madsen, W.~K. Hensinger, D.~Stick, J.~A. Rabchuk, and C.~Monroe.
\newblock Planar ion trap geometry for microfabrication.
\newblock {\em Appl. Phys. B}, 78:639--651, 2004.

\bibitem{03:Sasura}
M.~\v{S}a\v{s}ura and A.~M. Steane.
\newblock Fast quantum logic by selective displacement of trapped ions.
\newblock {\em Phys. Rev.}, A67(062318), 2003.

\bibitem{00:Sorenson}
A.~S{\o}renson and K.~M{\o}lmer.
\newblock Entanglement and quantum computation with ions in thermal motion.
\newblock {\em Phys. Rev.}, A62(022311), July 2000.

\bibitem{03:Leibfriedreview}
D. Leibfried, R. Blatt, C. Monroe, D.~J Wineland
\newblock Quantum dynamics of single trapped ions.
\newblock {\em Rev. Mod. Phys.}, 75, 281--324, 2003

\bibitem{67:DehmeltRadiofreq}
H.~G. Dehmelt.
\newblock Radiofrequency spectroscopy of stored ions.
\newblock {\em Adv. At. Mol. Phys}, 3(53), 1967.

\bibitem{Bk:Ghosh}
Prahdip~K. Ghosh.
\newblock {\em Ion Traps}.
\newblock Clarendon Press, Oxford, 1995.

\bibitem{04:Cruz}
D.~Cruz, J.P. Chang, and M.~G. Blain.
\newblock Field emission characteristics of a tungsten micro-electrical
  mechanical system device.
\newblock {\em preprint}, ., 2004.

\bibitem{Bk:Gomer}
R.~Gomer.
\newblock {\em Field Emission and Field Ionization}.
\newblock Harvard University Press, Cambridge, 1961.

\bibitem{pk:simion7}
Idaho~National Engineering and Environmental Laboratory.
\newblock Simion 3d version 7.0, 2000.

\bibitem{pk:CPO}
CPO Ltd.
\newblock Charged particle optics programs.

\bibitem{04:Reid}
J.R. Reid and R.T. Webster.
\newblock A 60 ghz branch line coupler fabricated using integrated rectangular
  coaxial lines.
\newblock {\em IEEE}, 2004.

\bibitem{04:Blain}
M.~G. Blain, L.~S. Riter, D.~Cruz, D.~E. Austin, G.~Wu, W.~R. Plass, and
  R.~Graham Cooks.
\newblock Towards the hand-held mass spectrometer: design considerations,
  simulation, and fabrication fo micrometer-scaled cylindrical ion traps.
\newblock {\em International Journal of Mass Spectrometry}, 236:91--104, 2004.

\end{thebibliography}

\end{document}